\documentclass[conference, draftcls, onecolumn]{IEEEtran} %, onecolumn, draft
\usepackage{cite}
\ifCLASSINFOpdf
\else
\fi
\usepackage{amsmath}
  \usepackage[caption=false,font=footnotesize]{subfig}
\usepackage{url}
\usepackage{amssymb}
\usepackage{amsthm}

\theoremstyle{plain}
\newtheorem{definition}{Definition}
\newtheorem{lemma}{Lemma}

\newtheorem{theorem}{Theorem}
\newtheorem{corollary}{Corollary}
\newtheorem{remark}{Remark}

\usepackage[pdftex]{graphicx}
\usepackage{subfig}
\usepackage{overpic}
\usepackage{pict2e}
\usepackage{fixltx2e}

\usepackage{color}
\definecolor{burgundy}{rgb}{0.545098,0,0}
\definecolor{navyblue}{rgb}{0.0, 0.0, 0.5}
\definecolor{leafgreen}{rgb}{0.290196, 0.470588, 0.0}
\definecolor{bluegreen}{rgb}{0, 0.470588, 0.415686}
\definecolor{zuhl}{rgb}{0.1875, 0.26171875, 0.46484375}
\definecolor{orange}{rgb}{1, 0.6470588235, 0}

\newcommand{\bvec}[1]{\mbox{\boldmath $#1$}}
\newcommand{\sbvec}[1]{\mbox{\scriptsize \boldmath $#1$}}
\newcommand{\sgn}{\operatorname{sgn}}

\allowdisplaybreaks[3]

\begin{document}
\title{Extremal Relations Between \\ Shannon Entropy and $\ell_{\alpha}$-Norm}

%\author{\IEEEauthorblockN{Yuta Sakai and Ken-ichi Iwata}
%\IEEEauthorblockA{Department of Information Science, University of Fukui, Japan, E-mail: \url{{ji140117, k-iwata}@u-fukui.ac.jp}}
%}
%\and
\author{\IEEEauthorblockN{Yuta Sakai and Ken-ichi Iwata}
\IEEEauthorblockA{Department of Information Science, University of Fukui, \\
3-9-1 Bunkyo, Fukui, Fukui, 910-8507, Japan, \\
E-mail: \url{{ji140117, k-iwata}@u-fukui.ac.jp}}
}
%\and
%\IEEEauthorblockN{James Kirk\\ and Montgomery Scott}
%\IEEEauthorblockA{Starfleet Academy\\
%San Francisco, California 96678--2391\\


%Telephone: (800) 555--1212, Fax: (888) 555--1212}


\maketitle

\begin{abstract}
%The abstract goes here.
%To be considered for the 2016 IEEE Jack Keil Wolf ISIT Student Paper Award.
The paper examines relationships between the Shannon entropy and the $\ell_{\alpha}$-norm for $n$-ary probability vectors, $n \ge 2$.
More precisely, we investigate the tight bounds of the $\ell_{\alpha}$-norm with a fixed Shannon entropy, and vice versa.
As applications of the results, we derive the tight bounds between the Shannon entropy and several information measures which are determined by the $\ell_{\alpha}$-norm. %, e.g., R\'{e}nyi entropy, Tsallis entropy, the $R$-norm information, and some diversity indices.
Moreover, we apply these results to uniformly focusing channels. %a particular class of discrete memoryless channels, i.e., uniformly focusing channels.
Then, we show the tight bounds of Gallager's $E_{0}$ functions with a fixed mutual information under a uniform input distribution.
\end{abstract}


% \fi
\IEEEpeerreviewmaketitle

\section{Introduction}

Information measures of random variables are used in several fields. %: information theory, probability theory, statistics, pattern recognition, cryptology, machine learning, and others.
The Shannon entropy \cite{shannon} is one of the famous measures of uncertainty for a given random variable.
On the studies of information measures, inequalities for information measures are commonly used in many applications.
As an instance, Fano's inequality \cite{fano} gives the tight upper bound of the conditional Shannon entropy with a fixed error probability.
Then, note that the \emph{tight} means the existence of the distribution which attains the equality of the bound.
Later, the reverse of Fano's inequality, i.e., the tight lower bound of the conditional Shannon entropy with a fixed error probability, are established \cite{kovalevsky, tebbe, feder}. %by Kovalevsky \cite{kovalevsky},  Tebbe and Dwyer \cite{tebbe}, and Feder and Merhav \cite{feder}.
%In addition, Ho and Verd\'{u} \cite{verdu} proved the tight lower bound of the Shannon entropy with a fixed error probability for random variables on the countably infinite alphabet by using majorization theory \cite{marshall}.
On the other hand, Harremo\"{e}s and Tops{\o}e \cite{topsoe} derived the exact range between the Shannon entropy and the index of coincidence (or the Simpson index) for all $n$-ary probability vectors, $n \ge 3$. %;
%moreover, their results \cite{topsoe} contains the exact range between the Shannon entropy and the R\'{e}nyi entropy \cite{renyi} of order 2.
In the above studies, note that the error probability and the index of coincidence are closely related to $\ell_{\infty}$-norm and $\ell_{2}$-norm, respectively.
Similarly, several axiomatic definitions of the entropies \cite{renyi, tsallis2, havrda, daroczy, behara, boekee} are also related to the $\ell_{\alpha}$-norm. %, e.g., R\'{e}nyi entropy \cite{renyi}, Tsallis entropy \cite{tsallis2}, entropy of type-$\beta$ \cite{havrda, daroczy}, $\gamma$-entropy \cite{behara}, and the $R$-norm information \cite{boekee},.
Furthermore, the $\ell_{\alpha}$-norm are also related to some diversity indices, such as the index of coincidence.

In this study, we examine extremal relations between the Shannon entropy and the $\ell_{\alpha}$-norm for $n$-ary probability vectors, $n \ge 2$.
More precisely, we establish the tight bounds of $\ell_{\alpha}$-norm with a fixed Shannon entropy in Theorem \ref{th:extremes}.
Similarly, we also derive the tight bounds of the Shannon entropy with a fixed $\ell_{\alpha}$-norm in Theorem \ref{th:extremes2}.
%Theorems \ref{th:extremes} and \ref{th:extremes2} also show the extremal distributions $\bvec{v}_{n}( \cdot )$ and $\bvec{w}_{n}( \cdot )$, defined in Section \ref{subsect:vw}, which attain the equalities of these bounds.
%These extremal distributions are defined by the $n$-ary probability vectors $\bvec{v}_{n}( \cdot )$ and $\bvec{w}_{n}( \cdot )$ in Section \ref{subsect:vw}.
Directly extending Theorem \ref{th:extremes} to Corollary \ref{cor:extremes}, we can obtain the tight bounds of several information measures which are determined by the $\ell_{\alpha}$-norm with a fixed Shannon entropy, as shown in Table \ref{table:extremes}.
%We describe this extension of Theorem \ref{th:extremes} in Corollary \ref{cor:extremes} and its applications are shown in Table \ref{table:extremes}.
In particular, we illustrate the exact feasible regions between the Shannon entropy and the R\'{e}nyi entropy in Fig. \ref{fig:Renyi} by using \eqref{eq:Renyi_bound1} and \eqref{eq:Renyi_bound2}.
In Section \ref{subsect:focusing}, we consider applications of Corollary \ref{cor:extremes} for a particular class of discrete memoryless channels, defined in Definition \ref{def:focusing}, which is called \emph{uniformly focusing} \cite{massey} or \emph{uniform from the output} \cite{fano2}.
%Corollary \ref{cor:RenyiDiv} of Section \ref{subsect:focusing} shows the tight bounds of the R\'{e}nyi divergence \cite{renyi} from a uniform distribution with a fixed relative entropy from a uniform distribution.
%Using Corollary \ref{cor:RenyiDiv}, we derive the tight bounds of the mutual information of order $\alpha$ \cite{arimoto} for uniformly focusing channels with a fixed (ordinary) mutual information under a uniform input distribution.
%Since Gallager's $E_{0}$ function \cite{gallager} is closely related to the mutual information of order $\alpha$, Theorem \ref{th:E0_focusing} provides the tight bounds of the $E_{0}$ function for uniformly focusing channels with a fixed mutual information under a uniform input distribution.

%All of the proofs of this paper will be submitted to arXiv.

%The rest of this paper is organized as follows:
%Section \ref{sect:pre} introduces the definitions which are used in this study.
%Section \ref{sect:result} shows the extremal relations between the Shannon entropy and the $\ell_{\alpha}$-norm and its applications.
%Section \ref{sect:conclusion} concludes this study.

\section{Preliminaries}
\label{sect:pre}

\subsection{$n$-ary probability vectors and its information measures}

%In this subsection, we introduce the probability distributions, the Shannon entropy, and the $\ell_{\alpha}$-norm, studied in the paper.
Let the set of all $n$-ary probability vectors be denoted by
\begin{align}
\mathcal{P}_{n} \!
\triangleq \!
\left\{ (p_{1}, p_{2}, \dots, p_{n}) \in \mathbb{R}^{n} \left| \; p_{j} \ge 0 \ \mathrm{and} \ \sum_{i=1}^{n} p_{i} = 1 \! \right. \right\}
\end{align}
for an integer $n \ge 2$.
For $\bvec{p} = (p_{1}, p_{2}, \dots, p_{n}) \in \mathcal{P}_{n}$, let
\begin{align}
p_{[1]} \ge p_{[2]} \ge \dots \ge p_{[n]}
\end{align}
denote the components of $\bvec{p}$ in decreasing order, and let
\begin{align}
\bvec{p}_{\downarrow}
\triangleq
(p_{[1]}, p_{[2]}, \dots, p_{[n]})
\label{def:rearrangement}
\end{align}
denote the decreasing rearrangement%
\footnote{This rearrangement is denoted by reference to the notation of \cite{marshall}.}
of $\bvec{p}$.
In particular, we define the following two $n$-ary probability vectors:
(i) an $n$-ary deterministic distribution
\begin{align}
\bvec{d}_{n}
\triangleq
(d_{1}, d_{2}, \dots, d_{n}) \in \mathcal{P}_{n}
\end{align}
is defined by $d_{1} = 1$ and $d_{i} = 0$ for $i \in \{ 2, 3, \dots, n \}$ and
(ii) the $n$-ary equiprobable distribution
\begin{align}
\bvec{u}_{n}
\triangleq
(u_{1}, u_{2}, \dots, u_{n}) \in \mathcal{P}_{n}
\end{align}
is defined by $u_{i} = \frac{1}{n}$ for $i \in \{ 1, 2, \dots, n \}$.

For an $n$-ary random variable $X \sim \bvec{p} \in \mathcal{P}_{n}$, we define the Shannon entropy \cite{shannon} as
\begin{align}
H( X )
=
H( \bvec{p} )
\triangleq
- \sum_{i=1}^{n} p_{i} \ln p_{i} ,
\end{align}
where $\ln$ denotes the natural logarithm and assume that
%\footnote{This assumption is based on the limiting value $\lim_{x \to 0^{+}} x \ln x = 0$.}
$0 \ln 0 = 0$.
%The Shannon entropy is used to a measure of uncertainly of a random variable.
Moreover, we define the $\ell_{\alpha}$-norm of $\bvec{p} \in \mathcal{P}_{n}$ as
\begin{align}
\| \bvec{p} \|_{\alpha}
\triangleq
\left( \sum_{i=1}^{n} p_{i}^{\alpha} \right)^{\frac{1}{\alpha}}
\end{align}
for $\alpha \in (0, \infty)$.
Note that $\lim_{\alpha \to \infty} \| \bvec{p} \|_{\alpha} = \| \bvec{p} \|_{\infty} \triangleq \max \{ p_{1}, p_{2}, \dots, p_{n} \}$ for $\bvec{p} \in \mathcal{P}_{n}$.
On the works of extending Shannon entropy, the $\ell_{\alpha}$-norm is appear in the several information measures.
As an instance, R\'{e}nyi \cite{renyi} generalized the Shannon entropy axiomatically to the R\'{e}nyi entropy of order $\alpha \in (0, 1) \cup (1, \infty)$, defined as
\begin{align}
H_{\alpha}( X )
=
H_{\alpha}( \bvec{p} )
\triangleq
\frac{ \alpha }{ 1 - \alpha } \ln \| \bvec{p} \|_{\alpha}
\label{def:renyi}
\end{align}
for $X \sim \bvec{p} \in \mathcal{P}_{n}$.
Note that it is usually defined that $H_{1}( X ) \triangleq H(X)$ since $\lim_{\alpha \to 1} H_{\alpha}(X) = H(X)$ by L'H\^{o}pital's rule.
In other axiomatic definitions of entropies \cite{tsallis2, havrda, daroczy, behara, boekee}, we can also define them by using the $\ell_{\alpha}$-norm, as with \eqref{def:renyi}.

In this study, we analyze relations between $H( \bvec{p} )$ and $\| \bvec{p} \|_{\alpha}$ to examine relationships between the Shannon entropy and several information measures.
Note that $H( \bvec{p} )$ and $\| \bvec{p} \|_{\alpha}$ are invariant for any permutation of the indices of $\bvec{p} \in \mathcal{P}_{n}$;
that is,
\begin{align}
H( \bvec{p} ) = H( \bvec{p}_{\downarrow} )
\qquad \mathrm{and} \qquad
\| \bvec{p} \|_{\alpha} = \| \bvec{p}_{\downarrow} \|_{\alpha}
\end{align}
for any $\bvec{p} \in \mathcal{P}_{n}$.
%Hence, instead of $H( \bvec{p} )$ and $\| \bvec{p} \|_{\alpha}$, it is enough to consider only $H( \bvec{p}_{\downarrow} )$ and $\| \bvec{p}_{\downarrow} \|_{\alpha}$ in this study.
Hence, we only consider $\bvec{p}_{\downarrow}$ for $\bvec{p} \in \mathcal{P}_{n}$ in the analyses of the study.
Since $\| \bvec{p} \|_{1} = 1$ for any $\bvec{p} \in \mathcal{P}_{n}$, we have no interest in the case $\alpha = 1$;
hence, we omit the case $\alpha = 1$ in this study.
Furthermore, since
%\footnote{The relations \eqref{eq:uniform} and \eqref{eq:deterministic} can be verified by the convexities of $H( \bvec{p} )$ and $\| \bvec{p} \|_{\alpha}$ in $\bvec{p} \in \mathcal{P}_{n}$.}
\begin{align}
H( \bvec{p} ) = \ln n
& \iff
\| \bvec{p} \|_{\alpha} = n^{\frac{1}{\alpha}-1}
\iff
\bvec{p}
=
\bvec{u}_{n} ,
\label{eq:uniform} \\
H( \bvec{p} ) = 0
& \iff
\| \bvec{p} \|_{\alpha} = 1
\iff
\bvec{p}_{\downarrow}
=
\bvec{d}_{n} ,
\label{eq:deterministic}
\end{align}
the cases $\bvec{p} = \bvec{u}_{n}$ and $\bvec{p}_{\downarrow} = \bvec{d}_{n}$ are trivial;
thus, we also omit these cases in the analyses of this study.

\subsection{Properties of two distributions $\bvec{v}_{n}( \cdot )$ and $\bvec{w}_{n}( \cdot )$}
\label{subsect:vw}

For a fixed $n \ge 2$, %we define the following two distributions:
let the $n$-ary distribution
$
\bvec{v}_{n}( p )
\triangleq
(v_{1}(p), v_{2}(p), \dots v_{n}(p)) \in \mathcal{P}_{n}
$
be defined by
\begin{align}
v_{i}( p )
=
\begin{cases}
1 - (n-1) p
& \mathrm{if} \ i = 1 , \\
p
& \mathrm{otherwise}
\end{cases}
\end{align}
for $p \in [0, \frac{1}{n-1}]$, and let the $n$-ary distribution%
\footnote{The definition of $\bvec{w}_{n}( \cdot )$ is similar to the definition of \cite[Eq. (26)]{verdu}.}
$
\bvec{w}_{n}( p )
\triangleq
(w_{1}( p ), w_{2}( p ), \dots, w_{n}( p )) \in \mathcal{P}_{n}
$
be defined by
\begin{align}
w_{i}( p )
=
\begin{cases}
p
& \mathrm{if} \ 1 \le i \le \lfloor p^{-1} \rfloor , \\
1 - \lfloor p^{-1} \rfloor p
& \mathrm{if} \ i = \lfloor p^{-1} \rfloor + 1 , \\
0
& \mathrm{otherwise}
\end{cases}
\end{align}
for $p \in [\frac{1}{n}, 1]$, where $\lfloor \cdot \rfloor$ denotes the floor function. %, i.e., the largest integer which is less that or equal to $x$.
Note that $\bvec{v}_{n}( p )_{\downarrow} = \bvec{w}_{n}( p )$ for $p \in [\frac{1}{n}, \frac{1}{n-1}]$.
In this subsection, we examine the properties of the Shannon entropies and the $\ell_{\alpha}$-norms for $\bvec{v}_{n}( \cdot )$ and $\bvec{w}_{n}( \cdot )$.
For simplicity, we define
\begin{align}
H_{\sbvec{v}_{n}}( p )
& \triangleq
H( \bvec{v}_{n}( p ) )
\\
& =
- (1 - (n-1) p ) \ln (1 - (n-1) p ) - (n-1) p \ln p ,
\\
H_{\sbvec{w}_{n}}( p )
& \triangleq
H( \bvec{w}_{n}( p ) )
\\
& =
- \lfloor p^{-1} \rfloor p \ln p - (1 - \lfloor p^{-1} \rfloor p ) \ln (1 - \lfloor p^{-1} \rfloor p ) .
\end{align}
Then, we first show the monotonicity of $H_{\sbvec{v}_{n}}( p )$ with respect to $p \in [0, \frac{1}{n}]$ in the following lemma.

\begin{lemma}
\label{lem:Hv}
%For any fixed $n \ge 2$, 
$H_{\sbvec{v}_{n}}( p )$ is strictly increasing for $p \in [0, \frac{1}{n}]$.
\end{lemma}

\begin{IEEEproof}[Proof of Lemma \ref{lem:Hv}]
It is easy to see that
\begin{align}
H_{\sbvec{v}_{n}}( p )
& =
- \sum_{i=1}^{n} v_{i}( p ) \ln v_{i}( p )
\\
& =
- v_{1}( p ) \ln v_{1}( p ) - \sum_{i=2}^{n} v_{i}( p ) \ln v_{i}( p )
\\
& =
- (1-(n-1)p) \ln (1-(n-1)p) - \sum_{i=2}^{n} v_{i}( p ) \ln v_{i}( p )
\\
& =
- (1-(n-1)p) \ln (1-(n-1)p) - (n-1) p \ln p .
\end{align}
Then, the first-order derivative of $H_{\sbvec{v}_{n}}( p )$ with respect to $p$ is
\begin{align}
\frac{ \partial H_{\sbvec{v}_{n}}( p ) }{ \partial p }
& =
\frac{ \partial }{ \partial p } \left( \vphantom{\sum} - (n-1) p \ln p - (1 - (n-1) p) \ln (1 - (n-1) p) \right)
\\
& =
- (n-1) \left( \frac{ \mathrm{d} }{ \mathrm{d} p } (p \ln p) \right) - \left( \frac{ \partial }{ \partial p } ((1 - (n-1) p) \ln (1 - (n-1) p)) \right)
\\
& =
- (n-1) \left( \vphantom{\sum} \ln p + 1 \right) + (n-1) \left( \vphantom{\sum} \ln (1 - (n-1)p) + 1 \right)
\\
& =
(n-1) \left( \vphantom{\sum} \ln (1 - (n-1)p) - \ln p \right)
\\
& =
(n-1) \ln \frac{1 - (n-1) p}{p} .
\label{eq:diff1_Hv}
\end{align}
Since $1 - (n-1) p > p > 0$ for $p \in (0, \frac{1}{n})$, it follows from \eqref{eq:diff1_Hv} that
\begin{align}
\frac{ \partial H_{\sbvec{v}_{n}}( p ) }{ \partial p } > 0
\end{align}
for $p \in (0, \frac{1}{n})$.
Note that $H_{\sbvec{v}_{n}}( p )$ is continuous for $p \in [0, \frac{1}{n}]$ since $\lim_{p \to \frac{1}{n}} H_{\sbvec{v}_{n}}( p ) = H_{\sbvec{v}_{n}}( \frac{1}{n} ) = \ln n$ and $\lim_{p \to 0^{+}} H_{\sbvec{v}_{n}}( p ) = H_{\sbvec{v}_{n}}( 0 ) = 0$ by the assumption $0 \ln 0 = 0$.
Therefore, $H_{\sbvec{v}_{n}}( p )$ is strictly increasing for $p \in [0, \frac{1}{n}]$.
\end{IEEEproof}

%It can be seen from the proof of Lemma \ref{lem:Hv} that $H_{\sbvec{v}_{n}}( p )$ takes a similar behavior to the binary entropy function.
Lemma \ref{lem:Hv} implies the existence of the inverse function of $H_{\sbvec{v}_{n}}( p )$ for $p \in [0, \frac{1}{n}]$.
We second show the monotonicity of $H_{\sbvec{w}_{n}}( p )$ with respect to $p \in [\frac{1}{n}, 1]$ as follows:

\begin{lemma}
\label{lem:Hw}
%For any fixed $n \ge 2$, 
$H_{\sbvec{w}_{n}}( p )$ is strictly decreasing for $p \in [\frac{1}{n}, 1]$.
\end{lemma}

\begin{IEEEproof}[Proof of Lemma \ref{lem:Hw}]
For an integer $m \in [2, n]$, assume that $p \in [\frac{1}{m}, \frac{1}{m-1}]$.
Then, note that $\lfloor p^{-1} \rfloor = m$.
It is easy to see that
\begin{align}
H_{\sbvec{w}_{n}}( p )
& =
- \sum_{i=1}^{n} w_{i}( p ) \ln w_{i}( p )
\\
& =
- \sum_{i=1}^{m} w_{i}( p ) \ln w_{i}( p ) - w_{m+1}( p ) \ln w_{m+1}( p ) - \sum_{j=m+2}^{n} w_{j}( p ) \ln w_{j}( p )
\\
& \overset{\text{(a)}}{=}
- \sum_{i=1}^{m} w_{i}( p ) \ln w_{i}( p ) - w_{m+1}( p ) \ln w_{m+1}( p )
\\
& =
- m \, p \ln p - w_{m+1}( p ) \ln w_{m+1}( p )
\\
& =
- m \, p \ln p - (1 - m \, p) \ln (1 - m \, p) ,
\end{align}
where (a) follows by the assumption $0 \ln 0 = 0$.
Then, the first-order derivative of $H_{\sbvec{w}_{n}}( p )$ with respect to $p$ is
\begin{align}
\frac{ \partial H_{\sbvec{w}_{n}}( p ) }{ \partial p }
& =
\frac{ \partial }{ \partial p } \left( \vphantom{\sum} - m \, p \ln p - (1 - m \, p) \ln (1 - m \, p) \right)
\\
& =
- m \left( \frac{ \mathrm{d} }{ \mathrm{d} p } (p \ln p) \right) - \left( \frac{ \partial }{ \partial p } ((1 - m \, p) \ln (1 - m \, p)) \right)
\\
& =
- m \left( \vphantom{\sum} \ln p + 1 \right) + m \left( \vphantom{\sum} \ln (1 - m \, p) + 1 \right)
\\
& =
m \left( \vphantom{\sum} \ln (1 - m \, p) - \ln p \right)
\\
& =
m \ln \frac{ 1 - m \, p }{ p } .
\label{eq:diff1_Hw}
\end{align}
Since $p > 1 - m \, p > 0$ for $p \in (\frac{1}{m}, \frac{1}{m-1})$, it follows from \eqref{eq:diff1_Hw} that
\begin{align}
\frac{ \partial H_{\sbvec{w}_{n}}( p ) }{ \partial p } < 0
\end{align}
for $p \in (\frac{1}{m}, \frac{1}{m-1})$.
On the other hand, we observe that
\begin{align}
\lim_{p \to (\frac{1}{m})^{-}} H_{\sbvec{w}_{n}}( p )
& =
\lim_{p \to (\frac{1}{m})^{-}} \left( \vphantom{\sum} - \lfloor p^{-1} \rfloor p \ln p - (1 - \lfloor p^{-1} \rfloor p) \ln (1 - \lfloor p^{-1} \rfloor p) \right)
\\
& =
\lim_{p \to (\frac{1}{m})^{-}} \left( \vphantom{\sum} - m \, p \ln p - (1 - m \, p) \ln (1 - m \, p) \right)
\\
& =
\ln m - \lim_{p \to (\frac{1}{m})^{-}} \left( \vphantom{\sum} (1 - m \, p) \ln (1 - m \, p) \right)
\\
& =
\ln m - \lim_{x \to 0^{+}} \left( \vphantom{\sum} x \ln x \right)
\\
& =
\ln m
\label{eq:Hw}
\end{align}
for an integer $m \in [1, n-1]$ and
\begin{align}
\lim_{p \to (\frac{1}{m})^{+}} H_{\sbvec{w}_{n}}( p )
& =
\lim_{p \to (\frac{1}{m})^{+}} \left( \vphantom{\sum} - \lfloor p^{-1} \rfloor p \ln p - (1 - \lfloor p^{-1} \rfloor p) \ln (1 - \lfloor p^{-1} \rfloor p) \right)
\\
& =
\lim_{p \to (\frac{1}{m})^{+}} \left( \vphantom{\sum} - (m-1) p \ln p - (1 - (m-1) p) \ln (1 - (m-1) p) \right)
\\
& =
\left( 1 - \frac{1}{m} \right) \ln m - \lim_{p \to (\frac{1}{m})^{+}} \left( \vphantom{\sum} (1 - (m-1) p) \ln (1 - (m-1) p) \right)
\\
& =
\left( 1 - \frac{1}{m} \right) \ln m - \left( - \frac{1}{m} \ln m \right)
\\
& =
\ln m
\end{align}
for an integer $m \in [2, n]$.
Note that $H_{\sbvec{w}_{n}}( \frac{1}{m} ) = \ln m$ from \eqref{eq:Hw} and the assumption $0 \ln 0 = 0$.
Hence, for any integer $m \in [2, n-1]$, we get that
\begin{align}
\lim_{p \to (\frac{1}{n})^{+}} H_{\sbvec{w}_{n}}( p ) & = H_{\sbvec{w}_{n}}( {\textstyle \frac{1}{n}} ) = \ln n
\\
\lim_{p \to \frac{1}{m}} H_{\sbvec{w}_{n}}( p ) & = H_{\sbvec{w}_{n}}( {\textstyle \frac{1}{m}} ) = \ln m ,
\\
\lim_{p \to 1^{-}} H_{\sbvec{w}_{n}}( p ) & = H_{\sbvec{w}_{n}}( 1 ) = 0 ,
\end{align}
which imply that $H_{\sbvec{w}_{n}}( p )$ is continuous for $p \in [\frac{1}{n}, 1]$.
Therefore, $H_{\sbvec{w}_{n}}( p )$ is strictly decreasing for $p \in [\frac{1}{n}, 1]$.
\end{IEEEproof}

%It can be seen from the proof of Lemma \ref{lem:Hw} that $H_{\sbvec{w}_{n}}( p )$ is a piecewise continuous function, composed of $n-1$ segments of the similar types to the binary entropy function.
As with Lemma \ref{lem:Hv}, Lemma \ref{lem:Hw} also implies the existence of the inverse function of $H_{\sbvec{w}_{n}}( p )$ for $p \in [\frac{1}{n}, 1]$.
Since $H_{\sbvec{v}_{n}}( 0 ) = 0$, $H_{\sbvec{v}_{n}}( \frac{1}{n} ) = \ln n$, $H_{\sbvec{w}_{n}}( \frac{1}{n} ) = \ln n$, and $H_{\sbvec{w}_{n}}( 1 ) = 0$, we can denote the inverse functions of $H_{\sbvec{v}_{n}}( p )$ and $H_{\sbvec{w}_{n}}( p )$ with respect to $p$ as follows:
%\begin{definition}
%\label{def:inverseH}
We denote by $H_{\sbvec{v}_{n}}^{-1} : [0, \ln n] \to [0, \frac{1}{n}]$ the inverse function of $H_{\sbvec{v}_{n}}( p )$ for $p \in [0, \frac{1}{n}]$.
Moreover, we also denote by $H_{\sbvec{w}_{n}}^{-1} : [0, \ln n] \to [\frac{1}{n}, 1]$ the inverse function of $H_{\sbvec{w}_{n}}( p )$ for $p \in [\frac{1}{n}, 1]$.
%\end{definition}

Now, we provide the monotonicity of $\| \bvec{v}_{n}( p ) \|_{\alpha}$ with respect to $H_{\sbvec{v}_{n}}( p )$ in the following lemma.

\begin{lemma}
\label{lem:mono_v}
For any fixed $n \ge 2$ and any fixed $\alpha \in (-\infty, 0) \cup(0, 1) \cup (1, \infty)$, if $p \in [0, \frac{1}{n}]$, the following monotonicity hold:
\begin{itemize}
\item[(i)]
if $\alpha > 1$, then $\| \bvec{v}_{n}( p ) \|_{\alpha}$ is strictly decreasing for $H_{\sbvec{v}_{n}}( p ) \in [0, \ln n]$ and
\item[(ii)]
if $\alpha < 1$, then $\| \bvec{v}_{n}( p ) \|_{\alpha}$ is strictly increasing for $H_{\sbvec{v}_{n}}( p ) \in [0, \ln n]$.
\end{itemize}
\end{lemma}

\begin{IEEEproof}[Proof of Lemma \ref{lem:mono_v}]
The proof of Lemma \ref{lem:mono_v} is given in a similar manner with \cite[Appendix I]{fabregas}.
By the chain rule of the derivation and the inverse function theorem, we have
\begin{align}
\frac{ \partial \| \bvec{v}_{n}( p ) \|_{\alpha} }{ \partial H_{\sbvec{v}_{n}}( p ) }
& =
\left( \frac{ \partial \| \bvec{v}_{n}( p ) \|_{\alpha} }{ \partial p } \right) \cdot \left( \frac{ \partial p }{ \partial H_{\sbvec{v}_{n}}( p ) } \right)
\\
& =
\left( \frac{ \partial \| \bvec{v}_{n}( p ) \|_{\alpha} }{ \partial p } \right) \cdot \left( \frac{ 1 }{ \frac{ \partial H( \sbvec{v}_{n}( p ) ) }{ \partial p } } \right) .
\label{eq:diff1}
\end{align}
Direct calculation shows
\begin{align}
\frac{ \partial \| \bvec{v}_{n}( p ) \|_{\alpha} }{ \partial p }
& =
\frac{ \partial }{ \partial p } \left( \vphantom{\sum} (n-1) \, p^{\alpha} + (1 - (n-1)p)^{\alpha} \right)^{\frac{1}{\alpha}}
\\
& =
\frac{1}{\alpha} \left( \vphantom{\sum} (n-1) \, p^{\alpha} + (1 - (n-1)p)^{\alpha} \right)^{\frac{1}{\alpha} - 1} \left( \frac{ \partial }{ \partial p } \left( \vphantom{\sum} (n-1) \, p^{\alpha} + (1 - (n-1)p)^{\alpha} \right) \right)
\\
& =
\frac{1}{\alpha} \left( \vphantom{\sum} (n-1) \, p^{\alpha} + (1 - (n-1)p)^{\alpha} \right)^{\frac{1}{\alpha} - 1} \left( \alpha (n-1) \left( \vphantom{\sum} p^{\alpha-1} - (1 - (n-1)p)^{\alpha-1} \right) \right)
\\
& =
(n-1) \left( \vphantom{\sum} (n-1) \, p^{\alpha} + (1 - (n-1)p)^{\alpha} \right)^{\frac{1}{\alpha} - 1} \left( \vphantom{\sum} p^{\alpha-1} - (1 - (n-1)p)^{\alpha-1} \right) .
\label{eq:norm_diff1}
\end{align}
Substituting \eqref{eq:diff1_Hv} and \eqref{eq:norm_diff1} into \eqref{eq:diff1}, we obtain
\begin{align}
&
\frac{ \partial \| \bvec{v}_{n}( p ) \|_{\alpha} }{ \partial H_{\sbvec{v}_{n}}( p ) }
\notag \\
& \quad =
(n-1) \left( \vphantom{\sum} (n-1) \, p^{\alpha} + (1 - (n-1)p)^{\alpha} \right)^{\frac{1}{\alpha} - 1} \left( \vphantom{\sum} p^{\alpha-1} - (1 - (n-1)p)^{\alpha-1} \right) \left( \frac{ 1 }{ (n-1) \ln \frac{ 1 - (n-1) p}{ p } } \right)
\\
& \quad =
\left( \vphantom{\sum} (n-1) \, p^{\alpha} + (1 - (n-1)p)^{\alpha} \right)^{\frac{1}{\alpha} - 1} \left( \vphantom{\sum} p^{\alpha-1} - (1 - (n-1)p)^{\alpha-1} \right) \frac{ 1 }{ \ln \frac{ 1 - (n-1) p}{ p } } .
\label{eq:diff1_N_H_v}
\end{align}
We now define the sign function as
\begin{align}
\sgn( x )
\triangleq
\begin{cases}
1
& \mathrm{if} \ x > 0 , \\
0
& \mathrm{if} \ x = 0 , \\
-1
& \mathrm{if} \ x < 0 .
\end{cases}
\end{align}
Since $0 < p < 1 - (n-1) p$ for $p \in (0, \frac{1}{n})$, we observe that
\begin{align}
\sgn \! \left( \left( \vphantom{\sum} (n-1) \, p^{\alpha} + (1 - (n-1)p)^{\alpha} \right)^{\frac{1}{\alpha} - 1} \right)
& =
1 ,
\\
\sgn \! \left( \vphantom{\sum} p^{\alpha-1} - (1 - (n-1)p)^{\alpha-1} \right)
& =
\begin{cases}
1
& \mathrm{if} \ \alpha < 1 , \\
0
& \mathrm{if} \ \alpha = 1 , \\
-1
& \mathrm{if} \ \alpha > 1 ,
\end{cases}
\\
\sgn \! \left( \frac{ 1 }{ \ln \frac{ 1 - (n-1) p}{ p } } \right)
& =
1
\end{align}
for $p \in (0, \frac{1}{n})$ and $\alpha \in (-\infty, 0) \cup (0, +\infty)$;
and therefore, we have
\begin{align}
&
\sgn \! \left( \frac{ \partial \| \bvec{v}_{n}( p ) \|_{\alpha} }{ \partial H_{\sbvec{v}_{n}}( p ) } \right)
\notag \\
& \quad \overset{\eqref{eq:diff1_N_H_v}}{=}
\sgn \! \left( \left( \vphantom{\sum} (n-1) \, p^{\alpha} + (1 - (n-1)p)^{\alpha} \right)^{\frac{1}{\alpha} - 1} \left( \vphantom{\sum} p^{\alpha-1} - (1 - (n-1)p)^{\alpha-1} \right) \frac{ 1 }{ \ln \frac{ 1 - (n-1) p}{ p } } \right)
\\
& \quad =
\sgn \! \left( \left( \vphantom{\sum} (n-1) \, p^{\alpha} + (1 - (n-1)p)^{\alpha} \right)^{\frac{1}{\alpha} - 1} \right) \! \cdot \sgn \! \left( \vphantom{\sum} p^{\alpha-1} - (1 - (n-1)p)^{\alpha-1} \right) \! \cdot \sgn \! \left( \frac{ 1 }{ \ln \frac{ 1 - (n-1) p}{ p } } \right)
\\
& \quad =
\begin{cases}
1
& \mathrm{if} \ \alpha < 1 , \\
0
& \mathrm{if} \ \alpha = 1 , \\
-1
& \mathrm{if} \ \alpha > 1 ,
\end{cases}
\label{eq:sign_diff1_N_H_v}
\end{align}
for $p \in (0, \frac{1}{n})$ and $\alpha \in (-\infty, 0) \cup (0, +\infty)$, which implies Lemma \ref{lem:mono_v}.
\end{IEEEproof}

It follows from Lemmas \ref{lem:Hv} and \ref{lem:mono_v} that, for each $\alpha \in (0, 1) \cup (1, \infty)$, $\| \bvec{v}_{n}( p ) \|_{\alpha}$ is bijective for $p \in [0, \frac{1}{n}]$.
Similarly, we also show the monotonicity of $\| \bvec{w}_{n}( p ) \|_{\alpha}$ with respect to $H_{\sbvec{w}_{n}}( p )$ in the following lemma.

\begin{lemma}
\label{lem:mono_w}
For any fixed $n \ge 2$ and any fixed $\alpha \in (0, 1) \cup (1, \infty)$, if $p \in [\frac{1}{n}, 1]$, the following monotonicity hold:
\begin{itemize}
\item[(i)]
if $\alpha > 1$, then $\| \bvec{w}_{n}( p ) \|_{\alpha}$ is strictly decreasing for $H_{\sbvec{w}_{n}}( p ) \in [0, \ln n]$ and
\item[(ii)]
if $\alpha < 1$, then $\| \bvec{w}_{n}( p ) \|_{\alpha}$ is strictly increasing for $H_{\sbvec{w}_{n}}( p ) \in [0, \ln n]$.
\end{itemize}
\end{lemma}

\begin{IEEEproof}[Proof of Lemma \ref{lem:mono_w}]
Since $\bvec{w}_{n}( p ) = \bvec{v}_{n}( p )_{\downarrow}$ for $p \in [\frac{1}{n}, \frac{1}{n-1}]$,
we can obtain immediately from \eqref{eq:diff1_N_H_v} that
\begin{align}
\frac{ \partial \| \bvec{w}_{n}( p ) \|_{\alpha} }{ \partial H_{\sbvec{w}_{n}}( p ) }
=
\left( \vphantom{\sum} (n-1) \, p^{\alpha} + (1 - (n-1)p)^{\alpha} \right)^{\frac{1}{\alpha} - 1} \left( \vphantom{\sum} p^{\alpha-1} - (1 - (n-1)p)^{\alpha-1} \right) \frac{ 1 }{ \ln \frac{ 1 - (n-1) p}{ p } }
\end{align}
for $p \in (\frac{1}{n}, \frac{1}{n-1})$.
Since $0 < 1 - (n-1) p < p$ for $p \in (\frac{1}{n}, \frac{1}{n-1})$, we observe that
\begin{align}
\sgn \! \left( \left( \vphantom{\sum} (n-1) \, p^{\alpha} + (1 - (n-1)p)^{\alpha} \right)^{\frac{1}{\alpha} - 1} \right)
& =
1 ,
\\
\sgn \! \left( \vphantom{\sum} p^{\alpha-1} - (1 - (n-1)p)^{\alpha-1} \right)
& =
\begin{cases}
1
& \mathrm{if} \ \alpha > 1 , \\
0
& \mathrm{if} \ \alpha = 1 , \\
-1
& \mathrm{if} \ \alpha < 1 ,
\end{cases}
\\
\sgn \! \left( \frac{ 1 }{ \ln \frac{ 1 - (n-1) p}{ p } } \right)
& =
-1
\end{align}
for $p \in (\frac{1}{n}, \frac{1}{n-1})$ and $\alpha \in (-\infty, 0) \cup (0, +\infty)$;
and therefore, we have
\begin{align}
\sgn \! \left( \frac{ \partial \| \bvec{w}_{n}( p ) \|_{\alpha} }{ \partial H_{\sbvec{w}_{n}}( p ) } \right)
& =
\begin{cases}
1
& \mathrm{if} \ \alpha < 1 , \\
0
& \mathrm{if} \ \alpha = 1 , \\
-1
& \mathrm{if} \ \alpha > 1 ,
\end{cases}
\label{eq:sign_diff1_N_H_w}
\end{align}
for $p \in (\frac{1}{n}, \frac{1}{n-1})$ and $\alpha \in (-\infty, 0) \cup (0, +\infty)$, as with \eqref{eq:sign_diff1_N_H_v}.
Hence, for $\alpha \in (-\infty, 0) \cup (0, +\infty)$, we have that
\begin{itemize}
\item
if $\alpha > 1$, then $\| \bvec{w}_{n}( p ) \|_{\alpha}$ is strictly decreasing for $H_{\sbvec{w}_{n}}( p ) \in [\ln (n-1), \ln n]$ and
\item
if $\alpha < 1$, then $\| \bvec{w}_{n}( p ) \|_{\alpha}$ is strictly increasing for $H_{\sbvec{w}_{n}}( p ) \in [\ln (n-1), \ln n]$.
\end{itemize}

%Finally, we note that, if $m = \lfloor p^{-1} \rfloor + 1$, then $H_{\sbvec{w}_{m}}( p ) = H_{\sbvec{w}_{n}}( p )$ and $\| \bvec{w}_{m}( p ) \|_{\alpha} = \| \bvec{w}_{n}( p ) \|_{\alpha}$ for $p \in [\frac{1}{n}, 1]$ and $\alpha \in (0, +\infty)$ since $0 \ln 0 = 0$ and $0^{\alpha} = 0$ for $\alpha \in (0, +\infty)$.
Finally, since $H_{\sbvec{w}_{m}}( p ) = H_{\sbvec{w}_{n}}( p )$ and $\| \bvec{w}_{m}( p ) \|_{\alpha} = \| \bvec{w}_{n}( p ) \|_{\alpha}$ for any integer $m \in [2, n-1]$, any $p \in [\frac{1}{m}, \frac{1}{m-1}]$, and any $\alpha \in (0, 1) \cup (1, +\infty)$, we can obtain that
\begin{itemize}
\item
if $\alpha > 1$, then $\| \bvec{w}_{n}( p ) \|_{\alpha}$ is strictly decreasing for $H_{\sbvec{w}_{n}}( p ) \in [\ln (m-1), \ln m]$ and
\item
if $\alpha < 1$, then $\| \bvec{w}_{n}( p ) \|_{\alpha}$ is strictly increasing for $H_{\sbvec{w}_{n}}( p ) \in [\ln (m-1), \ln m]$
\end{itemize}
for any integer $m \in [2, n]$ and any $\alpha \in (0, 1) \cup (1, \infty)$.
%Therefore, the monotonicity proved from \eqref{eq:sign_diff1_N_H_w} hold for $p \in [\frac{1}{n}, 1]$ and $\alpha \in (0, +\infty)$.
This completes the proof of Lemma \ref{lem:mono_w}.
\end{IEEEproof}

It also follows from Lemmas \ref{lem:Hw} and \ref{lem:mono_w} that, for each $\alpha \in (0, 1) \cup (1, \infty)$, $\| \bvec{w}_{n}( p ) \|_{\alpha}$ is also bijective for $p \in [\frac{1}{n}, 1]$.
%Since $\| \bvec{v}_{n}( 0 ) \|_{\alpha} = 1$, $\| \bvec{v}_{n}( \frac{1}{n} ) \|_{\alpha} = n^{\frac{1}{\alpha}-1}$, $\| \bvec{w}_{n}( 0 ) \|_{\alpha} = n^{\frac{1}{\alpha}-1}$, and $\| \bvec{w}_{n}( 1 ) \|_{\alpha} = 1$, for each $n \ge 2$ and each $\alpha \in (0, 1) \cup (1, +\infty)$, we can define the inverse functions of $\| \bvec{v}_{n}( p ) \|_{\alpha}$ and $\| \bvec{v}_{n}( p ) \|_{\alpha}$ with respect to $p$, as with Definition \ref{def:inverseH}, in the following definition.

\if0
\begin{definition}
\label{def:inverseN}
We denote by $N_{\alpha, \sbvec{v}_{n}}^{-1} : [\min\{ 1, n^{\frac{1}{\alpha}-1} \}, \max\{ 1, n^{\frac{1}{\alpha}-1} \}] \to [0, \frac{1}{n}]$ the inverse function of $\| \bvec{v}_{n}( p ) \|_{\alpha}$ for $p \in [0, \frac{1}{n}]$.
Moreover, we also denote by $N_{\alpha, \sbvec{w}_{n}}^{-1} : [\min\{ 1, n^{\frac{1}{\alpha}-1} \}, \max\{ 1, n^{\frac{1}{\alpha}-1} \}] \to [\frac{1}{n}, 1]$ the inverse function of $\| \bvec{w}_{n}( p ) \|_{\alpha}$ for $p \in [\frac{1}{n}, 1]$.
\end{definition}
\fi

\section{Results}
\label{sect:result}

In Section \ref{subsect:extremes}, we examine the extremal relations between the Shannon entropy and the $\ell_{\alpha}$-norm, as shown in Theorems \ref{th:extremes} and \ref{th:extremes2}.
%Accurately, we derive the tight bounds of the $\ell_{\alpha}$-norm with a fixed Shannon entropy in Theorem \ref{th:extremes}, and vice versa in Theorem \ref{th:extremes2}.
%Note that the tight bound means that there exists a distribution which attains the bound, as with the tightness of Fano's inequality.
Then, we can identify the exact feasible region of
\begin{align}
\mathcal{R}_{n}( \alpha )
\triangleq
\{ (H( \bvec{p} ), \| \bvec{p} \|_{\alpha}) \mid \bvec{p} \in \mathcal{P}_{n} \}
\label{def:region_Pn}
\end{align}
for any $n \ge 2$ and any $\alpha \in (0, 1) \cup (1, \infty)$.
Extending Theorems \ref{th:extremes} and \ref{th:extremes2} to Corollary \ref{cor:extremes}, we can obtain the tight bounds between the Shannon entropy and several information measures which are determined by the $\ell_{\alpha}$-norm, as shown in Table \ref{table:extremes}.
In Section \ref{subsect:focusing}, we apply the results of Section \ref{subsect:extremes} to uniformly focusing channels of Definition \ref{def:focusing}.
%Then, we derive the tight bounds of the mutual information of order $\alpha$ and the $E_{0}$ function (see Theorem \ref{th:E0_focusing}) by using the bounds of the R\'{e}nyi divergence from a uniform distribution (see Corollary \ref{cor:RenyiDiv}).

\subsection{Bounds on Shannon entropy and $\ell_{\alpha}$-norm}
\label{subsect:extremes}

%In this section, we derive the tight bounds of the Shannon entropy with a fixed $\ell_{\alpha}$-norm, and vice versa.
%We now define a generalized logarithmic function used in Tsallis statistics as follows:
Let the $\alpha$-logarithm function \cite{tsallis} be denoted by
\begin{align}
\ln_{\alpha} x
\triangleq
\frac{ x^{1-\alpha} - 1 }{ 1 - \alpha }
\end{align}
for $\alpha \neq 1$ and $x > 0$;
besides, since $\lim_{\alpha \to 1} \ln_{\alpha} x = \ln x$ by L'H\^{o}pital's rule, it is defined that $\ln_{1} x \triangleq \ln x$.
%In statistical physics, the Tsallis entropy \cite{tsallis2} of order $\alpha \in (0, 1) \cup (1, +\infty)$ is defined as
%$
%S_{\alpha}( \bvec{p} )
%\triangleq
%- \sum_{i=1}^{n} p_{i}^{\alpha} \ln_{\alpha} p_{i}
%=
%\sum_{i=1}^{n} p_{i} \ln_{\alpha} \frac{1}{p_{i}}
%$
%for $\bvec{p} \in \mathcal{P}_{n}$.
%It obviously holds that $S_{1}( \bvec{p} ) = H( \bvec{p} )$.
For the $\alpha$-logarithm function, we can see the following lemma.

\begin{lemma}
\label{lem:frac_qlog}
For $\alpha < \beta$ and $1 \le x \le y$ $(y \neq 1)$, we observe that
\begin{align}
\frac{ \ln_{\alpha} x }{ \ln_{\alpha} y }
\le
\frac{ \ln_{\beta} x }{ \ln_{\beta} y }
\end{align}
with equality if and only if $x \in \{ 1, y \}$.
\end{lemma}

\begin{IEEEproof}[Proof of Lemma \ref{lem:frac_qlog}]
For $1 \le x \le y$ $(y \neq 1)$, we consider the monotonicity of $\frac{ \ln_{\alpha} x }{ \ln_{\alpha} y }$ with respect to $\alpha$. %, as with the proof of Lemma \ref{lem:IT_ineq}.
Direct calculation shows
\begin{align}
\frac{ \partial }{ \partial \alpha } \left( \frac{ \ln_{\alpha} x }{ \ln_{\alpha} y } \right)
& =
\frac{ \partial }{ \partial \alpha } \left( \frac{ x^{1-\alpha} - 1 }{ y^{1-\alpha} - 1 } \right)
\\
& =
\left( \frac{ \partial }{ \partial \alpha } (x^{1-\alpha} - 1) \right) \left( \frac{ 1 }{ y^{1-\alpha} - 1 } \right) + (x^{1-\alpha} - 1) \left( \frac{ \partial }{ \partial \alpha } \left( \frac{ 1 }{ y^{1-\alpha} - 1 } \right) \right)
\\
& =
- \frac{ x^{1-\alpha} \ln x }{ y^{1-\alpha} - 1 } + (x^{1-\alpha} - 1) \left( - \frac{ 1 }{ (y^{1-\alpha} - 1)^{2} } \right) \left( \frac{ \partial }{ \partial \alpha } (y^{1-\alpha} - 1) \right)
\\
& =
- \frac{ x^{1-\alpha} \ln x }{ y^{1-\alpha} - 1 } + \frac{ y^{1-\alpha} (\ln y) (x^{1-\alpha} - 1) }{ (y^{1-\alpha} - 1)^{2} }
\\
& =
- \frac{ x^{1-\alpha} (\ln x) (y^{1-\alpha} - 1) - y^{1-\alpha} (\ln y) (x^{1-\alpha} - 1) }{ (y^{1-\alpha} - 1)^{2} }
\\
& =
- \frac{ 1 }{ (y^{1-\alpha} - 1)^{2} } \left( \vphantom{\sum} x^{1-\alpha} (\ln x) (y^{1-\alpha} - 1) - y^{1-\alpha} (\ln y) (x^{1-\alpha} - 1) \right) .
\label{eq:diff1_frac_qlog}
\end{align}
Then, we can see that
\begin{align}
\sgn \! \left( \frac{ \partial }{ \partial \alpha } \left( \frac{ \ln_{\alpha} x }{ \ln_{\alpha} y } \right) \right)
& \overset{\eqref{eq:diff1_frac_qlog}}{=}
\sgn \! \left( - \frac{ 1 }{ (y^{1-\alpha} - 1)^{2} } \left( \vphantom{\sum} x^{1-\alpha} (\ln x) (y^{1-\alpha} - 1) - y^{1-\alpha} (\ln y) (x^{1-\alpha} - 1) \right) \right)
\\
& =
\sgn \! \left( - \frac{ 1 }{ (y^{1-\alpha} - 1)^{2} } \right) \cdot \, \sgn \! \left( \vphantom{\sum} x^{1-\alpha} (\ln x) (y^{1-\alpha} - 1) - y^{1-\alpha} (\ln y) (x^{1-\alpha} - 1) \right)
\\
& \overset{\text{(a)}}{=}
- \sgn \! \left( \vphantom{\sum} x^{1-\alpha} (\ln x) (y^{1-\alpha} - 1) - y^{1-\alpha} (\ln y) (x^{1-\alpha} - 1) \right)
\\
& \overset{\text{(b)}}{=}
- \sgn \! \left( (\ln x) \frac{ y^{1-\alpha} - 1 }{ y^{1-\alpha} } - (\ln y) \frac{ x^{1-\alpha} - 1 }{ x^{1-\alpha} } \right)
\\
& =
\sgn \! \left( \vphantom{\sum} (y^{\alpha-1} - 1) \ln x - (x^{\alpha-1} - 1) \ln y \right)
\\
& =
\sgn \! \left( \frac{ (y^{\alpha-1} - 1) \ln x^{\alpha-1} - (x^{\alpha-1} - 1) \ln y^{\alpha-1} }{ \alpha - 1 } \right)
\\
& =
\sgn \! \left( \frac{1}{\alpha-1} \right) \cdot \, \sgn \! \left( \vphantom{\sum} (y^{\alpha-1} - 1) \ln x^{\alpha-1} - (x^{\alpha-1} - 1) \ln y^{\alpha-1} \right)
\label{eq:sign_frac_qlog} \\
& \overset{\text{(c)}}{=}
\sgn \! \left( \frac{1}{\alpha-1} \right) \cdot \, \sgn \! \left( \vphantom{\sum} (b - 1) \ln a - (a - 1) \ln b \right)
\end{align}
where
\begin{itemize}
\item
the equality (a) follows from the fact that
\begin{align}
\sgn \! \left( - \frac{ 1 }{ (y^{1-\alpha} - 1)^{2} } \right) = -1
\end{align}
for $y > 0 \ (y \neq 1)$ and $\alpha \in (-\infty, 1) \cup (1, +\infty)$,
\item
the equality (b) follows from the fact that $x^{1-\alpha}, y^{1-\alpha} > 0$ for $\alpha \in (-\infty, +\infty)$ and $x, y > 0$, and
\item
the equality (c) follows by the change of variables: $a = a(x, \alpha) \triangleq x^{\alpha-1}$ and $b = b(y, \alpha) \triangleq y^{\alpha-1}$.
\end{itemize}
Then, it can be easily seen that
\begin{align}
\sgn \! \left( \frac{1}{\alpha-1} \right)
=
\begin{cases}
1
& \mathrm{if} \ \alpha > 1 , \\
-1
& \mathrm{if} \ \alpha < 1 .
\end{cases}
\label{eq:sign_1_over_(1-a)}
\end{align}
Thus, to check the sign of $\frac{ \partial }{ \partial \alpha } \left( \frac{ \ln_{\alpha} x }{ \ln_{\alpha} y } \right)$, we now examine the function $(b - 1) \ln a - (a - 1) \ln b$.
We readily see that
\begin{align}
\left. \left( \vphantom{\sum} (b - 1) \ln a - (a - 1) \ln b \right) \right|_{a = 1}
=
\left. \left( \vphantom{\sum} (b - 1) \ln a - (a - 1) \ln b \right) \right|_{a = b}
=
0
\label{eq:gap_a=b}
\end{align}
for $b > 0$.
We calculate the second order derivative of $(b - 1) \ln a - (a - 1) \ln b$ with respect to $a$ as follows:
\begin{align}
\frac{ \partial^{2} }{ \partial a^{2} } \left( \vphantom{\sum} (b - 1) \ln a - (a - 1) \ln b \right)
& =
\frac{ \partial }{ \partial a } \left( \frac{ \partial }{ \partial a } \left( \vphantom{\sum} (b - 1) \ln a - (a - 1) \ln b \right) \right)
\\
& =
\frac{ \partial }{ \partial a } \left( (b-1) \left( \frac{ \mathrm{d} }{ \mathrm{d} a } (\ln a) \right) - \left( \frac{ \mathrm{d} }{ \mathrm{d} a } (a-1) \right) \ln b \right)
\\
& =
\frac{ \partial }{ \partial a } \left( \frac{b-1}{a} - \ln b \right)
\\
& =
(b-1) \left( \frac{ \mathrm{d} }{ \mathrm{d} a } \left( \frac{1}{a} \right) \right)
\\
& =
- \frac{ b-1 }{ a^{2} } .
\end{align}
Hence, we observe that
\begin{align}
\sgn \! \left( \frac{ \partial^{2} }{ \partial a^{2} } \left( \vphantom{\sum} (b - 1) \ln a - (a - 1) \ln b \right) \right)
& =
\sgn \! \left( - \frac{ b-1 }{ a^{2} } \right)
\\
& =
\begin{cases}
1
& \mathrm{if} \ 0 < b < 1 , \\
0
& \mathrm{if} \ b = 1 , \\
-1
& \mathrm{if} \ b > 1
\end{cases}
\end{align}
for $a > 0$, which implies that
\begin{itemize}
\item
if $b > 1$, then $(b - 1) \ln a - (a - 1) \ln b$ is strictly concave in $a > 0$ and
\item
if $0 < b < 1$, then $(b - 1) \ln a - (a - 1) \ln b$ is strictly convex in $a > 0$.
\end{itemize}
Therefore, it follows from \eqref{eq:gap_a=b} that
\begin{itemize}
\item
it $b > 1$, then
\begin{align}
\sgn \! \left( \vphantom{\sum} (b - 1) \ln a - (a - 1) \ln b \right)
=
\begin{cases}
1
& \mathrm{if} \ 1 < a < b , \\
0
& \mathrm{if} \ a = 1 \ \mathrm{or} \ a = b , \\
-1
& \mathrm{if} \ 0 < a < 1 \ \mathrm{or} \ a > b
\end{cases}
\end{align}
and
\item
it $0 < b < 1$, then
\begin{align}
\sgn \! \left( \vphantom{\sum} (b - 1) \ln a - (a - 1) \ln b \right)
=
\begin{cases}
1
& \mathrm{if} \ 0 < a < b \ \mathrm{or} \ a > 1 , \\
0
& \mathrm{if} \ a = b \ \mathrm{or} \ a = 1 , \\
-1
& \mathrm{if} \ b < a < 1 .
\end{cases}
\end{align}
\end{itemize}
Since $a = x^{\alpha-1}$ and $b = y^{\alpha-1}$, note that
\begin{itemize}
\item
if $\alpha > 1$, then $1 \le a \le b \ (b \neq 1)$ for $1 \le x \le y \ (y \neq 1)$ and
\item
if $\alpha < 1$, then $0 < b \le a \le 1 \ (b \neq 1)$ for $1 \le x \le y \ (y \neq 1)$.
\end{itemize}
Hence, we obtain
\begin{align}
\sgn \! \left( \vphantom{\sum} (y^{1-\alpha} - 1) \ln x^{1-\alpha} - (x^{1-\alpha} - 1) \ln y^{1-\alpha} \right)
=
\begin{cases}
1
& \mathrm{if} \ 1 < x < y \ \mathrm{and} \ \alpha > 1, \\
0
& \mathrm{if} \ x = 1 \ \mathrm{or} \ x = y \ \mathrm{or} \ \alpha = 1 , \\
-1
& \mathrm{if} \ 1 < x < y \ \mathrm{and} \ \alpha < 1
\end{cases}
\label{eq:sign_gap_(a-1)ln(b)}
\end{align}
for $\alpha \in (-\infty, +\infty)$ and $1 \le x \le y \ (y \neq 1)$.
Concluding the above analyses, we have
\begin{align}
\sgn \! \left( \frac{ \partial }{ \partial \alpha } \left( \frac{ \ln_{\alpha} x }{ \ln_{\alpha} y } \right) \right)
& \overset{\eqref{eq:sign_frac_qlog}}{=}
\sgn \! \left( \frac{1}{1-\alpha} \right) \cdot \, \sgn \! \left( \vphantom{\sum} (y^{1-\alpha} - 1) \ln x^{1-\alpha} - (x^{1-\alpha} - 1) \ln y^{1-\alpha} \right)
\\
& =
\begin{cases}
1
& \mathrm{if} \ 1 < x < y , \\
0
& \mathrm{if} \ x = 1 \ \mathrm{or} \ x = y
\end{cases}
\end{align}
for $\alpha \in (-\infty, 1) \cup (1, \infty)$, where the last equality follows from \eqref{eq:sign_1_over_(1-a)} and \eqref{eq:sign_gap_(a-1)ln(b)}.
Note that
\begin{align}
\lim_{\alpha \to 1} \left( \frac{ \ln_{\alpha} x }{ \ln_{\alpha} y } \right)
=
\frac{ \ln_{1} x }{ \ln_{1} y }
=
\frac{ \ln x }{ \ln y }
\end{align}
for $x, y > 0 \ (y \neq 1)$, which implies that $\frac{ \ln_{\alpha} x }{ \ln_{\alpha} y }$ is continuous at $\alpha = 1$.
Therefore, we have that, if $1 < x < y$, then $\frac{ \ln_{\alpha} x }{ \ln_{\alpha} y }$ is strictly increasing for $\alpha \in (-\infty, +\infty)$, which implies Lemma \ref{lem:frac_qlog}.
\end{IEEEproof}

The following two lemmas have important roles in the proving Theorem \ref{th:extremes}.

\begin{lemma}
\label{lem:vector_v}
For any $n \ge 2$ and any $\bvec{p} \in \mathcal{P}_{n}$, there exists $p \in [0, \frac{1}{n}]$ such that $H_{\sbvec{v}_{n}}( p ) = H( \bvec{p} )$ and $\| \bvec{v}_{n}( p ) \|_{\alpha} \ge \| \bvec{p} \|_{\alpha}$ for all $\alpha \in (0, \infty)$.
\end{lemma}

\begin{IEEEproof}[Proof of Lemma \ref{lem:vector_v}]
If $n = 2$, then it can be easily seen that $\bvec{p}_{\downarrow} = \bvec{v}_{2}( p )$ for any $\bvec{p} \in \mathcal{P}_{2}$ and some $p \in [0, \frac{1}{2}]$;
therefore, the lemma obviously holds when $n = 2$.
Moreover, since
\begin{align}
H( \bvec{p} ) & = \ln n
& \iff &&
\bvec{p} & = \bvec{u}_{n} = \bvec{v}_{n}( {\textstyle \frac{1}{n}} ),
\\
H( \bvec{p} ) & = 0
& \iff &&
\bvec{p}_{\downarrow} & = \bvec{d}_{n} = \bvec{v}_{n}( 0 ) ,
\end{align}
the lemma obviously holds if $H(\bvec{p}) \in \{ 0, \ln n \}$.
Thus, we omit the cases $n = 2$ and $H(\bvec{p}) \in \{ 0, \ln n \}$ in the analyses and consider $\bvec{p} \in P_{n}$ for $H( \bvec{p} ) \in (0, \ln n)$.
For a fixed $n \ge 3$ and a constant $A \in (0, \ln n)$, we assume for $\bvec{p} \in P_{n}$ that 
\begin{align}
H( \bvec{p} ) = A .
\label{eq:fixed_H}
\end{align}
For that $\bvec{p}$, let $k \in \{ 2, 3, \dots, n-1 \}$ be the index such that $p_{[k-1]} > p_{[k+1]} = p_{[n]}$;
namely, the index $k$ is chosen to satisfy the following inequalities:
\begin{align}
p_{[1]} \ge p_{[2]} \ge \dots \ge p_{[k-1]} \ge p_{[k]} \ge p_{[k+1]} = p_{[k+2]} = \dots = p_{[n]}
\qquad (p_{[k-1]} > p_{[k+1]}) .
\label{eq:equal_k+1_to_n}
\end{align}
Since
$
p_{1} + p_{2} + \dots + p_{n} = 1
$,
we observe that
\begin{align}
&&
\sum_{i=1}^{n} p_{i}
& =
1
\\
& \ \Longrightarrow \ &
\frac{ \mathrm{d} }{ \mathrm{d} p_{[k]} } \left( \sum_{i=1}^{n} p_{i} \right)
& =
\frac{ \mathrm{d} }{ \mathrm{d} p_{[k]} } (1)
\\
& \iff &
\frac{ \mathrm{d} }{ \mathrm{d} p_{[k]} } \left( \sum_{i=1}^{n} p_{[i]} \right)
& =
0
\\
& \iff &
\frac{ \mathrm{d} p_{[k]} }{ \mathrm{d} p_{[k]} } + \sum_{i = 1 : i \neq k}^{n} \frac{ \mathrm{d} p_{[i]} }{ \mathrm{d} p_{[k]} }
& =
0
\\
& \iff &
1 + \sum_{i = 1 : i \neq k}^{n} \frac{ \mathrm{d} p_{[i]} }{ \mathrm{d} p_{[k]} }
& =
0
\\
& \iff &
\sum_{i = 1 : i \neq k}^{n} \frac{ \mathrm{d} p_{[i]} }{ \mathrm{d} p_{[k]} }
& =
- 1 .
\label{eq:total_diff_prob}
\end{align}
In this proof, we further assume that
\begin{align}
\frac{ \mathrm{d} p_{[i]} }{ \mathrm{d} p_{[k]} }
=
0
\label{eq:hypo1}
\end{align}
for $i \in \{ 2, 3, \dots, k-1 \}$ and
\begin{align}
\frac{ \mathrm{d} p_{[j]} }{ \mathrm{d} p_{[k]} }
=
\frac{ \mathrm{d} p_{[n]} }{ \mathrm{d} p_{[k]} }
\label{eq:hypo2}
\end{align}
for $j \in \{ k+1, k+2, \dots, n-1 \}$.
By constraints \eqref{eq:hypo1} and \eqref{eq:hypo2}, we get
\begin{align}
&&
\sum_{i=1}^{n} p_{i}
& =
1
\\
& \ \overset{\eqref{eq:total_diff_prob}}{\Longrightarrow} \ &
\sum_{i = 1 : i \neq k}^{n} \frac{ \mathrm{d} p_{[i]} }{ \mathrm{d} p_{[k]} }
& =
- 1
\\
& \iff &
\sum_{i = 1}^{k-1} \frac{ \mathrm{d} p_{[i]} }{ \mathrm{d} p_{[k]} } + \sum_{j = k+1}^{n} \frac{ \mathrm{d} p_{[j]} }{ \mathrm{d} p_{[k]} }
& =
- 1
\\
& \overset{\eqref{eq:hypo1}}{\iff} &
\frac{ \mathrm{d} p_{[1]} }{ \mathrm{d} p_{[k]} } + \sum_{j = k+1}^{n} \frac{ \mathrm{d} p_{[j]} }{ \mathrm{d} p_{[k]} }
& =
-1
\\
& \overset{\eqref{eq:hypo2}}{\iff} &
\frac{ \mathrm{d} p_{[1]} }{ \mathrm{d} p_{[k]} } + (n-k) \frac{ \mathrm{d} p_{[n]} }{ \mathrm{d} p_{[k]} }
& =
-1
\\
& \iff &
\frac{ \mathrm{d} p_{[1]} }{ \mathrm{d} p_{[k]} }
& =
- 1 - (n-k) \frac{ \mathrm{d} p_{[n]} }{ \mathrm{d} p_{[k]} } .
\label{eq:total_prob_hypo}
\end{align}
Moreover, since
$
H( \bvec{p} )
=
A
$,
we observe that
\begin{align}
&&
- \sum_{i = 1}^{n} p_{i} \ln p_{i}
& =
A
\\
& \ \Longrightarrow \ &
\frac{ \mathrm{d} }{ \mathrm{d} p_{[k]} } \left( - \sum_{i = 1}^{n} p_{i} \ln p_{i} \right)
& =
\frac{ \mathrm{d} }{ \mathrm{d} p_{[k]} } (A)
\\
& \iff &
\frac{ \mathrm{d} }{ \mathrm{d} p_{[k]} } \left( - \sum_{i = 1}^{n} p_{[i]} \ln p_{[i]} \right)
& =
0
\\
& \iff &
- \sum_{i = 1}^{n} \frac{ \mathrm{d} }{ \mathrm{d} p_{[k]} } (p_{[i]} \ln p_{[i]})
& =
0
\\
& \iff &
- \frac{ \mathrm{d} }{ \mathrm{d} p_{[k]} } (p_{[k]} \ln p_{[k]}) - \sum_{i = 1 : i \neq k}^{n} \frac{ \mathrm{d} }{ \mathrm{d} p_{[k]} } (p_{[i]} \ln p_{[i]})
& =
0
\\
& \iff &
- (\ln p_{[k]} + 1) - \sum_{i = 1 : i \neq k}^{n} \frac{ \mathrm{d} }{ \mathrm{d} p_{[k]} } (p_{[i]} \ln p_{[i]})
& =
0
\\
& \iff &
- \sum_{i = 1 : i \neq k}^{n} \frac{ \mathrm{d} }{ \mathrm{d} p_{[k]} } (p_{[i]} \ln p_{[i]})
& =
\ln p_{[k]} + 1
\\
& \overset{\text{(a)}}{\iff} &
- \sum_{i = 1 : i \neq k}^{n} \left( \frac{ \mathrm{d} p_{[i]} }{ \mathrm{d} p_{[k]} } \right) \left( \frac{ \mathrm{d} }{ \mathrm{d} p_{[i]} } (p_{[i]} \ln p_{[i]}) \right)
& =
\ln p_{[k]} + 1
\\
& \iff &
- \sum_{i = 1 : i \neq k}^{n} \left( \frac{ \mathrm{d} p_{[i]} }{ \mathrm{d} p_{[k]} } \right) (\ln p_{[i]} + 1)
& =
\ln p_{[k]} + 1
\label{eq:diff1_H_halfway} \\
& \iff &
- \sum_{i = 1}^{k-1} \left( \frac{ \mathrm{d} p_{[i]} }{ \mathrm{d} p_{[k]} } \right) (\ln p_{[i]} + 1) - \sum_{j = k+1}^{n} \left( \frac{ \mathrm{d} p_{[j]} }{ \mathrm{d} p_{[k]} } \right) (\ln p_{[j]} + 1)
& =
\ln p_{[k]} + 1
\\
& \overset{\eqref{eq:equal_k+1_to_n}}{\iff} &
- \sum_{i = 1}^{k-1} \left( \frac{ \mathrm{d} p_{[i]} }{ \mathrm{d} p_{[k]} } \right) (\ln p_{[i]} + 1) - (\ln p_{[n]} + 1) \sum_{j = k+1}^{n} \left( \frac{ \mathrm{d} p_{[j]} }{ \mathrm{d} p_{[k]} } \right)
& =
\ln p_{[k]} + 1
\\
& \overset{\eqref{eq:hypo1}}{\iff} &
- \left( \frac{ \mathrm{d} p_{[1]} }{ \mathrm{d} p_{[k]} } \right) (\ln p_{[1]} + 1) - (\ln p_{[n]} + 1) \sum_{j = k+1}^{n} \left( \frac{ \mathrm{d} p_{[j]} }{ \mathrm{d} p_{[k]} } \right)
& =
\ln p_{[k]} + 1
\\
& \overset{\eqref{eq:hypo2}}{\iff} &
- \left( \frac{ \mathrm{d} p_{[1]} }{ \mathrm{d} p_{[k]} } \right) (\ln p_{[1]} + 1) - (\ln p_{[n]} + 1) (n-k) \left( \frac{ \mathrm{d} p_{[n]} }{ \mathrm{d} p_{[k]} } \right)
& =
\ln p_{[k]} + 1
\\
& \overset{\eqref{eq:total_prob_hypo}}{\iff} &
- \left( - 1 - (n-k) \frac{ \mathrm{d} p_{[n]} }{ \mathrm{d} p_{[k]} } \right) (\ln p_{[1]} + 1) - (n-k) \left( \frac{ \mathrm{d} p_{[n]} }{ \mathrm{d} p_{[k]} } \right) (\ln p_{[n]} + 1)
& =
\ln p_{[k]} + 1
\\
& \iff & \! \! \! \! \!
(\ln p_{[1]} + 1) + (n-k) \left( \frac{ \mathrm{d} p_{[n]} }{ \mathrm{d} p_{[k]} } \right) (\ln p_{[1]} + 1) - (n-k) \left( \frac{ \mathrm{d} p_{[n]} }{ \mathrm{d} p_{[k]} } \right) (\ln p_{[n]} + 1)
& =
\ln p_{[k]} + 1
\\
& \iff &
(n-k) \left( \frac{ \mathrm{d} p_{[n]} }{ \mathrm{d} p_{[k]} } \right) (\ln p_{[1]} + 1) - (n-k) \left( \frac{ \mathrm{d} p_{[n]} }{ \mathrm{d} p_{[k]} } \right) (\ln p_{[n]} + 1)
& =
\ln p_{[k]} - \ln p_{[1]}
\\
& \iff &
(n-k) \left( \frac{ \mathrm{d} p_{[n]} }{ \mathrm{d} p_{[k]} } \right) (\ln p_{[1]} - \ln p_{[n]})
& =
\ln p_{[k]} - \ln p_{[1]}
\\
& \iff &
(n-k) \left( \frac{ \mathrm{d} p_{[n]} }{ \mathrm{d} p_{[k]} } \right)
& =
\frac{ \ln p_{[k]} - \ln p_{[1]} }{ \ln p_{[1]} - \ln p_{[n]} }
\\
& \iff &
\frac{ \mathrm{d} p_{[n]} }{ \mathrm{d} p_{[k]} }
& =
- \frac{ 1 }{ n-k } \left( \frac{ \ln p_{[1]} - \ln p_{[k]} }{ \ln p_{[1]} - \ln p_{[n]} } \right) \!
\label{eq:total_entropy_hypo}
\end{align}
where the equivalence (a) follows by the chain rule.
We now check the sign of the right-hand side of \eqref{eq:total_entropy_hypo}.
If $1 > p_{[1]} > p_{[k]} \ge p_{[n]} > 0$, then
\begin{align}
0
<
\frac{ \ln p_{[1]} - \ln p_{[k]} }{ \ln p_{[1]} - \ln p_{[n]} }
<
1
\label{eq:cond1}
\end{align}
since
$
0 > \ln p_{[1]} > \ln p_{[k]} > \ln p_{[n]}
$;
therefore, we get from \eqref{eq:total_entropy_hypo} that
\begin{align}
- \frac{1}{n-k}
<
\frac{ \mathrm{d} p_{[n]} }{ \mathrm{d} p_{[k]} }
<
0
\label{eq:sign_dndk_1}
\end{align}
for $1 > p_{[1]} > p_{[k]} > p_{[n]} > 0$.
Note that $n - k \ge 1$.
Moreover, if $1 > p_{[1]} = p_{[k]} > p_{[n]} > 0$, then
\begin{align}
\frac{ \mathrm{d} p_{[n]} }{ \mathrm{d} p_{[k]} }
& =
- \frac{ 1 }{ n-k } \left( \frac{ \ln p_{[1]} - \ln p_{[k]} }{ \ln p_{[1]} - \ln p_{[n]} } \right)
\\
& =
- \frac{ 1 }{ n-k } \left( \frac{ 0 }{ \ln p_{[1]} - \ln p_{[n]} } \right)
\\
& =
0 .
\label{eq:sign_dndk_2}
\end{align}
Furthermore, if $1 > p_{[1]} > p_{[k]} = p_{[n]} > 0$, then
\begin{align}
\frac{ \mathrm{d} p_{[n]} }{ \mathrm{d} p_{[k]} }
& =
- \frac{ 1 }{ n-k } \left( \frac{ \ln p_{[1]} - \ln p_{[k]} }{ \ln p_{[1]} - \ln p_{[n]} } \right)
\\
& =
- \frac{ 1 }{ n-k } .
\label{eq:sign_dndk_3}
\end{align}
Combining \eqref{eq:sign_dndk_1}, \eqref{eq:sign_dndk_2}, and \eqref{eq:sign_dndk_3}, we get under the constraints \eqref{eq:fixed_H}, \eqref{eq:equal_k+1_to_n}, \eqref{eq:hypo1}, and \eqref{eq:hypo2} that
\begin{align}
\sgn \! \left( \frac{ \mathrm{d} p_{[n]} }{ \mathrm{d} p_{[k]} } \right)
=
\begin{cases}
0
& \mathrm{if} \ p_{[1]} = p_{[k]} , \\
-1
& \mathrm{otherwise}
\end{cases}
\label{eq:sign_dndk}
\end{align}
for $1 > p_{[1]} \ge p_{[k]} \ge p_{[n]} > 0 \ (p_{[1]} > p_{[n]})$.
Note for the constraint \eqref{eq:fixed_H} that
\begin{align}
\lim_{(p_{[k+1]}, p_{[k+2]}, \dots, p_{[n]}) \to (0^{+}, 0^{+}, \dots, 0^{+})} H( p_{[1]}, p_{[2]}, \dots, p_{[n]} )
=
H( p_{[1]}, p_{[2]}, \dots, p_{[k]}, 0, 0, \dots, 0 )
\end{align}
since $\lim_{x \to 0^{+}} x \ln x = 0 \ln 0$ by the assumption $0 \ln 0 = 0$.
Thus, it follows from \eqref{eq:sign_dndk} that, for all $j \in \{ k+1, k+2, \dots, n \}$, $p_{[j]}$ is strictly decreasing for $p_{[k]}$ under the constraints \eqref{eq:fixed_H}, \eqref{eq:equal_k+1_to_n}, \eqref{eq:hypo1}, and \eqref{eq:hypo2}.
Similarly, we check the sign of the right-hand side of \eqref{eq:total_prob_hypo}:
\begin{align}
\frac{ \mathrm{d} p_{[1]} }{ \mathrm{d} p_{[k]} }
& =
- 1 - (n-k) \frac{ \mathrm{d} p_{[n]} }{ \mathrm{d} p_{[k]} } .
\end{align}
By \eqref{eq:sign_dndk_1}, \eqref{eq:sign_dndk_2}, and \eqref{eq:sign_dndk_3}, we can see that
\begin{align}
-1 \le \frac{ \mathrm{d} p_{[1]} }{ \mathrm{d} p_{[k]} } < 0
\end{align}
for $1 > p_{[1]} \ge p_{[k]} > p_{[n]} > 0$ and
\begin{align}
\frac{ \mathrm{d} p_{[1]} }{ \mathrm{d} p_{[k]} }
& =
0
\end{align}
for $1 > p_{[1]} > p_{[k]} = p_{[n]} > 0$;
therefore, we also get under the constraints \eqref{eq:fixed_H}, \eqref{eq:equal_k+1_to_n}, \eqref{eq:hypo1}, and \eqref{eq:hypo2} that
\begin{align}
\sgn \! \left( \frac{ \mathrm{d} p_{[1]} }{ \mathrm{d} p_{[k]} } \right)
=
\begin{cases}
0
& \mathrm{if} \ p_{[k]} = p_{[n]} , \\
-1
& \mathrm{otherwise}
\end{cases}
\label{eq:sign_d1dk}
\end{align}
for $1 > p_{[1]} \ge p_{[k]} \ge p_{[n]} > 0 \ (p_{[1]} > p_{[n]})$.
As with \eqref{eq:sign_dndk}, it follows from \eqref{eq:sign_d1dk} that $p_{[1]}$ is strictly decreasing for $p_{[k]}$ under the constraints \eqref{eq:fixed_H}, \eqref{eq:equal_k+1_to_n}, \eqref{eq:hypo1}, and \eqref{eq:hypo2}.

On the other hand, for a fixed $\alpha \in (-\infty, 1) \cup (1, +\infty)$, we have
\begin{align}
\frac{ \mathrm{d} \| \bvec{p} \|_{\alpha} }{ \mathrm{d} p_{[k]} }
& =
\frac{ \mathrm{d} }{ \mathrm{d} p_{[k]} } \left( \sum_{i=1}^{n} p_{i}^{\alpha} \right)^{\frac{1}{\alpha}}
\\
& =
\frac{1}{\alpha} \left( \sum_{i=1}^{n} p_{i}^{\alpha} \right)^{\frac{1}{\alpha} - 1} \left( \frac{ \mathrm{d} }{ \mathrm{d} p_{[k]} } \sum_{i=1}^{n} p_{i}^{\alpha} \right)
\\
& =
\frac{1}{\alpha} \left( \sum_{i=1}^{n} p_{i}^{\alpha} \right)^{\frac{1}{\alpha} - 1} \left( \frac{ \mathrm{d} }{ \mathrm{d} p_{[k]} } \sum_{i=1}^{n} p_{[i]}^{\alpha} \right)
\\
& =
\frac{1}{\alpha} \left( \sum_{i=1}^{n} p_{i}^{\alpha} \right)^{\frac{1}{\alpha} - 1} \left( \sum_{i=1}^{n} \frac{ \mathrm{d} }{ \mathrm{d} p_{[k]} } (p_{[i]}^{\alpha}) \right)
\\
& =
\frac{1}{\alpha} \left( \sum_{i=1}^{n} p_{i}^{\alpha} \right)^{\frac{1}{\alpha} - 1} \left( \frac{ \mathrm{d} }{ \mathrm{d} p_{[k]} } (p_{[k]}^{\alpha}) + \sum_{i=1 : i \neq k}^{n} \frac{ \mathrm{d} }{ \mathrm{d} p_{[k]} } (p_{[i]}^{\alpha}) \right)
\\
& =
\frac{1}{\alpha} \left( \sum_{i=1}^{n} p_{i}^{\alpha} \right)^{\frac{1}{\alpha} - 1} \left( \alpha \, p_{[k]}^{\alpha-1} + \sum_{i=1 : i \neq k}^{n} \frac{ \mathrm{d} }{ \mathrm{d} p_{[k]} } (p_{[i]}^{\alpha}) \right)
\label{eq:diff_norm_pk_halfway} \\
& =
\frac{1}{\alpha} \left( \sum_{i=1}^{n} p_{i}^{\alpha} \right)^{\frac{1}{\alpha} - 1} \left( \alpha \, p_{[k]}^{\alpha-1} + \sum_{i=1 : i \neq k}^{n} \left( \frac{ \mathrm{d} p_{[i]} }{ \mathrm{d} p_{[k]} } \right) \left( \frac{ \mathrm{d} }{ \mathrm{d} p_{[i]} } (p_{[i]}^{\alpha}) \right) \right)
\\
& =
\frac{1}{\alpha} \left( \sum_{i=1}^{n} p_{i}^{\alpha} \right)^{\frac{1}{\alpha} - 1} \left( \alpha \, p_{[k]}^{\alpha-1} + \sum_{i=1 : i \neq k}^{n} \left( \frac{ \mathrm{d} p_{[i]} }{ \mathrm{d} p_{[k]} } \right) (\alpha \, p_{[i]}^{\alpha-1}) \right)
\\
& =
\left( \sum_{i=1}^{n} p_{i}^{\alpha} \right)^{\frac{1}{\alpha} - 1} \left( p_{[k]}^{\alpha-1} + \sum_{i=1 : i \neq k}^{n} \left( \frac{ \mathrm{d} p_{[i]} }{ \mathrm{d} p_{[k]} } \right) (p_{[i]}^{\alpha-1}) \right)
\\
& =
\left( \sum_{i=1}^{n} p_{i}^{\alpha} \right)^{\frac{1}{\alpha} - 1} \left( p_{[k]}^{\alpha-1} + \sum_{i=1}^{k-1} \left( \frac{ \mathrm{d} p_{[i]} }{ \mathrm{d} p_{[k]} } \right) (p_{[i]}^{\alpha-1}) + \sum_{j=k+1}^{n} \left( \frac{ \mathrm{d} p_{[j]} }{ \mathrm{d} p_{[k]} } \right) (p_{[j]}^{\alpha-1}) \right)
\\
& \overset{\eqref{eq:equal_k+1_to_n}}{=}
\left( \sum_{i=1}^{n} p_{i}^{\alpha} \right)^{\frac{1}{\alpha} - 1} \left( p_{[k]}^{\alpha-1} + \sum_{i=1}^{k-1} \left( \frac{ \mathrm{d} p_{[i]} }{ \mathrm{d} p_{[k]} } \right) (p_{[i]}^{\alpha-1}) + (p_{[n]}^{\alpha-1}) \sum_{j=k+1}^{n} \left( \frac{ \mathrm{d} p_{[j]} }{ \mathrm{d} p_{[k]} } \right) \right)
\\
& \overset{\eqref{eq:hypo1}}{=}
\left( \sum_{i=1}^{n} p_{i}^{\alpha} \right)^{\frac{1}{\alpha} - 1} \left( p_{[k]}^{\alpha-1} + \left( \frac{ \mathrm{d} p_{[1]} }{ \mathrm{d} p_{[k]} } \right) (p_{[1]}^{\alpha-1}) + (p_{[n]}^{\alpha-1}) \sum_{j=k+1}^{n} \left( \frac{ \mathrm{d} p_{[j]} }{ \mathrm{d} p_{[k]} } \right) \right)
\\
& \overset{\eqref{eq:hypo2}}{=}
\left( \sum_{i=1}^{n} p_{i}^{\alpha} \right)^{\frac{1}{\alpha} - 1} \left( p_{[k]}^{\alpha-1} + \left( \frac{ \mathrm{d} p_{[1]} }{ \mathrm{d} p_{[k]} } \right) (p_{[1]}^{\alpha-1}) + (p_{[n]}^{\alpha-1}) (n-k) \left( \frac{ \mathrm{d} p_{[n]} }{ \mathrm{d} p_{[k]} } \right) \right)
\\
& \overset{\eqref{eq:total_prob_hypo}}{=}
\left( \sum_{i=1}^{n} p_{i}^{\alpha} \right)^{\frac{1}{\alpha} - 1} \left( p_{[k]}^{\alpha-1} + \left( - 1 - (n-k) \frac{ \mathrm{d} p_{[n]} }{ \mathrm{d} p_{[k]} } \right) (p_{[1]}^{\alpha-1}) + (n-k) \left( \frac{ \mathrm{d} p_{[n]} }{ \mathrm{d} p_{[k]} } \right) (p_{[n]}^{\alpha-1}) \right)
\\
& =
\left( \sum_{i=1}^{n} p_{i}^{\alpha} \right)^{\frac{1}{\alpha} - 1} \left( p_{[k]}^{\alpha-1} - p_{[1]}^{\alpha-1} - (n-k) \left( \frac{ \mathrm{d} p_{[n]} }{ \mathrm{d} p_{[k]} } \right) (p_{[1]}^{\alpha-1}) + (n-k) \left( \frac{ \mathrm{d} p_{[n]} }{ \mathrm{d} p_{[k]} } \right) (p_{[n]}^{\alpha-1}) \right)
\\
& =
\left( \sum_{i=1}^{n} p_{i}^{\alpha} \right)^{\frac{1}{\alpha} - 1} \left( (p_{[k]}^{\alpha-1} - p_{[1]}^{\alpha-1}) + (n-k) \left( \frac{ \mathrm{d} p_{[n]} }{ \mathrm{d} p_{[k]} } \right) (p_{[n]}^{\alpha-1} - p_{[1]}^{\alpha-1}) \right)
\\
& \overset{\eqref{eq:total_entropy_hypo}}{=}
\left( \sum_{i=1}^{n} p_{i}^{\alpha} \right)^{\frac{1}{\alpha} - 1} \left( (p_{[k]}^{\alpha-1} - p_{[1]}^{\alpha-1}) + (n-k) \left( - \frac{ 1 }{ n-k } \left( \frac{ \ln p_{[1]} - \ln p_{[k]} }{ \ln p_{[1]} - \ln p_{[n]} } \right) \right) (p_{[n]}^{\alpha-1} - p_{[1]}^{\alpha-1}) \right)
\\
& =
\left( \sum_{i=1}^{n} p_{i}^{\alpha} \right)^{\frac{1}{\alpha} - 1} \left( (p_{[k]}^{\alpha-1} - p_{[1]}^{\alpha-1}) - \left( \frac{ \ln p_{[1]} - \ln p_{[k]} }{ \ln p_{[1]} - \ln p_{[n]} } \right) (p_{[n]}^{\alpha-1} - p_{[1]}^{\alpha-1}) \right)
\\
& =
\left( \sum_{i=1}^{n} p_{i}^{\alpha} \right)^{\frac{1}{\alpha} - 1} \left( p_{[n]}^{\alpha-1} - p_{[1]}^{\alpha-1} \right) \left( \frac{ p_{[k]}^{\alpha-1} - p_{[1]}^{\alpha-1} }{ p_{[n]}^{\alpha-1} - p_{[1]}^{\alpha-1} } - \frac{ \ln p_{[1]} - \ln p_{[k]} }{ \ln p_{[1]} - \ln p_{[n]} } \right)
\\
& =
\left( \sum_{i=1}^{n} p_{i}^{\alpha} \right)^{\frac{1}{\alpha} - 1} \left( p_{[n]}^{\alpha-1} - p_{[1]}^{\alpha-1} \right) \left( \frac{ p_{[1]}^{\alpha-1} \left( \left( \frac{ p_{[k]} }{ p_{[1]} } \right)^{\alpha-1} - 1 \right) }{ p_{[1]}^{\alpha-1} \left( \left( \frac{ p_{[n]} }{ p_{[1]} } \right)^{\alpha-1} - 1 \right) } - \frac{ \ln p_{[1]} - \ln p_{[k]} }{ \ln p_{[1]} - \ln p_{[n]} } \right)
\\
& =
\left( \sum_{i=1}^{n} p_{i}^{\alpha} \right)^{\frac{1}{\alpha} - 1} \left( p_{[n]}^{\alpha-1} - p_{[1]}^{\alpha-1} \right) \left( \frac{ \left( \frac{ p_{[k]} }{ p_{[1]} } \right)^{\alpha-1} - 1 }{ \left( \frac{ p_{[n]} }{ p_{[1]} } \right)^{\alpha-1} - 1 } - \frac{ \ln p_{[1]} - \ln p_{[k]} }{ \ln p_{[1]} - \ln p_{[n]} } \right)
\\
& =
\left( \sum_{i=1}^{n} p_{i}^{\alpha} \right)^{\frac{1}{\alpha} - 1} \left( p_{[n]}^{\alpha-1} - p_{[1]}^{\alpha-1} \right) \left( \frac{ \left( \frac{ p_{[1]} }{ p_{[k]} } \right)^{1 - \alpha} - 1 }{ \left( \frac{ p_{[1]} }{ p_{[n]} } \right)^{1 - \alpha} - 1 } - \frac{ \ln p_{[1]} - \ln p_{[k]} }{ \ln p_{[1]} - \ln p_{[n]} } \right)
\\
& =
\left( \sum_{i=1}^{n} p_{i}^{\alpha} \right)^{\frac{1}{\alpha} - 1} \left( p_{[n]}^{\alpha-1} - p_{[1]}^{\alpha-1} \right) \left( \frac{ \ln_{\alpha} \frac{ p_{[1]} }{ p_{[k]} } }{ \ln_{\alpha} \frac{ p_{[1]} }{ p_{[n]} } } - \frac{ \ln \frac{ p_{[1]} }{ p_{[k]} } }{ \ln \frac{ p_{[1]} }{ p_{[n]} } } \right) .
\end{align}
Hence, we can see that
\begin{align}
\sgn \! \left( \frac{ \mathrm{d} \| \bvec{p} \|_{\alpha} }{ \mathrm{d} p_{[k]} } \right)
& =
\sgn \! \left( \left( \sum_{i=1}^{n} p_{i}^{\alpha} \right)^{\frac{1}{\alpha} - 1} \left( p_{[n]}^{\alpha-1} - p_{[1]}^{\alpha-1} \right) \left( \frac{ \ln_{\alpha} \frac{ p_{[1]} }{ p_{[k]} } }{ \ln_{\alpha} \frac{ p_{[1]} }{ p_{[n]} } } - \frac{ \ln \frac{ p_{[1]} }{ p_{[k]} } }{ \ln \frac{ p_{[1]} }{ p_{[n]} } } \right) \right)
\\
& =
\underbrace{ \sgn \! \left( \left( \sum_{i=1}^{n} p_{i}^{\alpha} \right)^{\frac{1}{\alpha} - 1} \right) }_{ = 1 } \cdot \, \sgn \! \left( p_{[n]}^{\alpha-1} - p_{[1]}^{\alpha-1} \right) \cdot \sgn \! \left( \frac{ \ln_{\alpha} \frac{ p_{[1]} }{ p_{[k]} } }{ \ln_{\alpha} \frac{ p_{[1]} }{ p_{[n]} } } - \frac{ \ln \frac{ p_{[1]} }{ p_{[k]} } }{ \ln \frac{ p_{[1]} }{ p_{[n]} } } \right)
\\
& =
\sgn \! \left( p_{[n]}^{\alpha-1} - p_{[1]}^{\alpha-1} \right) \cdot \sgn \! \left( \frac{ \ln_{\alpha} \frac{ p_{[1]} }{ p_{[k]} } }{ \ln_{\alpha} \frac{ p_{[1]} }{ p_{[n]} } } - \frac{ \ln \frac{ p_{[1]} }{ p_{[k]} } }{ \ln \frac{ p_{[1]} }{ p_{[n]} } } \right)
\label{eq:diff1_norm_pk_1}
\end{align}
for $\alpha \in (-\infty, 0) \cup (0, +\infty)$.
%For $\bvec{p} \in \mathcal{P}_{n}$, note that $H( \bvec{p} ) < \ln n$ if and only if $p_{[1]} > p_{[n]}$ since $\bvec{p} \neq \bvec{u}_{n}$;
Since $\bvec{p} \neq \bvec{u}_{n}$, i.e., $p_{[1]} > p_{[n]}$, we readily see that
\begin{align}
\sgn \! \left( p_{[n]}^{\alpha-1} - p_{[1]}^{\alpha-1} \right)
& =
\begin{cases}
1
& \mathrm{if} \ \alpha < 1 , \\
0
& \mathrm{if} \ \alpha = 1 , \\
-1
& \mathrm{if} \ \alpha > 1 .
\end{cases}
\label{eq:sign_pn-p1}
\end{align}
Moreover, for $1 \le \frac{ p_{[1]} }{ p_{[k]} } \le \frac{ p_{[1]} }{ p_{[n]} } \ (\frac{ p_{[1]} }{ p_{[n]} } \neq 1)$, we observe from Lemma \ref{lem:frac_qlog} that
\begin{align}
\sgn \! \left( \frac{ \ln_{\alpha} \frac{ p_{[1]} }{ p_{[k]} } }{ \ln_{\alpha} \frac{ p_{[1]} }{ p_{[n]} } } - \frac{ \ln \frac{ p_{[1]} }{ p_{[k]} } }{ \ln \frac{ p_{[1]} }{ p_{[n]} } } \right)
& =
\begin{cases}
1
& \mathrm{if} \ \alpha > 1 \ \mathrm{and} \ p_{[1]} > p_{[k]} > p_{[n]} , \\
0
& \mathrm{if} \ \alpha = 1 \ \mathrm{or} \ p_{[1]} = p_{[k]} \ \mathrm{or} \ p_{[k]} = p_{[n]} , \\
-1
& \mathrm{if} \ \alpha < 1 \ \mathrm{and} \ p_{[1]} > p_{[k]} > p_{[n]} .
\end{cases}
\label{eq:sign_f(p1pkpn)}
\end{align}
Therefore, under the constraints \eqref{eq:fixed_H}, \eqref{eq:equal_k+1_to_n}, \eqref{eq:hypo1}, and \eqref{eq:hypo2}, we have
\begin{align}
\sgn \! \left( \frac{ \mathrm{d} \| \bvec{p} \|_{\alpha} }{ \mathrm{d} p_{[k]} } \right)
& \overset{\eqref{eq:diff1_norm_pk_1}}{=}
\sgn \! \left( p_{[n]}^{\alpha-1} - p_{[1]}^{\alpha-1} \right) \cdot \sgn \! \left( \frac{ \ln_{\alpha} \frac{ p_{[1]} }{ p_{[k]} } }{ \ln_{\alpha} \frac{ p_{[1]} }{ p_{[n]} } } - \frac{ \ln \frac{ p_{[1]} }{ p_{[k]} } }{ \ln \frac{ p_{[1]} }{ p_{[n]} } } \right)
\\
& =
\begin{cases}
0
& \mathrm{if} \ \alpha = 1 \ \mathrm{or} \ p_{[1]} = p_{[k]} \ \mathrm{or} \ p_{[k]} = p_{[n]} , \\
-1
& \mathrm{if} \ \alpha \neq 1 \ \mathrm{and} \ p_{[1]} > p_{[k]} > p_{[n]}
\end{cases}
\label{eq:sign_diff_norm}
\end{align}
for $\alpha \in (-\infty, 0) \cup (0, +\infty)$, where the last equality follows from \eqref{eq:sign_pn-p1} and \eqref{eq:sign_f(p1pkpn)}.
Hence, we have that $\| \bvec{p} \|_{\alpha}$ with a fixed $\alpha \in (-\infty, 0) \cup(0, 1) \cup (1, +\infty)$ is strictly decreasing for $p_{[k]}$ under the constraints \eqref{eq:fixed_H}, \eqref{eq:equal_k+1_to_n}, \eqref{eq:hypo1}, and \eqref{eq:hypo2}.

Using the above results, we now prove this lemma.
If $p_{[k]} = p_{[k+1]}$, then we reset the index $k \in \{ 3, 4, \dots, n-1 \}$ to $k - 1$;
namely, we now choose the index $k \in \{ 2, 3, \dots, n-1 \}$ to satisfy the following inequalities:
\begin{align}
p_{[1]} \ge p_{[2]} \ge \dots \ge p_{[k-1]} \ge p_{[k]} > p_{[k+1]} = p_{[k+2]} = \dots = p_{[n]} \ge 0 .
\label{eq:choose_k}
\end{align}
Then, we consider to decrease $p_{[k]}$ under the constraints of \eqref{eq:fixed_H}, \eqref{eq:equal_k+1_to_n}, \eqref{eq:hypo1}, and \eqref{eq:hypo2}.
It follows from \eqref{eq:sign_d1dk} that $p_{[1]}$ is strictly increased by according to decreasing $p_{[k]}$.
Hence, if $p_{[k]}$ is decreased, then the condition $p_{[1]} > p_{[2]}$ must be held.
Similarly, it follows from \eqref{eq:hypo2} and \eqref{eq:sign_dndk} that, for all $j \in \{ k+1, k+2, \dots, n \}$, $p_{[j]}$ is also strictly increased by according to decreasing $p_{[k]}$.
Hence, if $p_{[k]}$ is decreased, then the condition $p_{[k+1]} = p_{[k+2]} = \dots = p_{[n]} > 0$ must be held.
Let $\bvec{q} = (q_{1}, q_{2}, \dots, q_{n})$ denote the probability vector that made from $\bvec{p}$ by continuing the above operation until to satisfy $p_{[k]} = p_{[k+1]}$ under the conditions of \eqref{eq:fixed_H}, \eqref{eq:equal_k+1_to_n}, \eqref{eq:hypo1}, \eqref{eq:hypo2}, and \eqref{eq:choose_k}.
Namely, the probability vector $\bvec{q}$ satisfies the following inequalities:
\begin{align}
q_{[1]} > q_{[2]} \ge q_{[3]} \ge \dots \ge q_{[k-1]} > q_{[k]} = q_{[k+1]} = \dots = q_{[n]} > 0 .
\end{align}
Since $\bvec{q}$ is made from $\bvec{p}$ under the constraint \eqref{eq:fixed_H}, note that
\begin{align}
H( \bvec{p} ) = H( \bvec{q} ) .
\end{align}
Moreover, it follows from \eqref{eq:sign_diff_norm} that $\| \bvec{p} \|_{\alpha}$ with a fixed $\alpha \in (0, 1) \cup (1, +\infty)$ is also strictly increased by according to decreasing $p_{[k]}$;
that is, we observe that
\begin{align}
\| \bvec{p} \|_{\alpha} \le \| \bvec{q} \|_{\alpha}
\end{align}
for $\alpha \in (0, 1) \cup (1, +\infty)$.
Repeating these operation until to satisfy $k = 2$ and $p_{[k]} = p_{[n]}$, we have that
\begin{align}
H( \bvec{p} )
& =
H_{\sbvec{v}_{n}}( p ) ,
\\
\| \bvec{p} \|_{\alpha}
& \le
\| \bvec{v}_{n}( p ) \|_{\alpha}
\end{align}
for all $\alpha \in (0, 1) \cup (1, +\infty)$ and some $p \in [0, \frac{1}{n}]$.
That completes the proof of Lemma \ref{lem:vector_v}.
\end{IEEEproof}

\begin{lemma}
\label{lem:vector_w}
For any $n \ge 2$ and any $\bvec{p} \in \mathcal{P}_{n}$, there exists $p \in [\frac{1}{n}, 1]$ such that $H_{\sbvec{w}_{n}}( p ) = H( \bvec{p} )$ and $\| \bvec{w}_{n}( p ) \|_{\alpha} \le \| \bvec{p} \|_{\alpha}$ for all $\alpha \in (0, \infty)$.
\end{lemma}

\begin{IEEEproof}[Proof of Lemma \ref{lem:vector_w}]
This proof is similar to the proof of Lemma \ref{lem:vector_v}.
If $n = 2$, then it can be easily seen that $\bvec{p}_{\downarrow} = \bvec{w}_{2}( p )$ for any $\bvec{p} \in \mathcal{P}_{2}$ and some $p \in [\frac{1}{2}, 1]$;
therefore, the lemma obviously holds when $n = 2$.
Moreover, since
\begin{align}
H( \bvec{p} ) & = \ln n
& \iff &&
\bvec{p} & = \bvec{u}_{n} = \bvec{w}_{n}( {\textstyle \frac{1}{n}} ),
\\
H( \bvec{p} ) & = 0
& \iff &&
\bvec{p}_{\downarrow} & = \bvec{d}_{n} = \bvec{w}_{n}( 1 ) ,
\end{align}
the lemma obviously holds if $H(\bvec{p}) \in \{ 0, \ln n \}$.
Furthermore, if $\bvec{p}_{\downarrow} = \bvec{w}_{n}( \frac{1}{m} )$ for an integer $2 \le m \le n-1$, then the lemma also obviously holds.
Thus, we omit the cases $n=2$, $H(\bvec{p}) \in \{ 0, \ln n \}$, and $\bvec{p}_{\downarrow} = \bvec{w}_{n}( \frac{1}{m} )$ in the analyses.
For a fixed $n \ge 3$ and a constant $A \in (0, \ln n)$, we assume for $\bvec{p} \in P_{n}$ that 
\begin{align}
H( \bvec{p} ) = A .
\label{eq:fixed_H_w}
\end{align}
For that $\bvec{p}$, let $k, l \in \{ 2, 3, \dots, n \} \ (k < l)$ be the indices such that $p_{[1]} = p_{[k-1]} > p_{[k+1]}$ and $p_{[l]} > p_{[l+1]} = 0$;
namely, the indices $k, l$ are chosen to satisfy the following inequalities:
\begin{align}
p_{[1]} = \dots = p_{[k-1]} \ge p_{[k]} \ge p_{[k+1]} \ge \dots \ge p_{[l-1]} \ge p_{[l]} > p_{[l+1]} = \dots = p_{[n]} = 0
\quad (p_{[k-1]} > p_{[k+1]}) .
\label{eq:equal_1_to_k-1}
\end{align}
Since
$
p_{1} + p_{2} + \dots + p_{n} = 1
$,
we observe as with \eqref{eq:total_diff_prob} that
\begin{align}
\sum_{i=1}^{n} p_{i}
=
1
\qquad \Longrightarrow \qquad
\sum_{i = 1 : i \neq k}^{n} \frac{ \mathrm{d} p_{[i]} }{ \mathrm{d} p_{[k]} }
=
- 1 .
\label{eq:total_diff_prob_w}
\end{align}
In this proof, we further assume that
\begin{align}
\frac{ \mathrm{d} p_{[i]} }{ \mathrm{d} p_{[k]} }
=
\frac{ \mathrm{d} p_{[1]} }{ \mathrm{d} p_{[k]} }
\label{eq:hypo1_w}
\end{align}
for $i \in \{ 2, 3, \dots, k-1 \}$, 
\begin{align}
\frac{ \mathrm{d} p_{[j]} }{ \mathrm{d} p_{[k]} }
=
1
\label{eq:hypo2_w}
\end{align}
for $j \in \{ k+1, k+2, \dots, l-1 \}$, and
\begin{align}
\frac{ \mathrm{d} p_{[m]} }{ \mathrm{d} p_{[k]} }
=
0
\label{eq:hypo3_w}
\end{align}
for $m \in \{ l+1, l+2, \dots, n \}$.
Note that \eqref{eq:hypo2_w} implies that, for all $j \in \{ k+1, k+2, \dots, l-1 \}$, the increase/decrease rate of $p_{[j]}$ is equivalent to the increase/decrease rate of $p_{[k]}$.
By constraints \eqref{eq:hypo1_w}, \eqref{eq:hypo2_w}, and \eqref{eq:hypo3_w}, we get
\begin{align}
&&
\sum_{i=1}^{n} p_{i}
& =
1
\\
& \ \overset{\eqref{eq:total_diff_prob_w}}{\Longrightarrow} \ &
\sum_{i = 1 : i \neq k}^{n} \frac{ \mathrm{d} p_{[i]} }{ \mathrm{d} p_{[k]} }
& =
- 1
\\
& \iff &
\sum_{i = 1}^{k-1} \frac{ \mathrm{d} p_{[i]} }{ \mathrm{d} p_{[k]} } + \sum_{j = k+1}^{l-1} \frac{ \mathrm{d} p_{[j]} }{ \mathrm{d} p_{[k]} } + \frac{ \mathrm{d} p_{[l]} }{ \mathrm{d} p_{[k]} } + \sum_{m = l+1}^{n} \frac{ \mathrm{d} p_{[m]} }{ \mathrm{d} p_{[l]} }
& =
- 1
\\
& \overset{\eqref{eq:hypo1_w}}{\iff} &
(k-1) \frac{ \mathrm{d} p_{[1]} }{ \mathrm{d} p_{[k]} } + \sum_{j = k+1}^{l-1} \frac{ \mathrm{d} p_{[j]} }{ \mathrm{d} p_{[k]} } + \frac{ \mathrm{d} p_{[l]} }{ \mathrm{d} p_{[k]} } + \sum_{m = l+1}^{n} \frac{ \mathrm{d} p_{[m]} }{ \mathrm{d} p_{[l]} }
& =
- 1
\\
& \overset{\eqref{eq:hypo2_w}}{\iff} &
(k-1) \frac{ \mathrm{d} p_{[1]} }{ \mathrm{d} p_{[k]} } + (l - k - 1) + \frac{ \mathrm{d} p_{[l]} }{ \mathrm{d} p_{[k]} } + \sum_{m = l+1}^{n} \frac{ \mathrm{d} p_{[m]} }{ \mathrm{d} p_{[l]} }
& =
- 1
\\
& \overset{\eqref{eq:hypo3_w}}{\iff} &
(k-1) \frac{ \mathrm{d} p_{[1]} }{ \mathrm{d} p_{[k]} } + (l - k - 1) + \frac{ \mathrm{d} p_{[l]} }{ \mathrm{d} p_{[k]} }
& =
- 1
\\
& \iff &
(k-1) \frac{ \mathrm{d} p_{[1]} }{ \mathrm{d} p_{[k]} } + \frac{ \mathrm{d} p_{[l]} }{ \mathrm{d} p_{[k]} }
& =
- (l-k)
\\
& \iff &
(k-1) \frac{ \mathrm{d} p_{[1]} }{ \mathrm{d} p_{[k]} } 
& =
- (l-k) - \frac{ \mathrm{d} p_{[l]} }{ \mathrm{d} p_{[k]} }
\\
& \iff &
\frac{ \mathrm{d} p_{[1]} }{ \mathrm{d} p_{[k]} }
& =
- \frac{1}{k-1} \left( (l - k) +  \frac{ \mathrm{d} p_{[l]} }{ \mathrm{d} p_{[k]} } \right) ,
\label{eq:total_prob_hypo_w}
\end{align}
where note in \eqref{eq:total_prob_hypo_w} that $k \ge 2$.
Moreover, since
$
H( \bvec{p} )
=
A
$,
we observe that
\begin{align}
&&
- \sum_{i = 1}^{n} p_{i} \ln p_{i}
& =
A
\\
& \ \overset{\eqref{eq:diff1_H_halfway}}{\Longrightarrow} \ &
- \sum_{i = 1 : i \neq k}^{n} \left( \frac{ \mathrm{d} p_{[i]} }{ \mathrm{d} p_{[k]} } \right) (\ln p_{[i]} + 1)
& =
\ln p_{[k]} + 1
\\
& \iff &
- \sum_{i = 1 : i \neq k}^{l} \left( \frac{ \mathrm{d} p_{[i]} }{ \mathrm{d} p_{[k]} } \right) (\ln p_{[i]} + 1) - \sum_{m = l+1}^{n} \left( \frac{ \mathrm{d} p_{[m]} }{ \mathrm{d} p_{[k]} } \right) (\ln p_{[m]} + 1)
& =
\ln p_{[k]} + 1
\\
& \overset{\text{(a)}}{\iff} &
- \sum_{i = 1 : i \neq k}^{l} \left( \frac{ \mathrm{d} p_{[i]} }{ \mathrm{d} p_{[k]} } \right) (\ln p_{[i]} + 1)
& =
\ln p_{[k]} + 1
\\
& \iff &
- \sum_{i = 1}^{k-1} \left( \frac{ \mathrm{d} p_{[i]} }{ \mathrm{d} p_{[k]} } \right) (\ln p_{[i]} + 1) - \sum_{j = k+1}^{l-1} \left( \frac{ \mathrm{d} p_{[j]} }{ \mathrm{d} p_{[k]} } \right) (\ln p_{[j]} + 1) - \left( \frac{ \mathrm{d} p_{[l]} }{ \mathrm{d} p_{[k]} } \right) (\ln p_{[l]} + 1)
& =
\ln p_{[k]} + 1
\\
& \overset{\eqref{eq:equal_1_to_k-1}}{\iff} &
- (\ln p_{[1]} + 1) \sum_{i = 1}^{k-1} \left( \frac{ \mathrm{d} p_{[i]} }{ \mathrm{d} p_{[k]} } \right) - \sum_{j = k+1}^{l-1} \left( \frac{ \mathrm{d} p_{[j]} }{ \mathrm{d} p_{[k]} } \right) (\ln p_{[j]} + 1) - \left( \frac{ \mathrm{d} p_{[l]} }{ \mathrm{d} p_{[k]} } \right) (\ln p_{[l]} + 1)
& =
\ln p_{[k]} + 1
\\
& \overset{\eqref{eq:hypo2_w}}{\iff} &
- (\ln p_{[1]} + 1) \sum_{i = 1}^{k-1} \left( \frac{ \mathrm{d} p_{[i]} }{ \mathrm{d} p_{[k]} } \right) - \sum_{j = k+1}^{l-1} (\ln p_{[j]} + 1) - \left( \frac{ \mathrm{d} p_{[l]} }{ \mathrm{d} p_{[k]} } \right) (\ln p_{[l]} + 1)
& =
\ln p_{[k]} + 1
\\
& \overset{\eqref{eq:hypo1_w}}{\iff} &
- (k-1) (\ln p_{[1]} + 1) \left( \frac{ \mathrm{d} p_{[1]} }{ \mathrm{d} p_{[k]} } \right) - \sum_{j = k+1}^{l-1} (\ln p_{[j]} + 1) - \left( \frac{ \mathrm{d} p_{[l]} }{ \mathrm{d} p_{[k]} } \right) (\ln p_{[l]} + 1)
& =
\ln p_{[k]} + 1
\\
& \overset{\eqref{eq:total_prob_hypo_w}}{\iff} &
(\ln p_{[1]} + 1) \left( (l - k) +  \frac{ \mathrm{d} p_{[l]} }{ \mathrm{d} p_{[k]} } \right) - \sum_{j = k+1}^{l-1} (\ln p_{[j]} + 1) - \left( \frac{ \mathrm{d} p_{[l]} }{ \mathrm{d} p_{[k]} } \right) (\ln p_{[l]} + 1)
& =
\ln p_{[k]} + 1
\\
& \iff &
(l - k) (\ln p_{[1]} + 1) - \sum_{j = k+1}^{l-1} (\ln p_{[j]} + 1) + \left( \frac{ \mathrm{d} p_{[l]} }{ \mathrm{d} p_{[k]} } \right) (\ln p_{[1]} - \ln p_{[l]})
& =
\ln p_{[k]} + 1
\\
& \iff &
(l - k) (\ln p_{[1]} + 1) - \sum_{j = k}^{l-1} (\ln p_{[j]} + 1) + \left( \frac{ \mathrm{d} p_{[l]} }{ \mathrm{d} p_{[k]} } \right) (\ln p_{[1]} - \ln p_{[l]})
& =
0
\\
& \iff &
\sum_{j = k}^{l-1} (\ln p_{[1]} - \ln p_{[j]}) + \left( \frac{ \mathrm{d} p_{[l]} }{ \mathrm{d} p_{[k]} } \right) (\ln p_{[1]} - \ln p_{[l]})
& =
0 ,
\label{eq:total_entropy_hypo_w_halfway}
\end{align}
where (a) follows from the fact that $\left( \frac{ \mathrm{d} p_{[m]} }{ \mathrm{d} p_{[k]} } \right) (\ln p_{[m]} + 1) = 0$ for $m \in \{ l+1, l+2, \dots, n \}$ since $\frac{ \mathrm{d} p_{[m]} }{ \mathrm{d} p_{[k]} } = 0$ (see Eq. \eqref{eq:hypo3_w}), $p_{[m]} = 0$ (see Eq. \eqref{eq:equal_1_to_k-1}), and $0 \ln 0 = 0$.
Hence, under the constraints \eqref{eq:fixed_H_w}, \eqref{eq:equal_1_to_k-1}, \eqref{eq:hypo1_w}, \eqref{eq:hypo2_w}, and \eqref{eq:hypo3_w}, we observe that
\begin{align}
\frac{ \mathrm{d} p_{[l]} }{ \mathrm{d} p_{[k]} }
=
- \frac{ \sum_{j = k}^{l-1} (\ln p_{[1]} - \ln p_{[j]}) }{ \ln p_{[1]} - \ln p_{[l]} } .
\label{eq:total_entropy_hypo_w}
\end{align}
We now check the sign of the right-hand side of \eqref{eq:total_entropy_hypo_w}.
Note that
\begin{align}
- (l - k) \left( \frac{ \ln p_{[1]} - \ln p_{[l-1]} }{ \ln p_{[1]} - \ln p_{[l]} } \right)
\le
\frac{ \mathrm{d} p_{[l]} }{ \mathrm{d} p_{[k]} }
\le
- (l - k) \left( \frac{ \ln p_{[1]} - \ln p_{[k]} }{ \ln p_{[1]} - \ln p_{[l]} } \right)
\label{ineq:total_entropy_hypo_w}
\end{align}
since $\ln p_{[k]} \ge \ln p_{[j]} \ge \ln p_{[l-1]}$ for all $j \in \{ k, k+1, \dots, l-1 \}$.
If $1 > p_{[1]} > p_{[k]} \ge p_{[l]} > 0$, then
\begin{align}
\frac{ \ln p_{[1]} - \ln p_{[k]} }{ \ln p_{[1]} - \ln p_{[l]} }
> 0
\end{align}
since
$
0 > \ln p_{[1]} > \ln p_{[k]} \ge \ln p_{[l]}
$;
therefore, we get for the upper bound of \eqref{ineq:total_entropy_hypo_w} that
\begin{align}
- (l - k) \left( \frac{ \ln p_{[1]} - \ln p_{[k]} }{ \ln p_{[1]} - \ln p_{[l]} } \right)
<
0
\label{eq:sign_dndk_1_w}
\end{align}
for $1 > p_{[1]} > p_{[k]} \ge p_{[l]} > 0$, where note that $l - k \ge 1$.
Moreover, if $1 > p_{[1]} = p_{[k]} > p_{[l]} > 0$, then
\begin{align}
- (l - k) \left( \frac{ \ln p_{[1]} - \ln p_{[k]} }{ \ln p_{[1]} - \ln p_{[l]} } \right)
& =
- (l - k) \left( \frac{ 0 }{ \ln p_{[1]} - \ln p_{[l]} } \right)
\\
& =
0 .
\label{eq:sign_dndk_2_w}
\end{align}
Combining \eqref{eq:sign_dndk_1_w} and \eqref{eq:sign_dndk_2_w}, we see that the upper bound of \eqref{ineq:total_entropy_hypo_w} is always nonpositive for $1 > p_{[1]} \ge p_{[k]} \ge p_{[l]} > 0 \ (p_{[1]} > p_{[l]})$;
that is, we observe under the constraints \eqref{eq:fixed_H_w}, \eqref{eq:equal_1_to_k-1}, \eqref{eq:hypo1_w}, \eqref{eq:hypo2_w}, and \eqref{eq:hypo3_w} that
\begin{align}
\sgn \! \left( \frac{ \mathrm{d} p_{[l]} }{ \mathrm{d} p_{[k]} } \right)
& \overset{\eqref{ineq:total_entropy_hypo_w}}{\le}
\sgn \! \left( - (l - k) \left( \frac{ \ln p_{[1]} - \ln p_{[k]} }{ \ln p_{[1]} - \ln p_{[l]} } \right) \right)
\\
& =
\begin{cases}
0
& \mathrm{if} \ p_{[1]} = p_{[k]} , \\
-1
& \mathrm{otherwise}
\end{cases}
\label{eq:sign_dndk_w}
\end{align}
for $1 > p_{[1]} \ge p_{[k]} \ge p_{[l]} > 0 \ (p_{[1]} > p_{[l]})$.
Note for the constraint \eqref{eq:fixed_H_w} that
\begin{align}
\lim_{p_{[l]} \to 0^{+}} H( p_{[1]}, p_{[2]}, \dots, p_{[l-1]}, p_{[l]}, 0, 0, \dots, 0 )
=
H( p_{[1]}, p_{[2]}, \dots, p_{[l-1]}, 0, 0, \dots, 0 )
\end{align}
since $\lim_{x \to 0^{+}} x \ln x = 0 \ln 0$ by the assumption $0 \ln 0 = 0$.
Thus, it follows from \eqref{eq:sign_dndk_w} that $p_{[l]}$ is strictly decreasing for $p_{[k]}$ under the constraints \eqref{eq:fixed_H_w}, \eqref{eq:equal_1_to_k-1}, \eqref{eq:hypo1_w}, \eqref{eq:hypo2_w}, and \eqref{eq:hypo3_w}.
Similarly, we check the sign of the right-hand side of \eqref{eq:total_prob_hypo_w}.
Substituting the lower bound of \eqref{ineq:total_entropy_hypo_w} into the right-hand side of \eqref{eq:total_prob_hypo_w}, we observe that
\begin{align}
\frac{ \mathrm{d} p_{[1]} }{ \mathrm{d} p_{[k]} }
\le
- \frac{l-k}{k-1} \left( 1 - \frac{ \ln p_{[1]} - \ln p_{[l-1]} }{ \ln p_{[1]} - \ln p_{[l]} } \right) .
\label{eq:total_entropy_hypo_w_UB}
\end{align}
If $1 > p_{[1]} \ge p_{[l-1]} > p_{[l]} > 0$, then
\begin{align}
\frac{ \ln p_{[1]} - \ln p_{[l-1]} }{ \ln p_{[1]} - \ln p_{[l]} }
<
1
\end{align}
since
$
0 > \ln p_{[1]} \ge \ln p_{[l-1]} > \ln p_{[l]}
$;
therefore, we get for the upper bound of \eqref{eq:total_entropy_hypo_w_UB} that
\begin{align}
- \frac{l-k}{k-1} \left( 1 - \frac{ \ln p_{[1]} - \ln p_{[l-1]} }{ \ln p_{[1]} - \ln p_{[l]} } \right)
<
0
\label{eq:sign_dndk_1_w2}
\end{align}
for $1 > p_{[1]} \ge p_{[l-1]} > p_{[l]} > 0$, where note that $\frac{l-k}{k-1} > 0$.
Moreover, if $1 > p_{[1]} = p_{[l-1]} > p_{[l]} > 0$, then
\begin{align}
- \frac{l-k}{k-1} \left( 1 - \frac{ \ln p_{[1]} - \ln p_{[l-1]} }{ \ln p_{[1]} - \ln p_{[l]} } \right)
& =
- \frac{l-k}{k-1} \left( 1 - \frac{ \ln p_{[1]} - \ln p_{[l]} }{ \ln p_{[1]} - \ln p_{[l]} } \right)
\\
& =
- \frac{l-k}{k-1} (1 - 1)
\\
& =
0 .
\label{eq:sign_dndk_1_w3}
\end{align}
It follows from \eqref{eq:sign_dndk_1_w2} and \eqref{eq:sign_dndk_1_w3} that the upper bound of \eqref{eq:total_entropy_hypo_w_UB} is always nonpositive for $1 > p_{[1]} \ge p_{[l-1]} \ge p_{[l]} > 0 \ (p_{[1]} > p_{[l]})$;
that is, we observe under the constraints \eqref{eq:fixed_H_w}, \eqref{eq:equal_1_to_k-1}, \eqref{eq:hypo1_w}, \eqref{eq:hypo2_w}, and \eqref{eq:hypo3_w} that
\begin{align}
\sgn \! \left( \frac{ \mathrm{d} p_{[1]} }{ \mathrm{d} p_{[k]} } \right)
& \overset{\eqref{eq:total_entropy_hypo_w_UB}}{\le}
\sgn \! \left( - \frac{l-k}{k-1} \left( 1 - \frac{ \ln p_{[1]} - \ln p_{[l-1]} }{ \ln p_{[1]} - \ln p_{[l]} } \right) \right)
\\
& =
\begin{cases}
0
& \mathrm{if} \ p_{[l-1]} = p_{[l]} , \\
-1
& \mathrm{otherwise}
\end{cases}
\label{eq:sign_d1dk_w}
\end{align}
for $1 > p_{[1]} \ge p_{[l-1]} \ge p_{[l]} > 0 \ (p_{[1]} > p_{[l]})$.
As with \eqref{eq:sign_dndk_w}, it follows from \eqref{eq:sign_d1dk_w} that, for all $i \in \{ 1, 2, \dots, k-1 \}$, $p_{[i]}$ is strictly decreasing for $p_{[k]}$ under the constraints \eqref{eq:fixed_H_w}, \eqref{eq:equal_1_to_k-1}, \eqref{eq:hypo1_w}, \eqref{eq:hypo2_w}, and \eqref{eq:hypo3_w}.

On the other hand, for a fixed $\alpha \in (0, 1) \cup (1, +\infty)$, we have
\begin{align}
\frac{ \mathrm{d} \| \bvec{p} \|_{\alpha} }{ \mathrm{d} p_{[k]} }
& \overset{\eqref{eq:diff_norm_pk_halfway}}{=}
\frac{1}{\alpha} \left( \sum_{i=1}^{n} p_{i}^{\alpha} \right)^{\frac{1}{\alpha} - 1} \left( \alpha \, p_{[k]}^{\alpha-1} + \sum_{i=1 : i \neq k}^{n} \frac{ \mathrm{d} }{ \mathrm{d} p_{[k]} } (p_{[i]}^{\alpha}) \right)
\\
& =
\frac{1}{\alpha} \left( \sum_{i=1}^{n} p_{i}^{\alpha} \right)^{\frac{1}{\alpha} - 1} \left( \alpha \, p_{[k]}^{\alpha-1} + \sum_{i=1 : i \neq k}^{l} \frac{ \mathrm{d} }{ \mathrm{d} p_{[k]} } (p_{[i]}^{\alpha}) + \sum_{m=l+1}^{n} \frac{ \mathrm{d} }{ \mathrm{d} p_{[k]} } (p_{[m]}^{\alpha}) \right)
\\
& \overset{\text{(a)}}{=}
\frac{1}{\alpha} \left( \sum_{i=1}^{n} p_{i}^{\alpha} \right)^{\frac{1}{\alpha} - 1} \left( \alpha \, p_{[k]}^{\alpha-1} + \sum_{i=1 : i \neq k}^{l} \frac{ \mathrm{d} }{ \mathrm{d} p_{[k]} } (p_{[i]}^{\alpha}) \right)
\\
& =
\frac{1}{\alpha} \left( \sum_{i=1}^{n} p_{i}^{\alpha} \right)^{\frac{1}{\alpha} - 1} \left( \alpha \, p_{[k]}^{\alpha-1} + \sum_{i=1 : i \neq k}^{l} \left( \frac{ \mathrm{d} p_{[i]} }{ \mathrm{d} p_{[k]} } \right) \left( \frac{ \mathrm{d} }{ \mathrm{d} p_{[i]} } (p_{[i]}^{\alpha}) \right) \right)
\\
& =
\frac{1}{\alpha} \left( \sum_{i=1}^{n} p_{i}^{\alpha} \right)^{\frac{1}{\alpha} - 1} \left( \alpha \, p_{[k]}^{\alpha-1} + \sum_{i=1 : i \neq k}^{l} \left( \frac{ \mathrm{d} p_{[i]} }{ \mathrm{d} p_{[k]} } \right) (\alpha \, p_{[i]}^{\alpha-1}) \right)
\\
& =
\left( \sum_{i=1}^{n} p_{i}^{\alpha} \right)^{\frac{1}{\alpha} - 1} \left( p_{[k]}^{\alpha-1} + \sum_{i=1 : i \neq k}^{l} \left( \frac{ \mathrm{d} p_{[i]} }{ \mathrm{d} p_{[k]} } \right) p_{[i]}^{\alpha-1} \right)
\\
& =
\left( \sum_{i=1}^{n} p_{i}^{\alpha} \right)^{\frac{1}{\alpha} - 1} \left( p_{[k]}^{\alpha-1} + \sum_{i=1}^{k-1} \left( \frac{ \mathrm{d} p_{[i]} }{ \mathrm{d} p_{[k]} } \right) p_{[i]}^{\alpha-1} + \sum_{j=k+1}^{l-1} \left( \frac{ \mathrm{d} p_{[j]} }{ \mathrm{d} p_{[k]} } \right) p_{[j]}^{\alpha-1} + \left( \frac{ \mathrm{d} p_{[l]} }{ \mathrm{d} p_{[k]} } \right) p_{[l]}^{\alpha-1} \right)
\\
& \overset{\eqref{eq:equal_1_to_k-1}}{=}
\left( \sum_{i=1}^{n} p_{i}^{\alpha} \right)^{\frac{1}{\alpha} - 1} \left( p_{[k]}^{\alpha-1} + p_{[1]}^{\alpha-1} \sum_{i=1}^{k-1} \left( \frac{ \mathrm{d} p_{[i]} }{ \mathrm{d} p_{[k]} } \right) + \sum_{j=k+1}^{l-1} \left( \frac{ \mathrm{d} p_{[j]} }{ \mathrm{d} p_{[k]} } \right) p_{[j]}^{\alpha-1} + \left( \frac{ \mathrm{d} p_{[l]} }{ \mathrm{d} p_{[k]} } \right) p_{[l]}^{\alpha-1} \right)
\\
& \overset{\eqref{eq:hypo1_w}}{=}
\left( \sum_{i=1}^{n} p_{i}^{\alpha} \right)^{\frac{1}{\alpha} - 1} \left( p_{[k]}^{\alpha-1} + p_{[1]}^{\alpha-1} (k-1) \left( \frac{ \mathrm{d} p_{[1]} }{ \mathrm{d} p_{[k]} } \right) + \sum_{j=k+1}^{l-1} \left( \frac{ \mathrm{d} p_{[j]} }{ \mathrm{d} p_{[k]} } \right) p_{[j]}^{\alpha-1} + \left( \frac{ \mathrm{d} p_{[l]} }{ \mathrm{d} p_{[k]} } \right) p_{[l]}^{\alpha-1} \right)
\\
& \overset{\eqref{eq:hypo2_w}}{=}
\left( \sum_{i=1}^{n} p_{i}^{\alpha} \right)^{\frac{1}{\alpha} - 1} \left( p_{[k]}^{\alpha-1} + p_{[1]}^{\alpha-1} (k-1) \left( \frac{ \mathrm{d} p_{[1]} }{ \mathrm{d} p_{[k]} } \right) + \sum_{j=k+1}^{l-1} p_{[j]}^{\alpha-1} + \left( \frac{ \mathrm{d} p_{[l]} }{ \mathrm{d} p_{[k]} } \right) p_{[l]}^{\alpha-1} \right)
\\
& \overset{\eqref{eq:total_prob_hypo_w}}{=}
\left( \sum_{i=1}^{n} p_{i}^{\alpha} \right)^{\frac{1}{\alpha} - 1} \left( p_{[k]}^{\alpha-1} - p_{[1]}^{\alpha-1} \left( (l - k) + \frac{ \mathrm{d} p_{[l]} }{ \mathrm{d} p_{[k]} } \right) + \sum_{j=k+1}^{l-1} p_{[j]}^{\alpha-1} + \left( \frac{ \mathrm{d} p_{[l]} }{ \mathrm{d} p_{[k]} } \right) p_{[l]}^{\alpha-1} \right)
\\
& =
\left( \sum_{i=1}^{n} p_{i}^{\alpha} \right)^{\frac{1}{\alpha} - 1} \left( \sum_{j=k}^{l-1} p_{[j]}^{\alpha-1} - p_{[1]}^{\alpha-1} (l - k)  + \left( \frac{ \mathrm{d} p_{[l]} }{ \mathrm{d} p_{[k]} } \right) (p_{[l]}^{\alpha-1} - p_{[1]}^{\alpha-1}) \right)
\\
& =
\left( \sum_{i=1}^{n} p_{i}^{\alpha} \right)^{\frac{1}{\alpha} - 1} \left( \sum_{j=k}^{l-1} (p_{[j]}^{\alpha-1} - p_{[1]}^{\alpha-1})  + \left( \frac{ \mathrm{d} p_{[l]} }{ \mathrm{d} p_{[k]} } \right) (p_{[l]}^{\alpha-1} - p_{[1]}^{\alpha-1}) \right)
\\
& \overset{\eqref{eq:total_entropy_hypo_w}}{=}
\left( \sum_{i=1}^{n} p_{i}^{\alpha} \right)^{\frac{1}{\alpha} - 1} \left( \sum_{j=k}^{l-1} (p_{[j]}^{\alpha-1} - p_{[1]}^{\alpha-1})  + \left( - \frac{ \sum_{j = k}^{l-1} (\ln p_{[1]} - \ln p_{[j]}) }{ \ln p_{[1]} - \ln p_{[l]} } \right) (p_{[l]}^{\alpha-1} - p_{[1]}^{\alpha-1}) \right)
\\
& =
\left( \sum_{i=1}^{n} p_{i}^{\alpha} \right)^{\frac{1}{\alpha} - 1} \left( \vphantom{\sum} p_{[l]}^{\alpha-1} - p_{[1]}^{\alpha-1} \right) \left( \frac{ \sum_{j=k}^{l-1} (p_{[j]}^{\alpha-1} - p_{[1]}^{\alpha-1}) }{ p_{[l]}^{\alpha-1} - p_{[1]}^{\alpha-1} } - \frac{ \sum_{j = k}^{l-1} (\ln p_{[1]} - \ln p_{[j]}) }{ \ln p_{[1]} - \ln p_{[l]} } \right)
\\
& =
\left( \sum_{i=1}^{n} p_{i}^{\alpha} \right)^{\frac{1}{\alpha} - 1} \left( \vphantom{\sum} p_{[l]}^{\alpha-1} - p_{[1]}^{\alpha-1} \right) \sum_{j=k}^{l-1} \left( \frac{ p_{[j]}^{\alpha-1} - p_{[1]}^{\alpha-1} }{ p_{[l]}^{\alpha-1} - p_{[1]}^{\alpha-1} } - \frac{ \ln p_{[1]} - \ln p_{[j]} }{ \ln p_{[1]} - \ln p_{[l]} } \right)
\\
& =
\left( \sum_{i=1}^{n} p_{i}^{\alpha} \right)^{\frac{1}{\alpha} - 1} \left( \vphantom{\sum} p_{[l]}^{\alpha-1} - p_{[1]}^{\alpha-1} \right) \sum_{j=k}^{l-1} \left( \frac{ \left( \frac{ p_{[1]} }{ p_{[j]} } \right)^{1-\alpha} - 1 }{ \left( \frac{ p_{[1]} }{ p_{[l]} } \right)^{1-\alpha} - 1 } - \frac{ \ln p_{[1]} - \ln p_{[j]} }{ \ln p_{[1]} - \ln p_{[l]} } \right)
\\
& =
\left( \sum_{i=1}^{n} p_{i}^{\alpha} \right)^{\frac{1}{\alpha} - 1} \left( \vphantom{\sum} p_{[l]}^{\alpha-1} - p_{[1]}^{\alpha-1} \right) \sum_{j=k}^{l-1} \left( \frac{ \ln_{\alpha} \frac{ p_{[1]} }{ p_{[j]} } }{ \ln_{\alpha} \frac{ p_{[1]} }{ p_{[l]} } } - \frac{ \ln \frac{ p_{[1]} }{ p_{[j]} } }{ \ln \frac{ p_{[1]} }{ p_{[l]} } } \right)
\label{eq:diff_norm_pk_w}
\end{align}
where (a) holds since the constraint \eqref{eq:hypo3_w} implies that $p_{[m]}$ is constant for $p_{[k]}$.
Hence, we can see that
\begin{align}
\sgn \! \left( \frac{ \mathrm{d} \| \bvec{p} \|_{\alpha} }{ \mathrm{d} p_{[k]} } \right)
& =
\sgn \! \left( \left( \sum_{i=1}^{n} p_{i}^{\alpha} \right)^{\frac{1}{\alpha} - 1} \left( \vphantom{\sum} p_{[l]}^{\alpha-1} - p_{[1]}^{\alpha-1} \right) \sum_{j=k}^{l-1} \left( \frac{ \ln_{\alpha} \frac{ p_{[1]} }{ p_{[j]} } }{ \ln_{\alpha} \frac{ p_{[1]} }{ p_{[l]} } } - \frac{ \ln \frac{ p_{[1]} }{ p_{[j]} } }{ \ln \frac{ p_{[1]} }{ p_{[l]} } } \right) \right)
\\
& =
\underbrace{ \sgn \! \left( \left( \sum_{i=1}^{n} p_{i}^{\alpha} \right)^{\frac{1}{\alpha} - 1} \right) }_{ = 1 } \cdot \, \sgn \! \left( \vphantom{\sum} p_{[l]}^{\alpha-1} - p_{[1]}^{\alpha-1} \right) \cdot \, \sgn \! \left( \sum_{j=k}^{l-1} \left( \frac{ \ln_{\alpha} \frac{ p_{[1]} }{ p_{[j]} } }{ \ln_{\alpha} \frac{ p_{[1]} }{ p_{[l]} } } - \frac{ \ln \frac{ p_{[1]} }{ p_{[j]} } }{ \ln \frac{ p_{[1]} }{ p_{[l]} } } \right) \right)
\\
& =
\sgn \! \left( \vphantom{\sum} p_{[l]}^{\alpha-1} - p_{[1]}^{\alpha-1} \right) \cdot \, \sgn \! \left( \sum_{j=k}^{l-1} \left( \frac{ \ln_{\alpha} \frac{ p_{[1]} }{ p_{[j]} } }{ \ln_{\alpha} \frac{ p_{[1]} }{ p_{[l]} } } - \frac{ \ln \frac{ p_{[1]} }{ p_{[j]} } }{ \ln \frac{ p_{[1]} }{ p_{[l]} } } \right) \right)
\label{eq:diff1_norm_pk_1_w}
\end{align}
for $\alpha \in (0, 1) \cup (1, \infty)$.
As with \eqref{eq:sign_pn-p1}, we readily see that
\begin{align}
\sgn \! \left( p_{[l]}^{\alpha-1} - p_{[1]}^{\alpha-1} \right)
& =
\begin{cases}
1
& \mathrm{if} \ \alpha < 1 , \\
0
& \mathrm{if} \ \alpha = 1 , \\
-1
& \mathrm{if} \ \alpha > 1
\end{cases}
\label{eq:sign_pl-p1_w}
\end{align}
for $p_{[1]} > p_{[l]} > 0$.
Moreover, since $1 \le \frac{ p_{[1]} }{ p_{[j]} } \le \frac{ p_{[1]} }{ p_{[l]} } \ (\frac{ p_{[1]} }{ p_{[l]} } \neq 1)$ for $j \in \{ k, k+1, \dots, l-1 \}$, we observe from Lemma \ref{lem:frac_qlog} that
\begin{align}
\sgn \! \left( \frac{ \ln_{\alpha} \frac{ p_{[1]} }{ p_{[j]} } }{ \ln_{\alpha} \frac{ p_{[1]} }{ p_{[l]} } } - \frac{ \ln \frac{ p_{[1]} }{ p_{[j]} } }{ \ln \frac{ p_{[1]} }{ p_{[l]} } } \right)
& =
\begin{cases}
1
& \mathrm{if} \ \alpha > 1 \ \mathrm{and} \ p_{[1]} > p_{[j]} > p_{[l]} , \\
0
& \mathrm{if} \ \alpha = 1 \ \mathrm{or} \ p_{[1]} = p_{[j]} \ \mathrm{or} \ p_{[j]} = p_{[l]} , \\
-1
& \mathrm{if} \ \alpha < 1 \ \mathrm{and} \ p_{[1]} > p_{[j]} > p_{[l]} .
\end{cases}
\end{align}
for $j \in \{ k, k+1, \dots, l-1 \}$;
and therefore, we have
\begin{align}
\sgn \! \left( \sum_{j=k}^{l-1} \left( \frac{ \ln_{\alpha} \frac{ p_{[1]} }{ p_{[j]} } }{ \ln_{\alpha} \frac{ p_{[1]} }{ p_{[l]} } } - \frac{ \ln \frac{ p_{[1]} }{ p_{[j]} } }{ \ln \frac{ p_{[1]} }{ p_{[l]} } } \right) \right)
& =
\begin{cases}
1
& \mathrm{if} \ \alpha > 1 \ \mathrm{and} \ (p_{[1]} > p_{[k]} \ge p_{[l]} \ \mathrm{or} \ p_{[1]} \ge p_{[k]} > p_{[l]}) , \\
0
& \mathrm{if} \ \alpha = 1 \ \mathrm{or} \ (p_{[1]} = p_{[k]} \ \mathrm{and} \ p_{[k+1]} = p_{[l]})  \ \mathrm{or} \ p_{[k]} = p_{[l]} , \\
-1
& \mathrm{if} \ \alpha < 1 \ \mathrm{and} \ (p_{[1]} > p_{[k]} \ge p_{[l]} \ \mathrm{or} \ p_{[1]} \ge p_{[k]} > p_{[l]})
\end{cases}
\label{eq:sign_f(p1pkpn)_w}
\end{align}
for $\bvec{p} \in \mathcal{P}_{n}$ under the constraint \eqref{eq:equal_1_to_k-1}.
Therefore, under the constraints \eqref{eq:fixed_H_w}, \eqref{eq:equal_1_to_k-1}, \eqref{eq:hypo1_w}, \eqref{eq:hypo2_w}, and \eqref{eq:hypo3_w}, we obtain
\begin{align}
\sgn \! \left( \frac{ \mathrm{d} \| \bvec{p} \|_{\alpha} }{ \mathrm{d} p_{[k]} } \right)
& \overset{\eqref{eq:diff1_norm_pk_1_w}}{=}
\sgn \! \left( \vphantom{\sum} p_{[l]}^{\alpha-1} - p_{[1]}^{\alpha-1} \right) \cdot \, \sgn \! \left( \sum_{j=k}^{l-1} \left( \frac{ \ln_{\alpha} \frac{ p_{[1]} }{ p_{[j]} } }{ \ln_{\alpha} \frac{ p_{[1]} }{ p_{[l]} } } - \frac{ \ln \frac{ p_{[1]} }{ p_{[j]} } }{ \ln \frac{ p_{[1]} }{ p_{[l]} } } \right) \right)
\\
& =
\begin{cases}
0
& \mathrm{if} \ \alpha = 1 \ \mathrm{or} \ (p_{[1]} = p_{[k]} \ \mathrm{and} \ p_{[k+1]} = p_{[l]})  \ \mathrm{or} \ p_{[k]} = p_{[l]} , \\
-1
& \mathrm{if} \ \alpha \neq 1 \ \mathrm{and} \ (p_{[1]} > p_{[k]} \ge p_{[l]} \ \mathrm{or} \ p_{[1]} \ge p_{[k]} > p_{[l]})
\end{cases}
\label{eq:sign_diff_norm_w}
\end{align}
for $\alpha \in (0, 1) \cup (1, +\infty)$, where the last equality follows from \eqref{eq:sign_pl-p1_w} and \eqref{eq:sign_f(p1pkpn)_w}.
Hence, we have that $\| \bvec{p} \|_{\alpha}$ with a fixed $\alpha \in (0, 1) \cup (1, +\infty)$ is strictly decreasing for $p_{[k]}$ under the constraints \eqref{eq:fixed_H_w}, \eqref{eq:equal_1_to_k-1}, \eqref{eq:hypo1_w}, \eqref{eq:hypo2_w}, and \eqref{eq:hypo3_w}.

Using the above results, we now prove this lemma.
Note that, if $p_{[k-1]} = p_{[k]}$ and $k = l-1$, then $\bvec{p}_{\downarrow} = \bvec{w}_{n}( p )$ for some $p \in [\frac{1}{n}, 1]$.
If $p_{[k-1]} = p_{[k]}$ and $k < l-1$, then we reset the index $k \in \{ 2, 3, \dots, n-2 \}$ to $k + 1$;
namely;
we now choose the indices $k, l \in \{ 2, 3, \dots, n \} \ (k < l)$ to satisfy the following inequalities:
\begin{align}
p_{[1]} = p_{[2]} = \dots = p_{[k-1]} > p_{[k]} \ge p_{[k+1]} \ge \dots \ge p_{[l-1]} \ge p_{[l]} > p_{[l+1]} = p_{[l+2]} = \dots = p_{[n]} = 0 .
\label{eq:choose_k2}
\end{align}
Then, we consider to increase $p_{[k]}$ under the constraints of \eqref{eq:fixed_H_w}, \eqref{eq:equal_1_to_k-1}, \eqref{eq:hypo1_w}, \eqref{eq:hypo2_w}, and \eqref{eq:hypo3_w}.
Note that the constraint \eqref{eq:hypo2_w} implies that, for all $j \in \{ k+1, k+2, \dots, l-1 \}$, $p_{[j]}$ is strictly increased with the same speed of increasing $p_{[k]}$.
It follows from \eqref{eq:hypo1_w} and \eqref{eq:sign_d1dk_w} that, for all $i \in \{ 1, 2, \dots, k-1 \}$, $p_{[i]}$ is strictly decreased by according to increasing $p_{[k]}$.
Hence, if $p_{[k]}$ is decreased, then there is a possibility that $p_{[1]} = \dots = p_{[k-1]} = p_{[k]}$.
Similarly, it follows from \eqref{eq:sign_dndk_w} that $p_{[l]}$ is also strictly decreased by according to increasing $p_{[k]}$.
Hence, if $p_{[k]}$ is decreased, then there is a possibility that $p_{[l]} = p_{[l+1]} = \dots = p_{[n]} = 0$.
Let $\bvec{q} = (q_{1}, q_{2}, \dots, q_{n})$ denotes the probability vector that made from $\bvec{p}$ by continuing the above operation until to satisfy $p_{[1]} = p_{[k]}$ or $p_{[l]} = 0$ under the conditions of \eqref{eq:fixed_H_w}, \eqref{eq:hypo1_w}, \eqref{eq:hypo2_w}, and \eqref{eq:hypo3_w}.
Namely, the probability vector $\bvec{q}$ satisfies either
\begin{align}
q_{[1]} = q_{[2]} = \dots = q_{[k-1]} = q_{[k]} \ge q_{[k+1]} \ge q_{[k+2]} \ge \dots \ge q_{[l-1]} > q_{[l]} \ge q_{[l+1]} = q_{[l+2]} = \dots = q_{[n]} = 0
\label{ineq:q_w_1}
\end{align}
or
\begin{align}
q_{[1]} = q_{[2]} = \dots = q_{[k-1]} \ge q_{[k]} \ge q_{[k+1]} \ge q_{[k+2]} \ge \dots \ge q_{[l-1]} > q_{[l]} = q_{[l+1]} = q_{[l+2]} = \dots = q_{[n]} = 0 .
\label{ineq:q_w_2}
\end{align}
Note that there is a possibility that both of \eqref{ineq:q_w_1} and \eqref{ineq:q_w_2} hold;
that is,
\begin{align}
q_{[1]} = q_{[2]} = \dots = q_{[k-1]} = q_{[k]} \ge q_{[k+1]} \ge q_{[k+2]} \ge \dots \ge q_{[l-1]} > q_{[l]} = q_{[l+1]} = q_{[l+2]} = \dots = q_{[n]} = 0
\end{align}
holds.
Since $\bvec{q}$ is made under the constraint \eqref{eq:fixed_H_w}, note that
\begin{align}
H( \bvec{p} ) = H( \bvec{q} ) .
\end{align}
Moreover, it follows from \eqref{eq:sign_diff_norm_w} that $\| \bvec{p} \|_{\alpha}$ with a fixed $\alpha \in (0, 1) \cup (1, +\infty)$ is also strictly decreased by according to increasing $p_{[k]}$;
therefore, we observe that
\begin{align}
\| \bvec{p} \|_{\alpha} \ge \| \bvec{q} \|_{\alpha}
\end{align}
for $\alpha \in (0, 1) \cup (1, +\infty)$.
Repeating these operation until to satisfy $k = l-1$ and $p_{[1]} = p_{[k]} > p_{[l]} \ge p_{[l-1]} = p_{[n]} = 0$, we have that
\begin{align}
H( \bvec{p} )
& =
H_{\sbvec{w}_{n}}( p ) ,
\\
\| \bvec{p} \|_{\alpha}
& \ge
\| \bvec{w}_{n}( p ) \|_{\alpha}
\end{align}
for all $\alpha \in (0, 1) \cup (1, +\infty)$ and some $p \in [\frac{1}{n}, 1]$.
That completes the proof of Lemma \ref{lem:vector_w}.
\end{IEEEproof}

Lemmas \ref{lem:vector_v} and \ref{lem:vector_w} are derived by using Lemma \ref{lem:frac_qlog}.
Lemmas \ref{lem:vector_v} and \ref{lem:vector_w} imply that the distributions $\bvec{v}_{n}( \cdot )$ and $\bvec{w}_{n}( \cdot )$ have extremal properties in the sense of a relation between the Shannon entropy and the $\ell_{\alpha}$-norm.
Then, we can derive tight bounds of $\ell_{\alpha}$-norms with a fixed Shannon entropy as follows:

\begin{theorem}
\label{th:extremes}
Let $\bar{\bvec{v}}_{n}( \bvec{p} ) \triangleq \bvec{v}_{n}( H_{\sbvec{v}_{n}}^{-1}( H( \bvec{p} ) ) )$ and $\bar{\bvec{w}}_{n}( \bvec{p} ) \triangleq \bvec{w}_{n}( H_{\sbvec{w}_{n}}^{-1}( H( \bvec{p} ) ) )$ for $\bvec{p} \in \mathcal{P}_{n}$. %, where $H_{\sbvec{v}_{n}}^{-1}( \cdot )$ and $H_{\sbvec{w}_{n}}^{-1}( \cdot )$ are defined in Definition \ref{def:inverseH}.
Then, we observe that
\begin{align}
\| \bar{\bvec{w}}_{n}( \bvec{p} ) \|_{\alpha} \le \| \bvec{p} \|_{\alpha} \le \| \bar{\bvec{v}}_{n}( \bvec{p} ) \|_{\alpha}
\label{ineq:extremes}
\end{align}
for any $n \ge 2$, any $\bvec{p} \in \mathcal{P}_{n}$, and any $\alpha \in (0, \infty)$.
\end{theorem}

\begin{IEEEproof}[Proof of Theorem \ref{th:extremes}]
It follows from Lemmas \ref{lem:vector_v} and \ref{lem:vector_w} that, for any $n \ge 2$ and any $\bvec{p} \in \mathcal{P}_{n}$, there exist $p \in [0, \frac{1}{n}]$ and $p^{\prime} \in [\frac{1}{n}, 1]$ such that
\begin{align}
H_{\sbvec{w}_{n}}( p^{\prime} )
& =
H( \bvec{p} )
=
H_{\sbvec{v}_{n}}( p ) ,
\\
\| \bvec{w}_{n}( p^{\prime} ) \|_{\alpha}
& \le
\| \bvec{p} \|_{\alpha}
\le
\| \bvec{v}_{n}( p ) \|_{\alpha}
\end{align}
for all $\alpha \in (0, +\infty)$.
Then, we now consider $\bvec{q}, \bvec{q}^{\prime} \in \mathcal{P}_{n}$ such that
\begin{align}
H( \bvec{q}^{\prime} )
=
H_{\sbvec{w}_{n}}( p^{\prime} )
& =
H_{\sbvec{v}_{n}}( p )
=
H( \bvec{q} ) ,
\label{eq:sameH_q_1} \\
\| \bvec{q}^{\prime} \|_{\alpha}
\le
\| \bvec{w}_{n}( p^{\prime} ) \|_{\alpha}
& \le
\| \bvec{v}_{n}( p ) \|_{\alpha}
\le
\| \bvec{q} \|_{\alpha}
\end{align}
for $\alpha \in (0, +\infty)$.
It also follows from Lemmas \ref{lem:vector_v} and \ref{lem:vector_w} that there exist $q \in [0, \frac{1}{n}]$ and $q^{\prime} \in [\frac{1}{n}, 1]$ such that
\begin{align}
H_{\sbvec{w}_{n}}( q^{\prime} )
=
H( \bvec{q}^{\prime} )
& =
H( \bvec{q} )
=
H_{\sbvec{v}_{n}}( q ) ,
\label{eq:sameH_q_2} \\
\| \bvec{w}_{n}( q^{\prime} ) \|_{\alpha}
\le
\| \bvec{q}^{\prime} \|_{\alpha}
& \le
\| \bvec{q} \|_{\alpha}
\le
\| \bvec{v}_{n}( q ) \|_{\alpha}
\end{align}
for $\alpha \in (0, +\infty)$.
Note from \eqref{eq:sameH_q_1} and \eqref{eq:sameH_q_2} that
\begin{align}
H_{\sbvec{v}_{n}}( p )
& =
H_{\sbvec{v}_{n}}( q ) ,
\\
H_{\sbvec{w}_{n}}( p^{\prime} )
& =
H_{\sbvec{w}_{n}}( q^{\prime} ) .
\end{align}
Note that it follows from Lemmas \ref{lem:Hv} and \ref{lem:Hw} that $H_{\sbvec{v}_{n}}( p )$ and $H_{\sbvec{w}_{n}}( p^{\prime} )$ are both bijective functions of $p \in [0, \frac{1}{n}]$ and $p^{\prime} \in [\frac{1}{n}, 1]$, respectively.
Therefore, we get
\begin{align}
p
& =
q ,
\\
p^{\prime}
& =
q^{\prime} ,
\end{align}
which imply that, for $\bvec{q}$ and $\bvec{q}^{\prime}$, the following equalities must be held:
\begin{align}
\| \bvec{v}_{n}( p ) \|_{\alpha}
& =
\| \bvec{q} \|_{\alpha}
=
\| \bvec{v}_{n}( q ) \|_{\alpha} ,
\\
\| \bvec{w}_{n}( p^{\prime} ) \|_{\alpha}
& =
\| \bvec{q}^{\prime} \|_{\alpha}
=
\| \bvec{w}_{n}( q^{\prime} ) \|_{\alpha} .
\end{align}
That completes the proof of Theorem \ref{th:extremes}.
\end{IEEEproof}

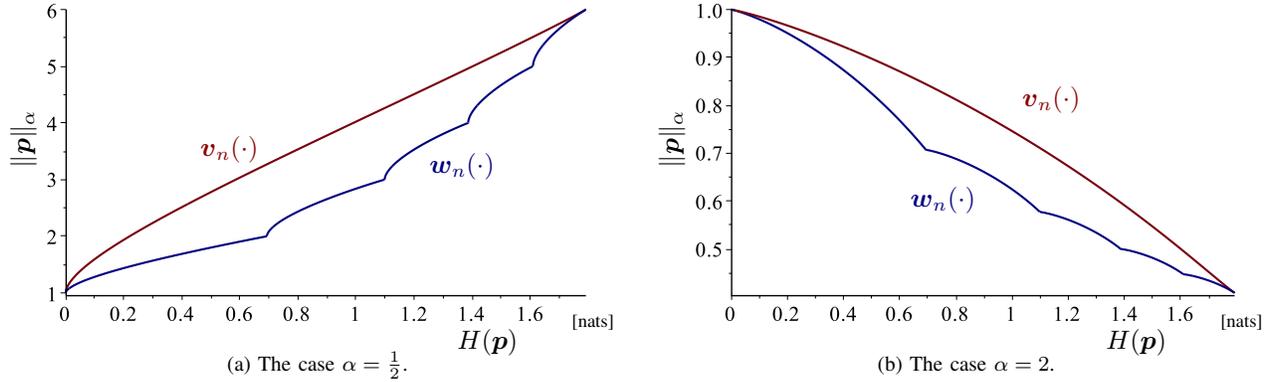
\begin{figure}[!t]
\centering
\subfloat[The case $\alpha = \frac{1}{2}$.]{
\begin{overpic}[width = 0.45\hsize, clip]{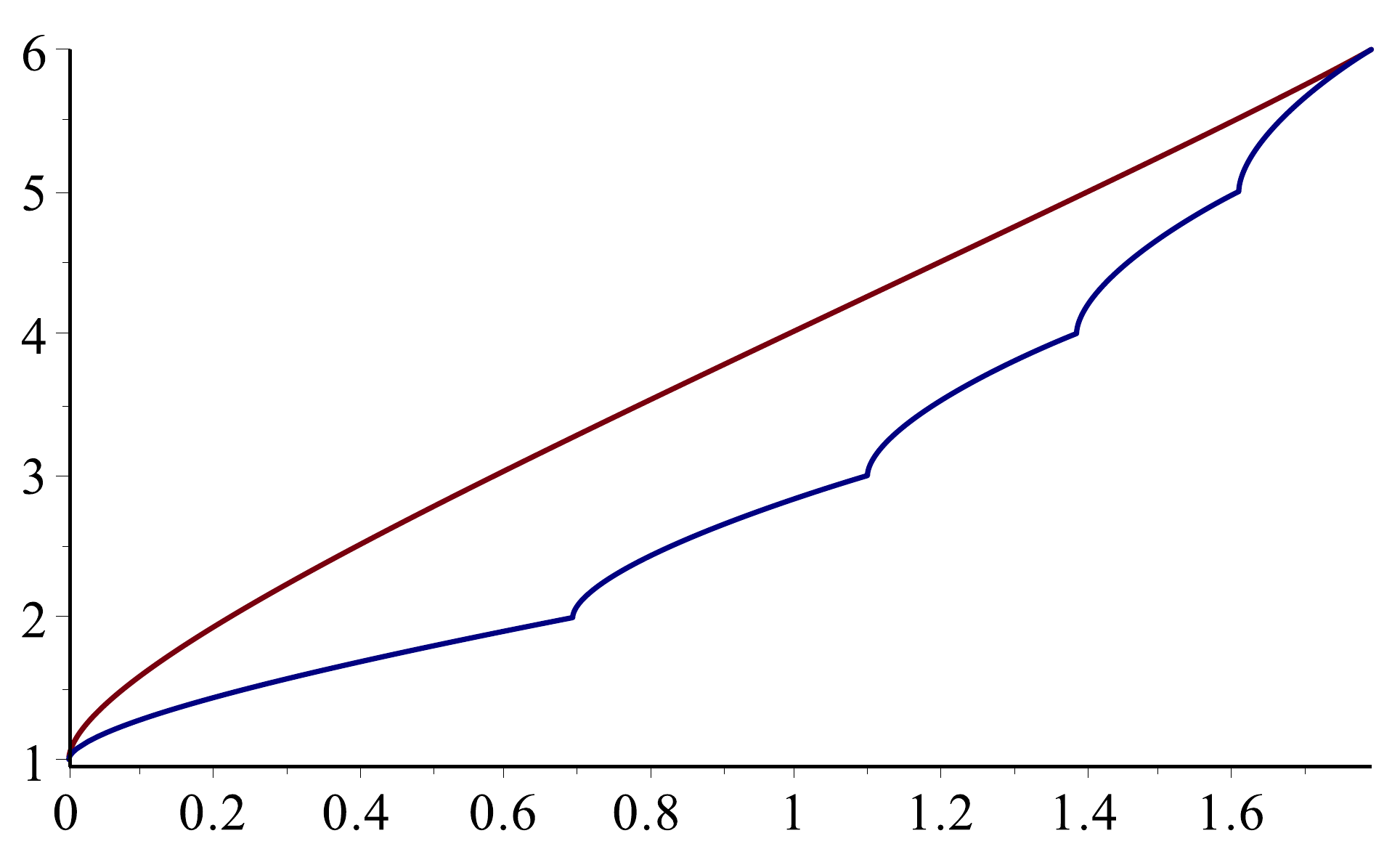}
\put(-5, 30){\rotatebox{90}{$\| \bvec{p} \|_{\alpha}$}}
\put(75, -2.5){$H( \bvec{p} )$}
\put(95.5, 1.5){\scriptsize [nats]}
\put(29, 32){\color{burgundy} $\bvec{v}_{n}( \cdot )$}
\put(70, 29){\color{navyblue} $\bvec{w}_{n}( \cdot )$}
%\put(75, 12){\small $H_{\sbvec{v}_{n}}( p ) = \chi_{n}( \alpha )$}
%\put(74, 13){\vector(-2, -1){11}}
%\put(45, 48){inflection}
%\put(51, 43){point}
\end{overpic}
}\hspace{0.05\hsize}
\subfloat[The case $\alpha = 2$.]{
\begin{overpic}[width = 0.45\hsize, clip]{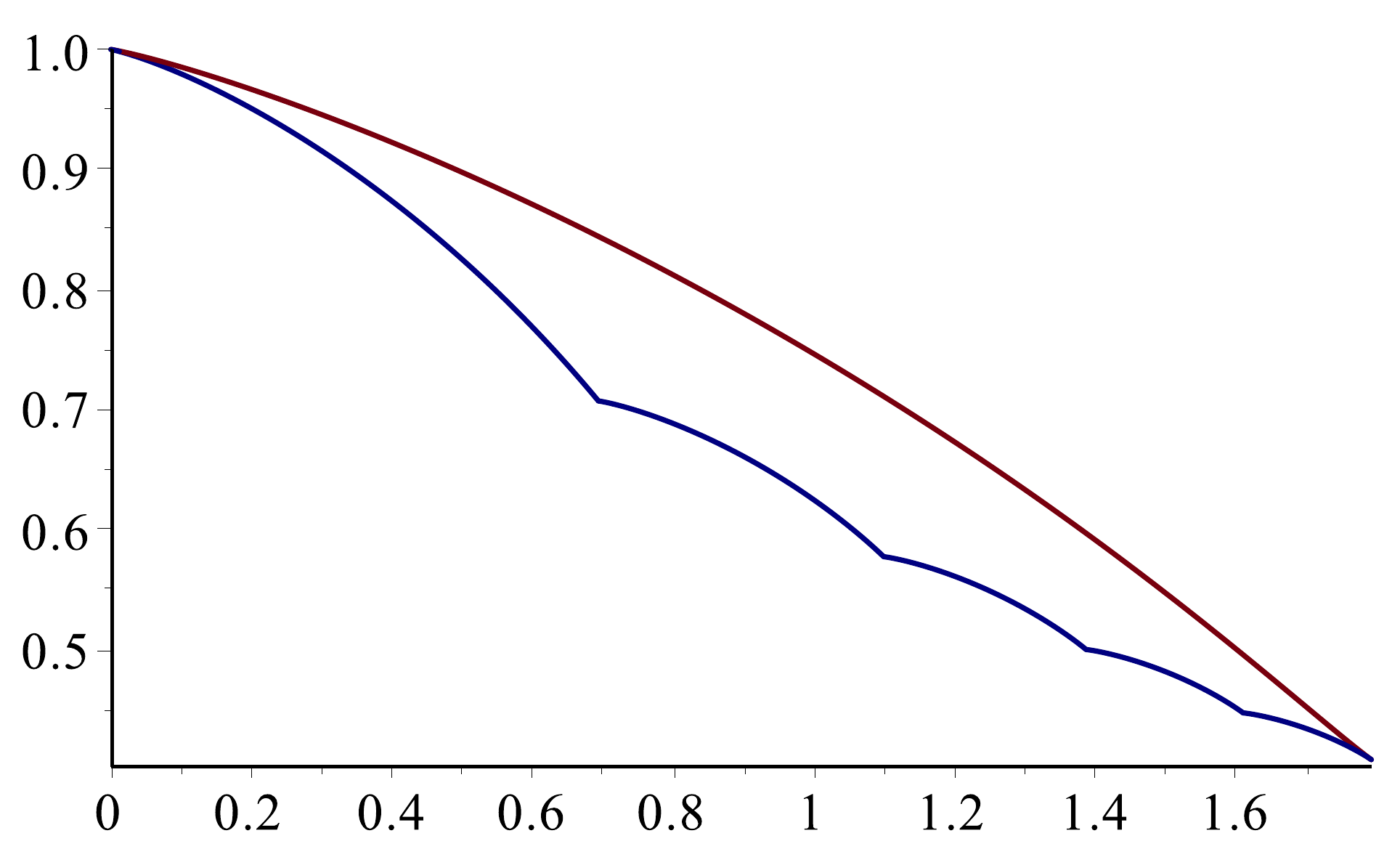}
\put(-5, 30){\rotatebox{90}{$\| \bvec{p} \|_{\alpha}$}}
\put(75, -2.5){$H( \bvec{p} )$}
\put(95.5, 1.5){\scriptsize [nats]}
\put(60, 41){\color{burgundy} $\bvec{v}_{n}( \cdot )$}
\put(40, 23){\color{navyblue} $\bvec{w}_{n}( \cdot )$}
%\put(46, 11){\small $H_{\sbvec{v}_{n}}( p ) = \chi_{n}( \alpha )$}
%\put(76, 12){\vector(4, -1){17}}
%\put(88, 20){inflection}
%\put(95, 15){point}
\end{overpic}
}
\caption{
Plot of the boundary of $\mathcal{R}_{n}( \alpha )$ with $n = 6$.
The upper- and lower-boundaries correspond to distributions $\bvec{v}_{n}( \cdot )$ and $\bvec{w}_{n}( \cdot )$, respectively.}
\label{fig:region_P6_half}
\end{figure}

Note that the distributions $\bar{\bvec{v}}_{n}( \bvec{p} )$ and $\bar{\bvec{w}}_{n}( \bvec{p} )$ denote $\bvec{v}_{n}( p )$ and $\bvec{w}_{n}( q )$, respectively, such that $H_{\sbvec{v}_{n}}( p ) = H_{\sbvec{w}_{n}}( q ) = H( \bvec{p} )$ for a given $\bvec{p} \in \mathcal{P}_{n}$.
Theorem \ref{th:extremes} shows that, among all $n$-ary probability vectors with a fixed Shannon entropy, the distributions $\bvec{v}_{n}( \cdot )$ and $\bvec{w}_{n}( \cdot )$ take the maximum and the minimum $\ell_{\alpha}$-norm, respectively.
Thus, the bounds \eqref{ineq:extremes} of Theorem \ref{th:extremes} are tight in the sense of the existences of the distributions $\bvec{v}_{n}( \cdot )$ and $\bvec{w}_{n}( \cdot )$ which attain both equalities of the bounds \eqref{ineq:extremes}. %, as with Fano's inequality \cite{fano}%
%\footnote{Fano's inequality \cite{fano} also holds with equality by the distribution $\bvec{v}_{n}( \cdot )$.}.
In other words, Theorem \ref{th:extremes} implies that the boundaries of $\mathcal{R}_{n}( \alpha )$, defined in \eqref{def:region_Pn}, can be attained by $\bvec{v}_{n}( \cdot )$ and $\bvec{w}_{n}( \cdot )$.
We illustrate the graphs of the boundaries of $\mathcal{R}_{n}( \alpha )$ in Fig. \ref{fig:region_P6_half}.
Note that $\| \bar{\bvec{v}}_{2}( \bvec{p} ) \|_{\alpha} = \| \bar{\bvec{w}}_{2}( \bvec{p} ) \|_{\alpha}$ for any $\bvec{p} \in \mathcal{P}_{2}$ and any $\alpha \in (0, \infty)$ since $\bvec{v}_{2}( p ) = \bvec{w}_{2}( 1 - p )$ for $p \in [0, \frac{1}{2}]$.
Therefore, Theorem \ref{th:extremes} becomes meaningful for $n \ge 3$.

On the other hand, the following theorem shows that, among all $n$-ary probability vectors with a fixed $\ell_{\alpha}$-norm, the distributions $\bvec{v}_{n}( \cdot )$ and $\bvec{w}_{n}( \cdot )$ also take the extreme values of the Shannon entropy.

\begin{theorem}
\label{th:extremes2}
Let $p \in [0, \frac{1}{n}]$ and $p^{\prime} \in [\frac{1}{n}, 1]$ be chosen to satisfy
\begin{align}
\| \bvec{v}_{n}( p ) \|_{\alpha}
=
\| \bvec{p} \|_{\alpha}
=
\| \bvec{w}_{n}( p^{\prime} ) \|_{\alpha}
\label{ineq:norm_v_to_w}
\end{align}
for a fixed $\alpha \in (0, 1) \cup (1, \infty)$.
Then, we observe that
\begin{align}
0 < \alpha < 1
& \ \Longrightarrow \
H_{\sbvec{v}_{n}}( p ) \le H( \bvec{p} ) \le H_{\sbvec{w}_{n}}( p^{\prime} ) ,
\label{ineq:H_0to1} \\
\alpha > 1
& \ \Longrightarrow \
H_{\sbvec{w}_{n}}( p^{\prime} ) \le H( \bvec{p} ) \le H_{\sbvec{v}_{n}}( p )
\label{ineq:H_1toinfty}
\end{align}
for any $n \ge 2$ and any $\bvec{p} \in \mathcal{P}_{n}$.
\end{theorem}

\begin{IEEEproof}[Proof of Theorem \ref{th:extremes2}]
From Theorem \ref{th:extremes}, for a fixed $n \ge 2$, we consider $\bvec{p} \in \mathcal{P}_{n}$, $p \in [0, \frac{1}{n}]$, and $p^{\prime} \in [\frac{1}{n}, 1]$ such that
\begin{align}
H_{\sbvec{w}_{n}}( p^{\prime} )
& =
H( \bvec{p} )
=
H_{\sbvec{v}_{n}}( p ) ,
\\
\| \bvec{w}_{n}( p^{\prime} ) \|_{\alpha}
& \le
\| \bvec{p} \|_{\alpha}
\le
\| \bvec{v}_{n}( p ) \|_{\alpha}
\end{align}
for $\alpha \in (0, 1) \cup (1, +\infty)$.
Note that $p$ and $p^{\prime}$ are uniquely determined for a given $\bvec{p} \in \mathcal{P}_{n}$.
It follows from Lemmas \ref{lem:Hv} and \ref{lem:Hw} that $H_{\sbvec{v}_{n}}( p ) \in [0, \ln n]$ and $H_{\sbvec{w}_{n}}( p^{\prime} ) \in [0, \ln n]$ are strictly increasing for $p \in [0, \frac{1}{n}]$ and strictly decreasing for $p^{\prime} \in [\frac{1}{n}, 1]$, respectively.
Moreover, it follows from Lemmas \ref{lem:mono_v} and \ref{lem:mono_w} that, if $\alpha \in (0, 1)$, then $\| \bvec{v}_{n}( p ) \|_{\alpha}$ and $\| \bvec{w}_{n}( p^{\prime} ) \|_{\alpha}$ are strictly increasing for $p \in [0, \frac{1}{n}]$ and strictly decreasing for $p^{\prime} \in [\frac{1}{n}, 1]$, respectively.
Therefore, decreasing both $p \in [0, \frac{1}{n}]$ and $p^{\prime} \in [\frac{1}{n}, 1]$, we can obtain $q \in [0, \frac{1}{n}]$ and $q^{\prime} \in [\frac{1}{n}, 1]$ such that
\begin{align}
H_{\sbvec{w}_{n}}( q^{\prime} )
& \ge
H( \bvec{p} )
\ge
H_{\sbvec{v}_{n}}( q ) ,
\\
\| \bvec{w}_{n}( q^{\prime} ) \|_{\alpha}
& =
\| \bvec{p} \|_{\alpha}
=
\| \bvec{v}_{n}( q ) \|_{\alpha}
\end{align}
for a fixed $\alpha \in (0, 1)$.

On the other hand, it follows from Lemmas \ref{lem:mono_v} and \ref{lem:mono_w} that, if $\alpha \in (1, +\infty)$, then $\| \bvec{v}_{n}( p ) \|_{\alpha}$ and $\| \bvec{w}_{n}( p^{\prime} ) \|_{\alpha}$ are strictly decreasing for $p \in [0, \frac{1}{n}]$ and strictly increasing for $p \in [\frac{1}{n}, 1]$, respectively.
Therefore, increasing both $p \in [0, \frac{1}{n}]$ and $p^{\prime} \in [\frac{1}{n}, 1]$, we can obtain $q \in [0, \frac{1}{n}]$ and $q^{\prime} \in [\frac{1}{n}, 1]$ such that
\begin{align}
H_{\sbvec{w}_{n}}( q^{\prime} )
& \le
H( \bvec{p} )
\le
H_{\sbvec{v}_{n}}( q ) ,
\\
\| \bvec{w}_{n}( q^{\prime} ) \|_{\alpha}
& =
\| \bvec{p} \|_{\alpha}
=
\| \bvec{v}_{n}( q ) \|_{\alpha}
\end{align}
for a fixed $\alpha \in (1, +\infty)$.

Finally, we note that the strict monotonicity of Lemmas \ref{lem:mono_v} and \ref{lem:mono_w} prove the uniquenesses of the values $q \in [0, \frac{1}{n}]$ and $q^{\prime} \in [\frac{1}{n}, 1]$.
In fact, it follows from Lemmas \ref{lem:Hv}, \ref{lem:Hw}, \ref{lem:mono_v}, and \ref{lem:mono_w} that, for a fixed $n \ge 2$ and a fixed $\alpha \in (0, 1) \cup (1, +\infty)$, $\| \bvec{v}_{n}( p ) \|_{\alpha}$ and $\| \bvec{w}_{n}( p^{\prime} ) \|_{\alpha}$ are both bijective function of $p \in [0, \frac{1}{n}]$ and $p^{\prime} \in [\frac{1}{n}, 1]$, respectively.
That completes the proof of Theorem \ref{th:extremes2}.
\end{IEEEproof}

In Theorem \ref{th:extremes2}, note that the values $p \in [0, \frac{1}{n}]$ and $p^{\prime} \in [\frac{1}{n}, 1]$ are uniquely determined by the value of $\| \bvec{p} \|_{\alpha}$.
Theorems \ref{th:extremes} and \ref{th:extremes2} show that extremality between the Shannon entropy and the $\ell_{\alpha}$-norm can be attained by the distributions $\bvec{v}_{n}( \cdot )$ and $\bvec{w}_{n}( \cdot )$.

Following a same manner with \cite[Theorem 2]{fabregas}, we extend the bounds of Theorem \ref{th:extremes} from the $\ell_{\alpha}$-norm to several information measures, which are related to $\ell_{\alpha}$-norm, as follows:

\begin{corollary}
\label{cor:extremes}
Let $f( \cdot )$ be a strictly monotonic function.
Then, we observe that:
(i) if $f( \cdot )$ is strictly increasing, then
\begin{align}
f( \vphantom{\sum} \| \bar{\bvec{w}}_{n}( \bvec{p} ) \|_{\alpha} )
\le
f( \vphantom{\sum} \| \bvec{p} \|_{\alpha} )
\le
f( \vphantom{\sum} \| \bar{\bvec{v}}_{n}( \bvec{p} ) \|_{\alpha} )
\label{ineq:increasing}
\end{align}
and (ii) if $f( \cdot )$ is strictly decreasing, then
\begin{align}
f( \vphantom{\sum} \| \bar{\bvec{v}}_{n}( \bvec{p} ) \|_{\alpha} )
\le
f( \vphantom{\sum} \| \bvec{p} \|_{\alpha} )
\le
f( \vphantom{\sum} \| \bar{\bvec{w}}_{n}( \bvec{p} ) \|_{\alpha} )
\label{ineq:decreasing}
\end{align}
for any $n \ge 2$, any $\bvec{p} \in \mathcal{P}_{n}$, and any $\alpha \in (0, \infty)$. %, where $\bar{\bvec{v}}_{n}( \cdot )$ and $\bar{\bvec{w}}_{n}( \cdot )$ are cited in Theorem \ref{th:extremes}.
%
%Moreover, let $p \in [0, \frac{1}{n}]$ and $p^{\prime} \in [\frac{1}{n}, 1]$ be chosen to satisfy
%\begin{align}
%f( \vphantom{\sum} \| \bvec{p} \|_{\alpha} )
%=
%f( \vphantom{\sum} \| \bvec{v}_{n}( p ) \|_{\alpha} )
%=
%f( \vphantom{\sum} \| \bvec{w}_{n}( p^{\prime} ) \|_{\alpha} )
%\label{eq:f(norm)_equal}
%\end{align}
%for a fixed $\alpha \in (0, 1) \cup (1, \infty)$;
%then, we observe that
%\begin{align}
%0 < \alpha < 1
%& \ \Longrightarrow \
%H_{\sbvec{v}_{n}}( p ) \le H( \bvec{p} ) \le H_{\sbvec{w}_{n}}( p^{\prime} ) ,
%\label{ineq:H_0to1_cor} \\
%\alpha > 1
%& \ \Longrightarrow \
%H_{\sbvec{w}_{n}}( p^{\prime} ) \le H( \bvec{p} ) \le H_{\sbvec{v}_{n}}( p )
%\label{ineq:H_1toinfty_cor}
%\end{align}
%for any $n \ge 2$ and any $\bvec{p} \in \mathcal{P}_{n}$.
\end{corollary}

\begin{IEEEproof}[Proof of Corollary \ref{cor:extremes}]
Since any strictly increasing function $f( \cdot )$ satisfies $f( x ) < f( y )$ for $x < y$, it is easy to see that \eqref{ineq:increasing} from \eqref{ineq:extremes} of Theorem \ref{th:extremes}.
Similarly, since any strictly decreasing function $f( \cdot )$ satisfies $f( x ) > f( y )$ for $x < y$, it is also easy to see that \eqref{ineq:decreasing} from \eqref{ineq:extremes} of Theorem \ref{th:extremes}.
%Finally, since any function $f( \cdot )$ satisfies $f( x ) = f( y )$ for $x = y$, the statements of \eqref{eq:f(norm)_equal}--\eqref{ineq:H_1toinfty_cor} hold immediately from Theorem \ref{th:extremes2}.
\end{IEEEproof}

\begin{figure}[!t]
\centering
\subfloat[The case $\alpha = \frac{1}{2}$.]{
\begin{overpic}[width = 0.45\hsize, clip]{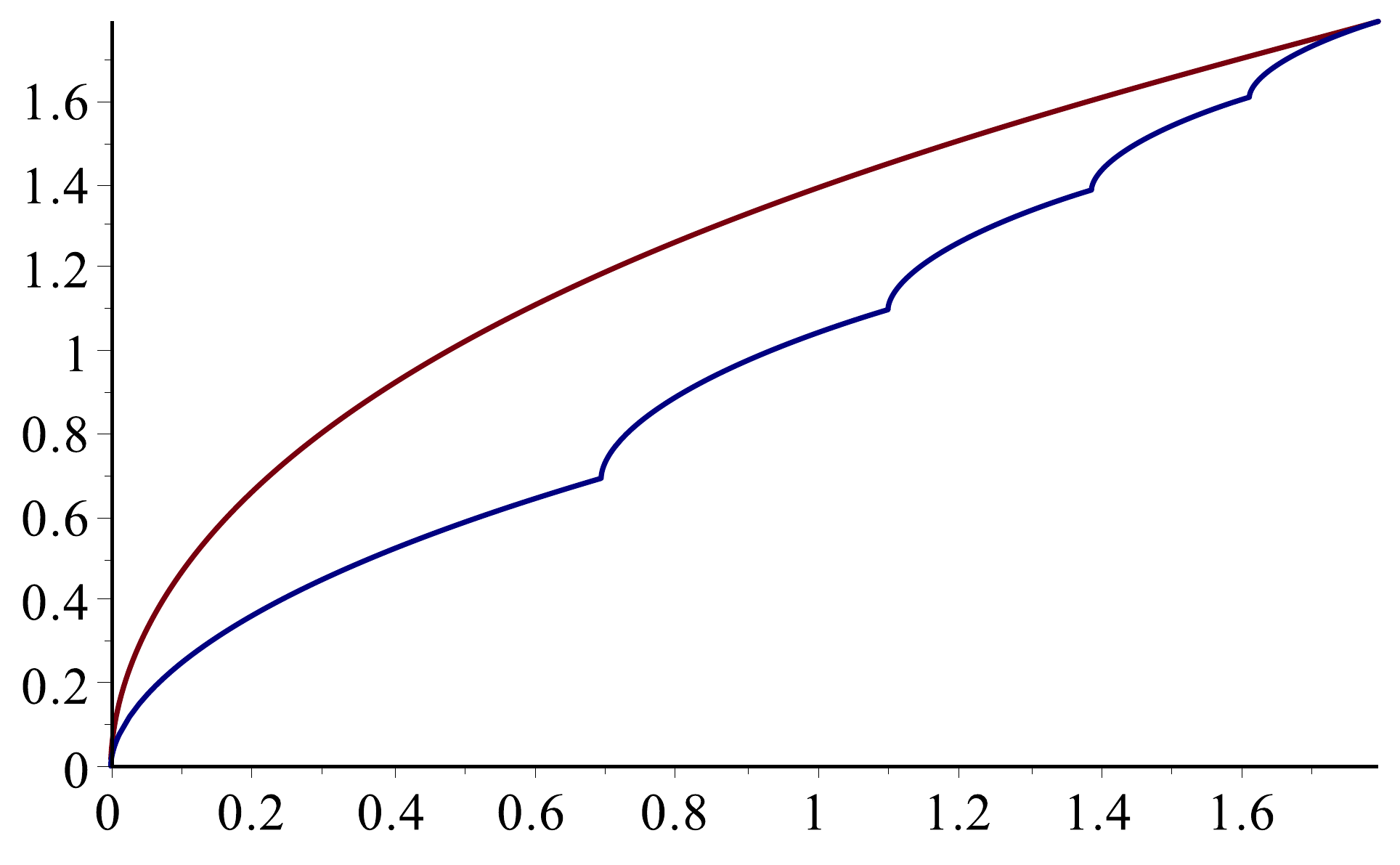}
\put(-5, 25){\rotatebox{90}{$H_{\alpha}( \bvec{p} )$}}
\put(0, 59){\scriptsize [nats]}
\put(75, -2.5){$H( \bvec{p} )$}
\put(95.5, 1.5){\scriptsize [nats]}
\put(30, 45){\color{burgundy} $\bvec{v}_{n}( \cdot )$}
\put(60, 32){\color{navyblue} $\bvec{w}_{n}( \cdot )$}
\end{overpic}
}\hspace{0.05\hsize}
\subfloat[The case $\alpha = 2$.]{
\begin{overpic}[width = 0.45\hsize, clip]{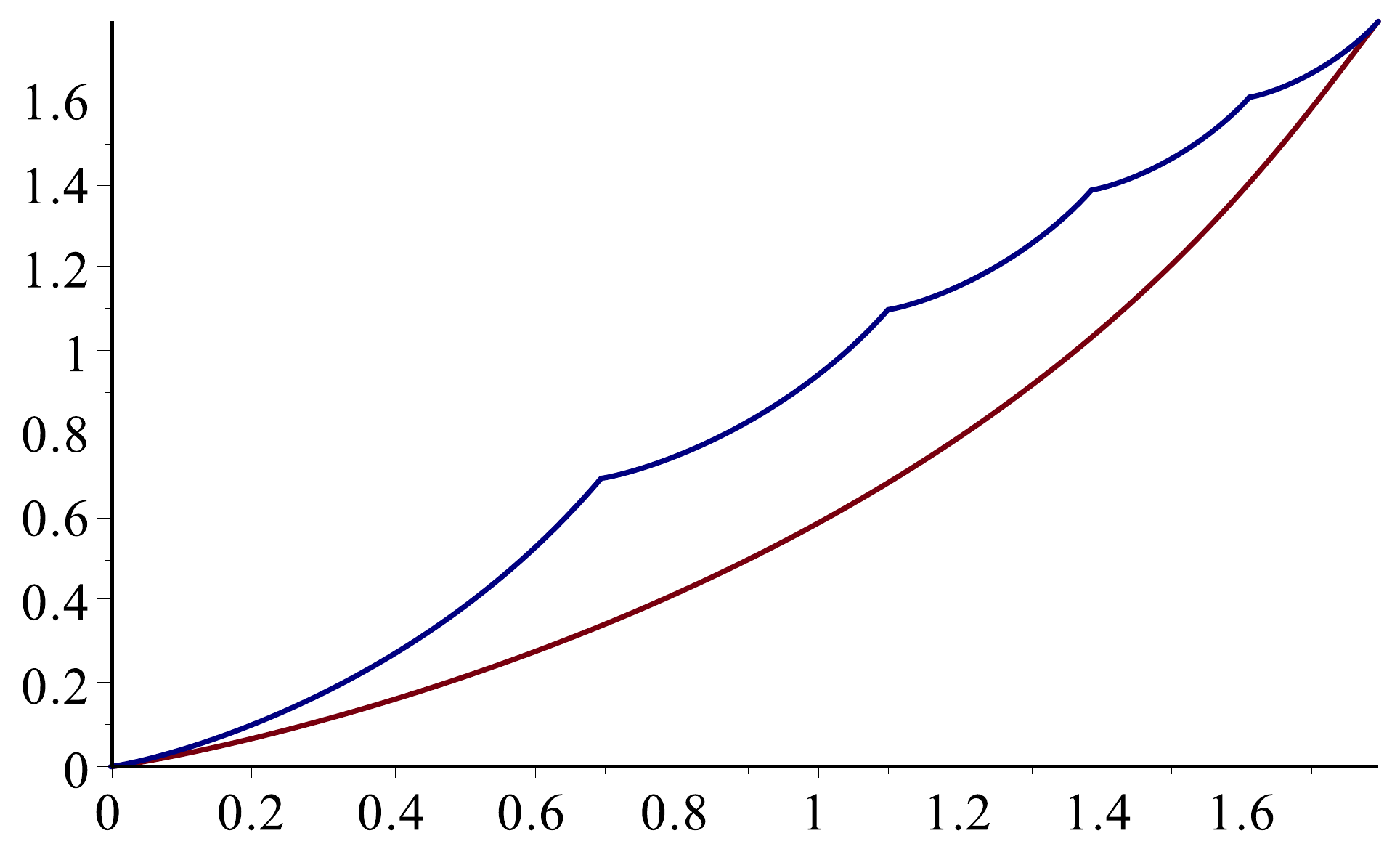}
\put(-5, 25){\rotatebox{90}{$H_{\alpha}( \bvec{p} )$}}
\put(0, 59){\scriptsize [nats]}
\put(75, -2.5){$H( \bvec{p} )$}
\put(95.5, 1.5){\scriptsize [nats]}
\put(75, 30){\color{burgundy} $\bvec{v}_{n}( \cdot )$}
\put(45, 37){\color{navyblue} $\bvec{w}_{n}( \cdot )$}
\end{overpic}
}
\caption{
Plots of the boundaries of $\mathcal{R}_{n}^{\text{R\'{e}nyi}}( \alpha )$ with $n = 6$.
If $0 < \alpha < 1$, then the upper- and lower-boundaries correspond to distributions $\bvec{v}_{n}( \cdot )$ and $\bvec{w}_{n}( \cdot )$, respectively.
If $\alpha > 1$, then these correspondences are reversed.}
\label{fig:Renyi}
\end{figure}

\begin{figure}[!t]
\centering
\subfloat[The case $R = \frac{1}{2}$.]{
\begin{overpic}[width = 0.45\hsize, clip]{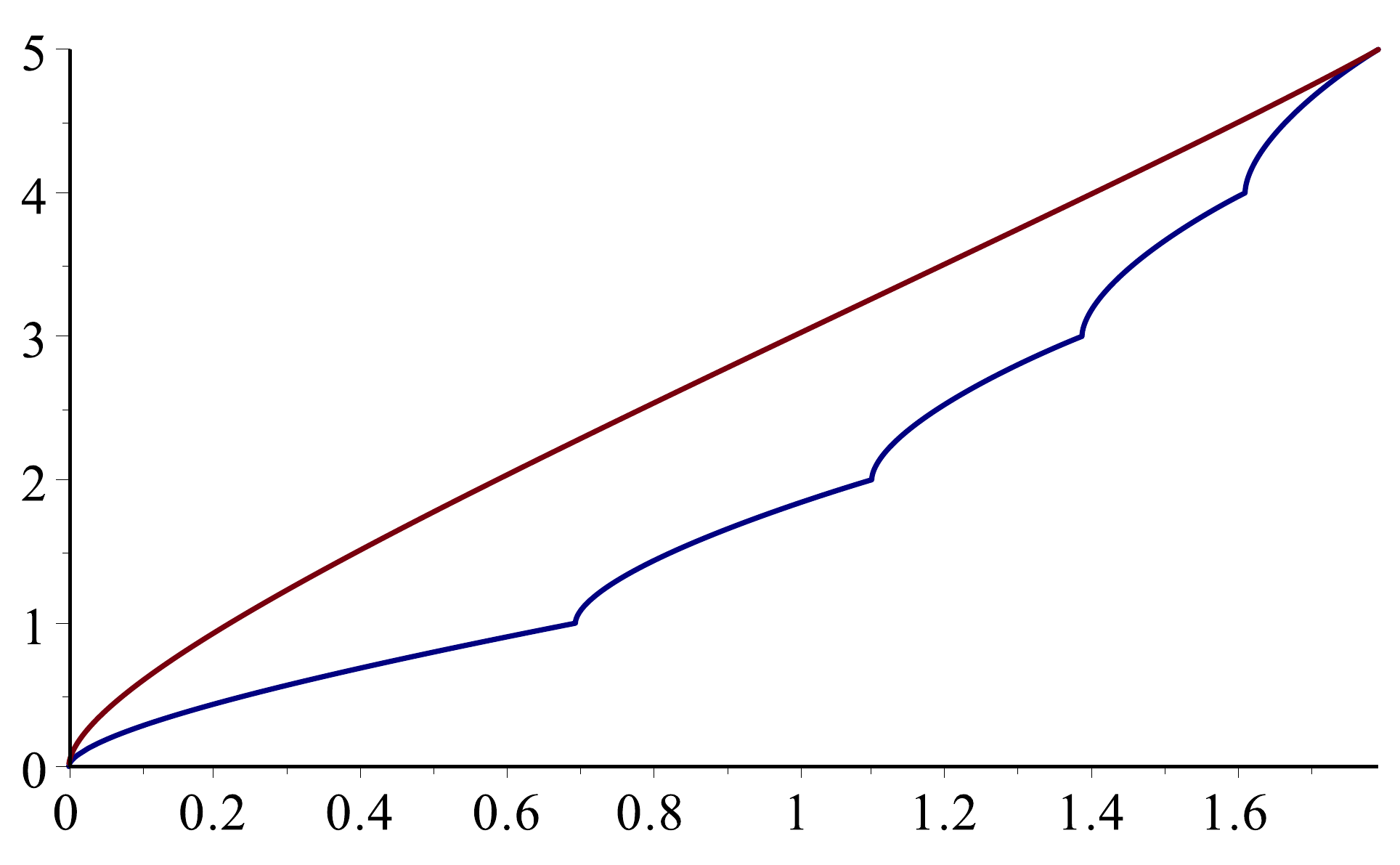}
\put(-5, 25){\rotatebox{90}{$H_{R}( \bvec{p} )$}}
\put(75, -2.5){$H( \bvec{p} )$}
\put(95.5, 1.5){\scriptsize [nats]}
\put(30, 35){\color{burgundy} $\bvec{v}_{n}( \cdot )$}
\put(60, 22){\color{navyblue} $\bvec{w}_{n}( \cdot )$}
\end{overpic}
}\hspace{0.05\hsize}
\subfloat[The case $R = 2$.]{
\begin{overpic}[width = 0.45\hsize, clip]{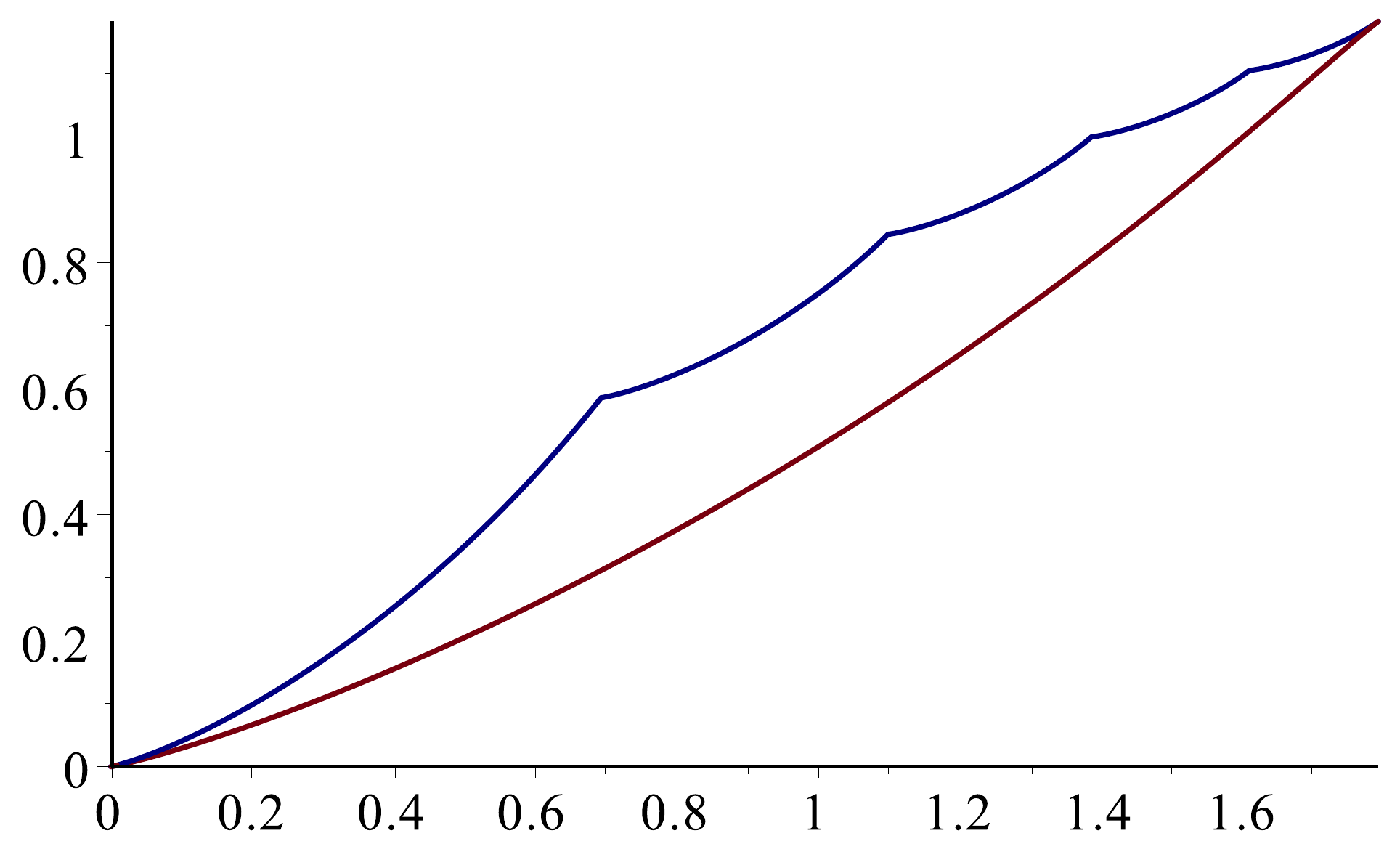}
\put(-5, 25){\rotatebox{90}{$H_{R}( \bvec{p} )$}}
\put(75, -2.5){$H( \bvec{p} )$}
\put(95.5, 1.5){\scriptsize [nats]}
\put(75, 35){\color{burgundy} $\bvec{v}_{n}( \cdot )$}
\put(45, 44){\color{navyblue} $\bvec{w}_{n}( \cdot )$}
\end{overpic}
}
\caption{
Plots of the boundaries of $\{ (H( \bvec{p} ), H_{R}( \bvec{p} )) \mid \bvec{p} \in \mathcal{P}_{n} \}$ with $n = 6$.
If $0 < \alpha < 1$, then the upper- and lower-boundaries correspond to distributions $\bvec{v}_{n}( \cdot )$ and $\bvec{w}_{n}( \cdot )$, respectively.
If $\alpha > 1$, then these correspondences are reversed.}
\label{fig:R}
\end{figure}

Therefore, we can obtain tight bounds of several information measures, which are determined by $\ell_{\alpha}$-norm, with a fixed Shannon entropy.
As an instance, we introduce the application of Corollary \ref{cor:extremes} to the R\'{e}nyi entropy as follows:
Let $f_{\alpha}( x ) = \frac{\alpha}{1-\alpha} \ln x$.
Then, we readily see that $H_{\alpha}( \bvec{p} ) = f_{\alpha}( \| \bvec{p} \|_{\alpha} )$.
It can be easily seen that $f_{\alpha}( x )$ is strictly increasing for $x \ge 0$ when $\alpha \in (0, 1)$ and strictly decreasing for $x \ge 0$ when $\alpha \in (1, \infty)$.
Hence, it follows from Corollary \ref{cor:extremes} that
\begin{align}
0 < \alpha < 1
& \, \Longrightarrow \,
H_{\alpha}( \bar{\bvec{w}}_{n}( \bvec{p} ) ) \le H_{\alpha}( \bvec{p} ) \le H_{\alpha}( \bar{\bvec{v}}_{n}( \bvec{p} ) ) ,
\label{eq:Renyi_bound1} \\
\alpha > 1
& \, \Longrightarrow \,
H_{\alpha}( \bar{\bvec{v}}_{n}( \bvec{p} ) ) \le H_{\alpha}( \bvec{p} ) \le H_{\alpha}( \bar{\bvec{w}}_{n}( \bvec{p} ) )
\label{eq:Renyi_bound2}
\end{align}
for any $n \ge 2$ and any $\bvec{p} \in \mathcal{P}_{n}$.
Moreover, if $p \in [0, \frac{1}{n}]$ and $p^{\prime} \in [\frac{1}{n}, 1]$ are chosen to satisfy
$
H_{\alpha}( \bvec{p} )
=
H_{\alpha}( \bvec{v}_{n}( p ) )
=
H_{\alpha}( \bvec{w}_{n}( p^{\prime} ) )
%\label{eq:Renyi_bound3}
$
for a fixed $\alpha \in (0, 1) \cup (1, \infty)$, then \eqref{ineq:H_0to1} and \eqref{ineq:H_1toinfty} hold for any $n \ge 2$ and any $\bvec{p} \in \mathcal{P}_{n}$ from Theorem \ref{th:extremes2}.
These bounds between the Shannon entropy and the R\'{e}nyi entropy imply the boundary of the region
$
\mathcal{R}_{n}^{\text{R\'{e}nyi}}( \alpha )
\triangleq
\{ (H( \bvec{p} ), H_{\alpha}( \bvec{p} )) \mid \bvec{p} \in \mathcal{P}_{n} \}
$
for any $n \ge 2$ and any $\alpha \in (0, 1) \cup (1, \infty)$.
We illustrate the boundaries of $\mathcal{R}_{n}^{\text{R\'{e}nyi}}( \alpha )$ in Fig. \ref{fig:Renyi}.
Similarly, we can apply Corollary \ref{cor:extremes} to several entropies as shown in Table \ref{table:extremes}, and we illustrate these exact feasible region in Figs. \ref{fig:Renyi}--\ref{fig:R}.
%Note in Table \ref{table:extremes} that the $\gamma$-entropy uses the $\ell_{1/\gamma}$-norm;
%namely, the inequalities \eqref{ineq:increasing} and \eqref{ineq:decreasing} are reversed in the $\gamma$-entropy.

\begin{table*}[t]
\centering
\caption{Applications of Corollary \ref{cor:extremes}}
\label{table:extremes}
\begin{tabular}{|c|l|c|c|}\hline
Entropies & \centering function $f_{t}( \cdot )$ & monotonicity ($0 < t < 1$) & monotonicity ($t > 1$) \\ \hline
R\'{e}nyi entropy \cite{renyi} $H_{\alpha}( \bvec{p} ) = f_{\alpha}( \| \bvec{p} \|_{\alpha} )$ & $f_{t}( x ) = \frac{ t }{ 1 - t } \ln x$ & strictly increasing for $x > 0$ & strictly decreasing for $x > 0$ \\ \hline
Tsallis entropy \cite{tsallis2} $S_{q}( \bvec{p} ) = f_{q}( \| \bvec{p} \|_{q} )$ & $f_{t}( x ) = \frac{ 1 }{ 1 - t } (x^{t}-1)$ & strictly increasing for $x > 0$ & strictly decreasing for $x > 0$ \\ \hline
Entropy of type-$\beta$ \cite{havrda, daroczy} $H_{\beta}( \bvec{p} ) = f_{\beta}( \| \bvec{p} \|_{\beta} )$ & $f_{t}( x ) = \frac{ 1 }{ 2^{1-t} - 1 } (x^{t}-1)$ & strictly increasing for $x > 0$ & strictly decreasing for $x > 0$ \\ \hline
$\gamma$-entropy \cite{behara} $H_{\gamma}( \bvec{p} ) = f_{\gamma}( \| \bvec{p} \|_{1/\gamma} )$ & $f_{t}( x ) = \frac{ 1 }{ 1 - 2^{t-1} }( 1 - x )$ & strictly decreasing for $x > 0$ & strictly increasing for $x > 0$ \\ \hline
The $R$-norm information \cite{boekee} $H_{R}( \bvec{p} ) = f_{R}( \| \bvec{p} \|_{R} )$ & $f_{t}( x ) = \frac{ t }{ t - 1 } ( 1 - x )$ & strictly increasing for $x > 0$ & strictly decreasing for $x > 0$ \\ \hline
\end{tabular}
\end{table*}

\begin{figure}[!t]
\centering
\subfloat[The case $q = \frac{1}{2}$.]{
\begin{overpic}[width = 0.45\hsize, clip]{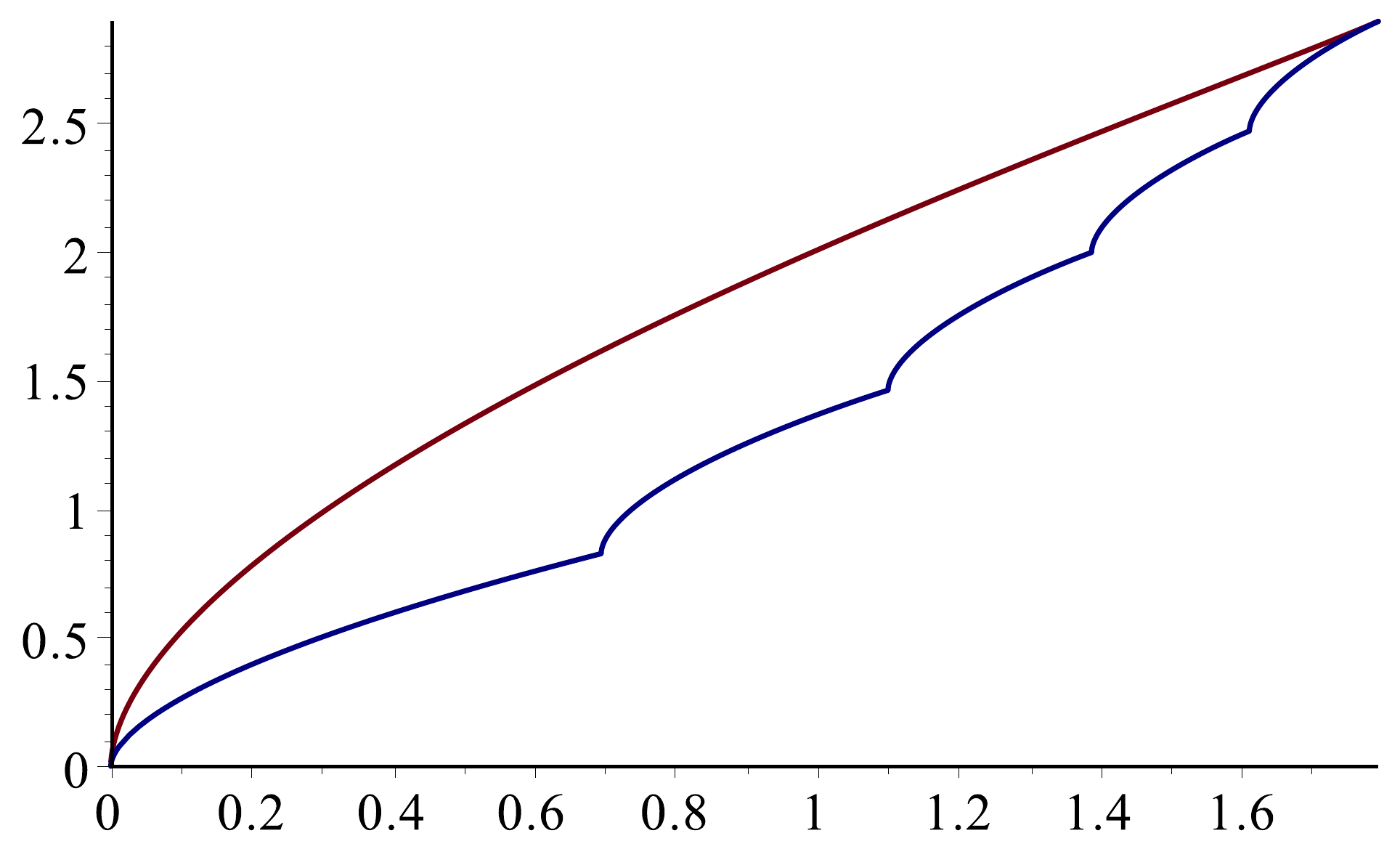}
\put(-5, 25){\rotatebox{90}{$S_{q}( \bvec{p} )$}}
\put(75, -2.5){$H( \bvec{p} )$}
\put(95.5, 1.5){\scriptsize [nats]}
\put(30, 38){\color{burgundy} $\bvec{v}_{n}( \cdot )$}
\put(60, 27){\color{navyblue} $\bvec{w}_{n}( \cdot )$}
\end{overpic}
}\hspace{0.05\hsize}
\subfloat[The case $q = 2$.]{
\begin{overpic}[width = 0.45\hsize, clip]{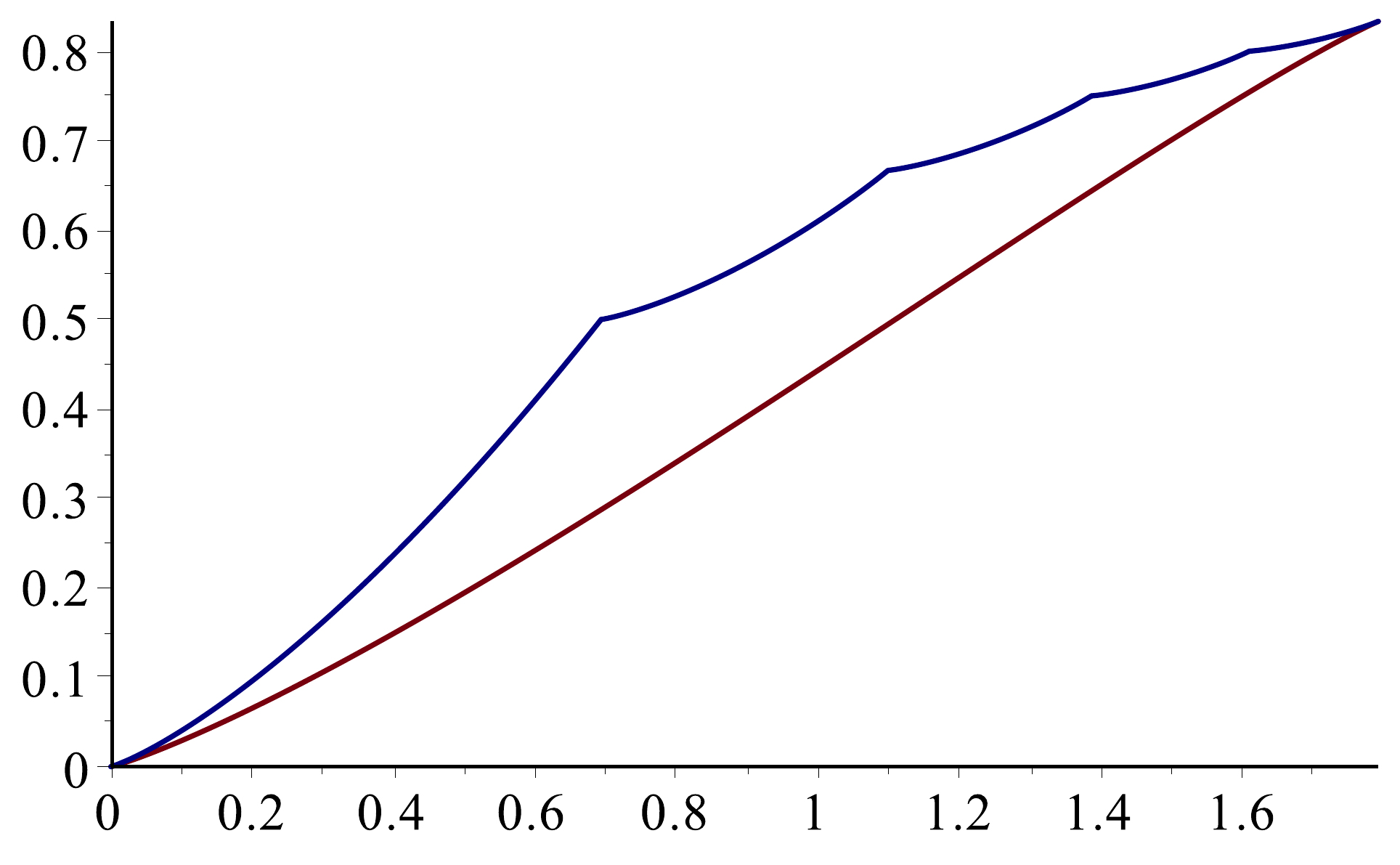}
\put(-5, 25){\rotatebox{90}{$S_{q}( \bvec{p} )$}}
\put(75, -2.5){$H( \bvec{p} )$}
\put(95.5, 1.5){\scriptsize [nats]}
\put(65, 33){\color{burgundy} $\bvec{v}_{n}( \cdot )$}
\put(35, 43){\color{navyblue} $\bvec{w}_{n}( \cdot )$}
\end{overpic}
}
\caption{
Plots of the boundaries of $\{ (H( \bvec{p} ), S_{q}( \bvec{p} )) \mid \bvec{p} \in \mathcal{P}_{n} \}$ with $n = 6$.
If $0 < \alpha < 1$, then the upper- and lower-boundaries correspond to distributions $\bvec{v}_{n}( \cdot )$ and $\bvec{w}_{n}( \cdot )$, respectively.
If $\alpha > 1$, then these correspondences are reversed.}
\end{figure}

\begin{figure}[!t]
\centering
\subfloat[The case $\beta = \frac{1}{2}$.]{
\begin{overpic}[width = 0.45\hsize, clip]{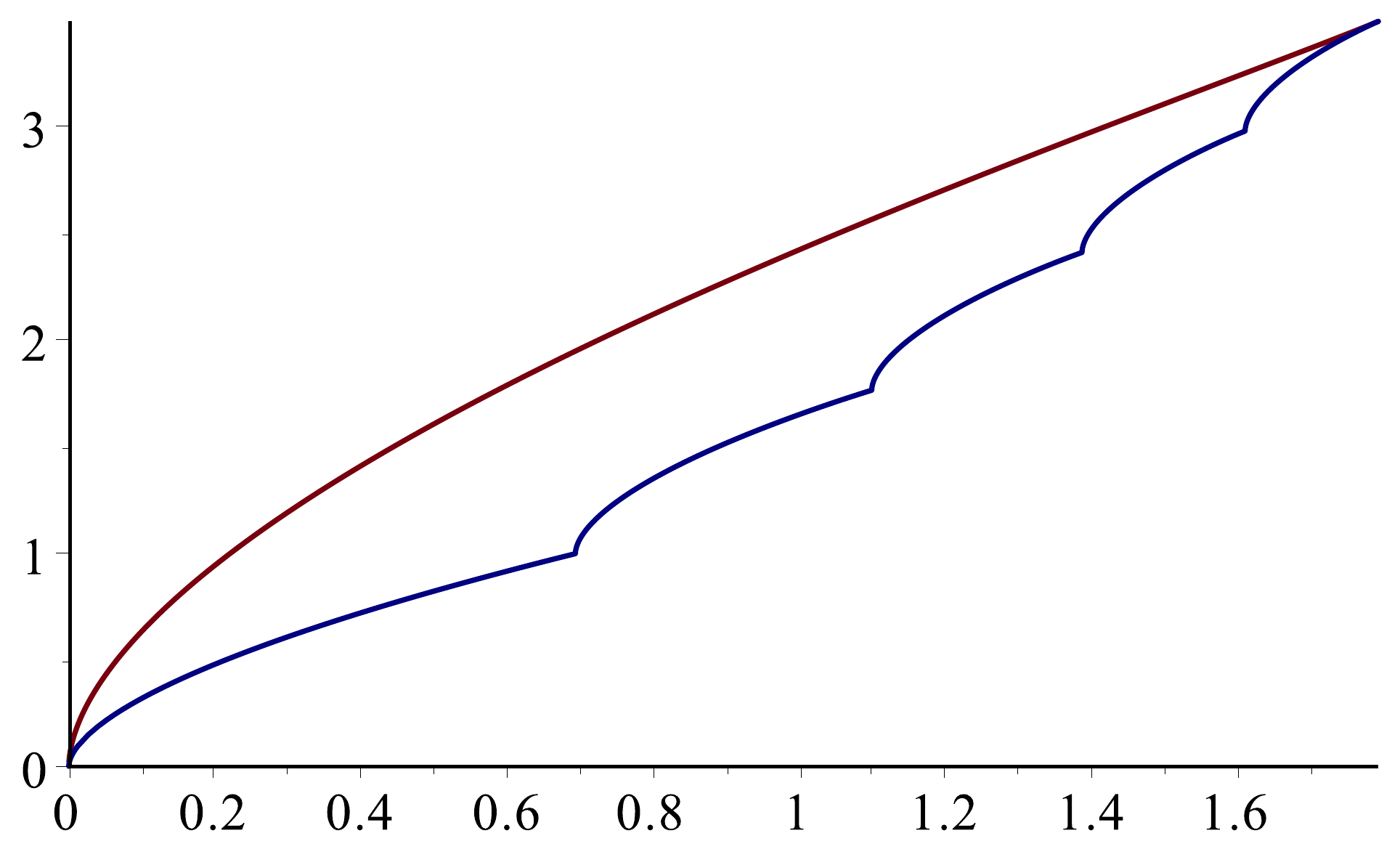}
\put(-5, 25){\rotatebox{90}{$H_{\beta}( \bvec{p} )$}}
\put(75, -2.5){$H( \bvec{p} )$}
\put(95.5, 1.5){\scriptsize [nats]}
\put(30, 38){\color{burgundy} $\bvec{v}_{n}( \cdot )$}
\put(60, 27){\color{navyblue} $\bvec{w}_{n}( \cdot )$}
\end{overpic}
}\hspace{0.05\hsize}
\subfloat[The case $\beta = 2$.]{
\begin{overpic}[width = 0.45\hsize, clip]{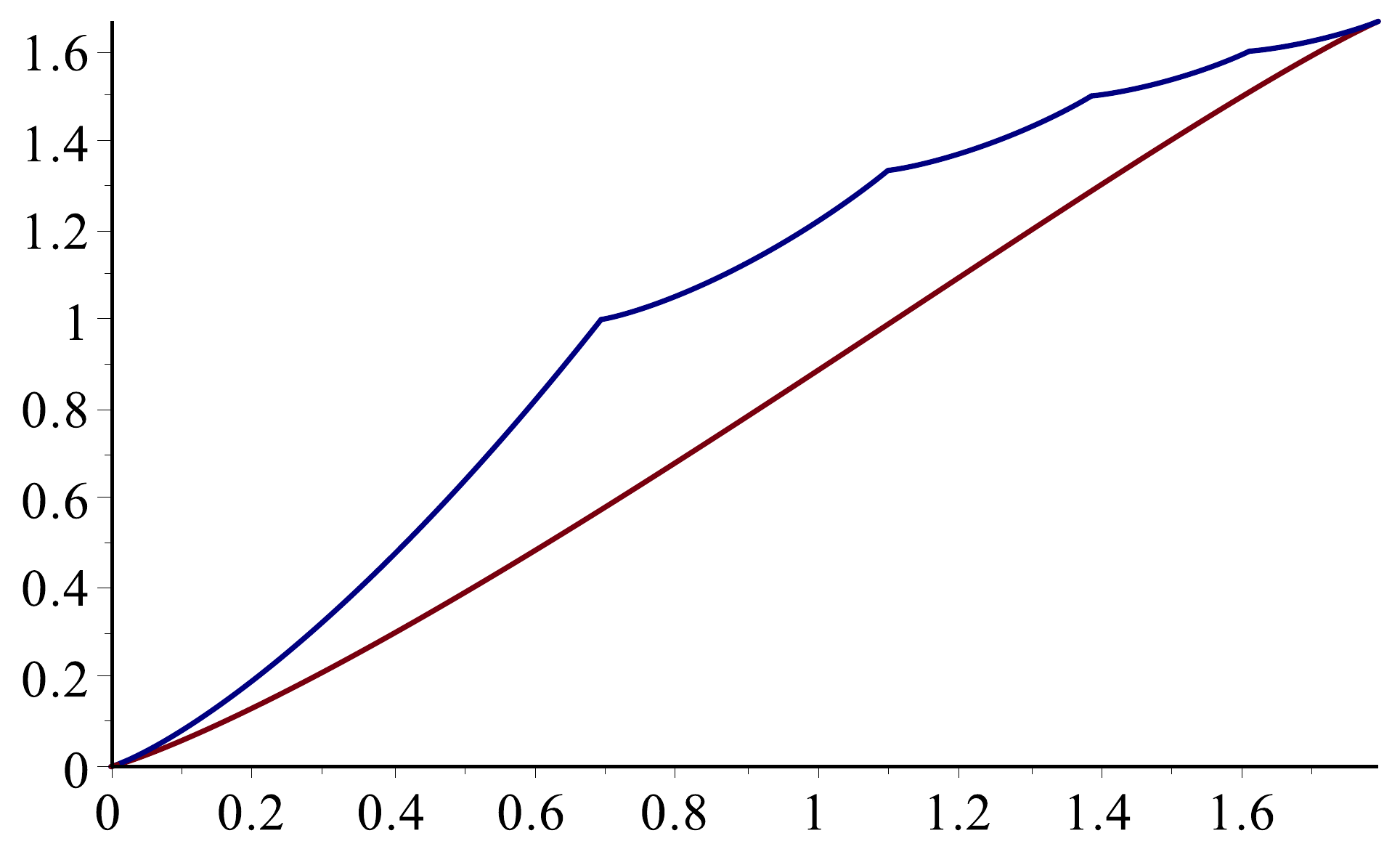}
\put(-5, 25){\rotatebox{90}{$H_{\beta}( \bvec{p} )$}}
\put(75, -2.5){$H( \bvec{p} )$}
\put(95.5, 1.5){\scriptsize [nats]}
\put(65, 33){\color{burgundy} $\bvec{v}_{n}( \cdot )$}
\put(35, 43){\color{navyblue} $\bvec{w}_{n}( \cdot )$}
\end{overpic}
}
\caption{
Plots of the boundaries of $\{ (H( \bvec{p} ), H_{\beta}( \bvec{p} )) \mid \bvec{p} \in \mathcal{P}_{n} \}$ with $n = 6$.
If $0 < \alpha < 1$, then the upper- and lower-boundaries correspond to distributions $\bvec{v}_{n}( \cdot )$ and $\bvec{w}_{n}( \cdot )$, respectively.
If $\alpha > 1$, then these correspondences are reversed.}
\end{figure}

\begin{figure}[!t]
\centering
\subfloat[The case $\gamma = \frac{1}{2}$.]{
\begin{overpic}[width = 0.45\hsize, clip]{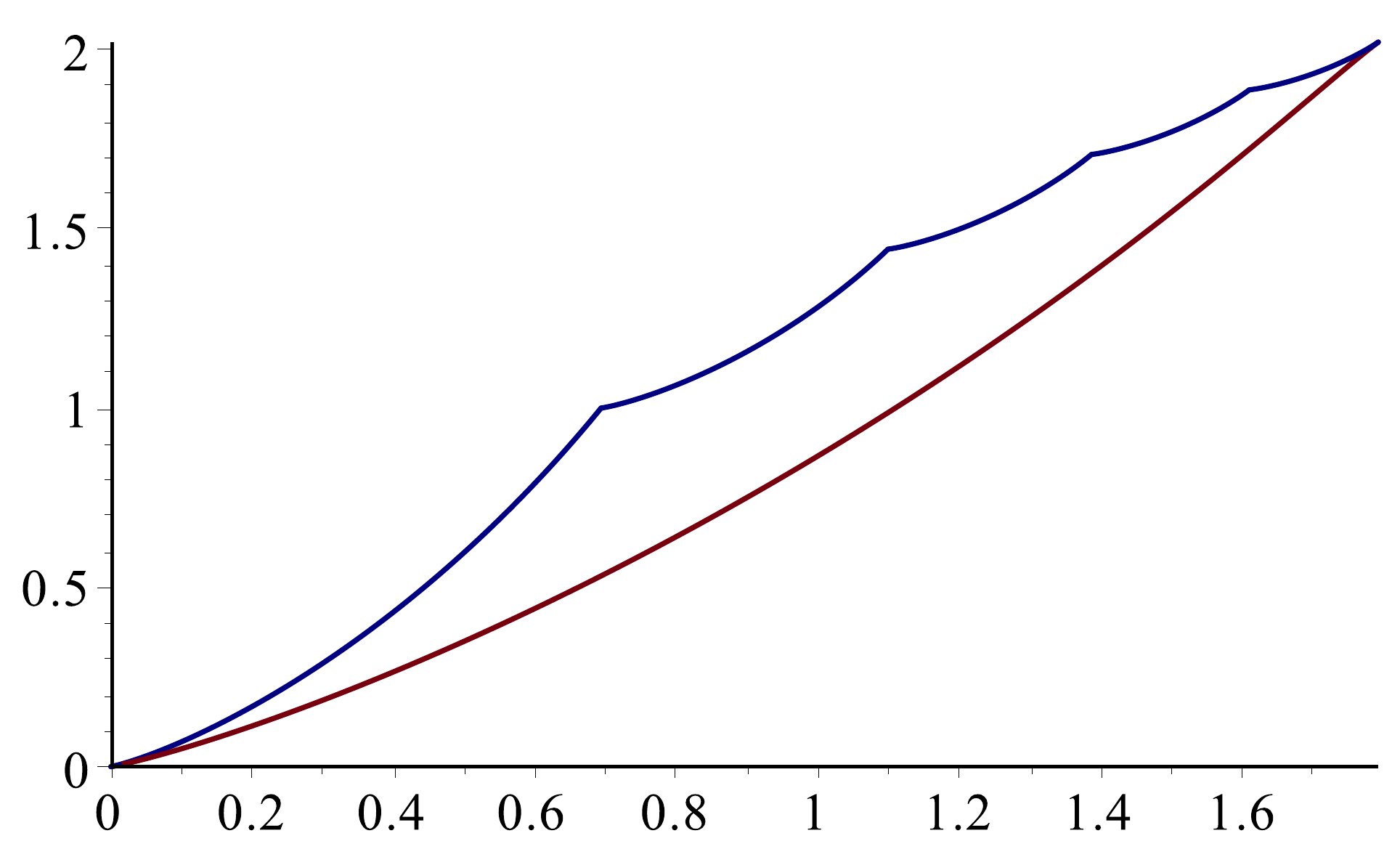}
\put(-5, 25){\rotatebox{90}{$H_{\gamma}( \bvec{p} )$}}
\put(75, -2.5){$H( \bvec{p} )$}
\put(95.5, 1.5){\scriptsize [nats]}
\put(65, 26){\color{burgundy} $\bvec{v}_{n}( \cdot )$}
\put(35, 39){\color{navyblue} $\bvec{w}_{n}( \cdot )$}
\end{overpic}
}\hspace{0.05\hsize}
\subfloat[The case $\gamma = 2$.]{
\begin{overpic}[width = 0.45\hsize, clip]{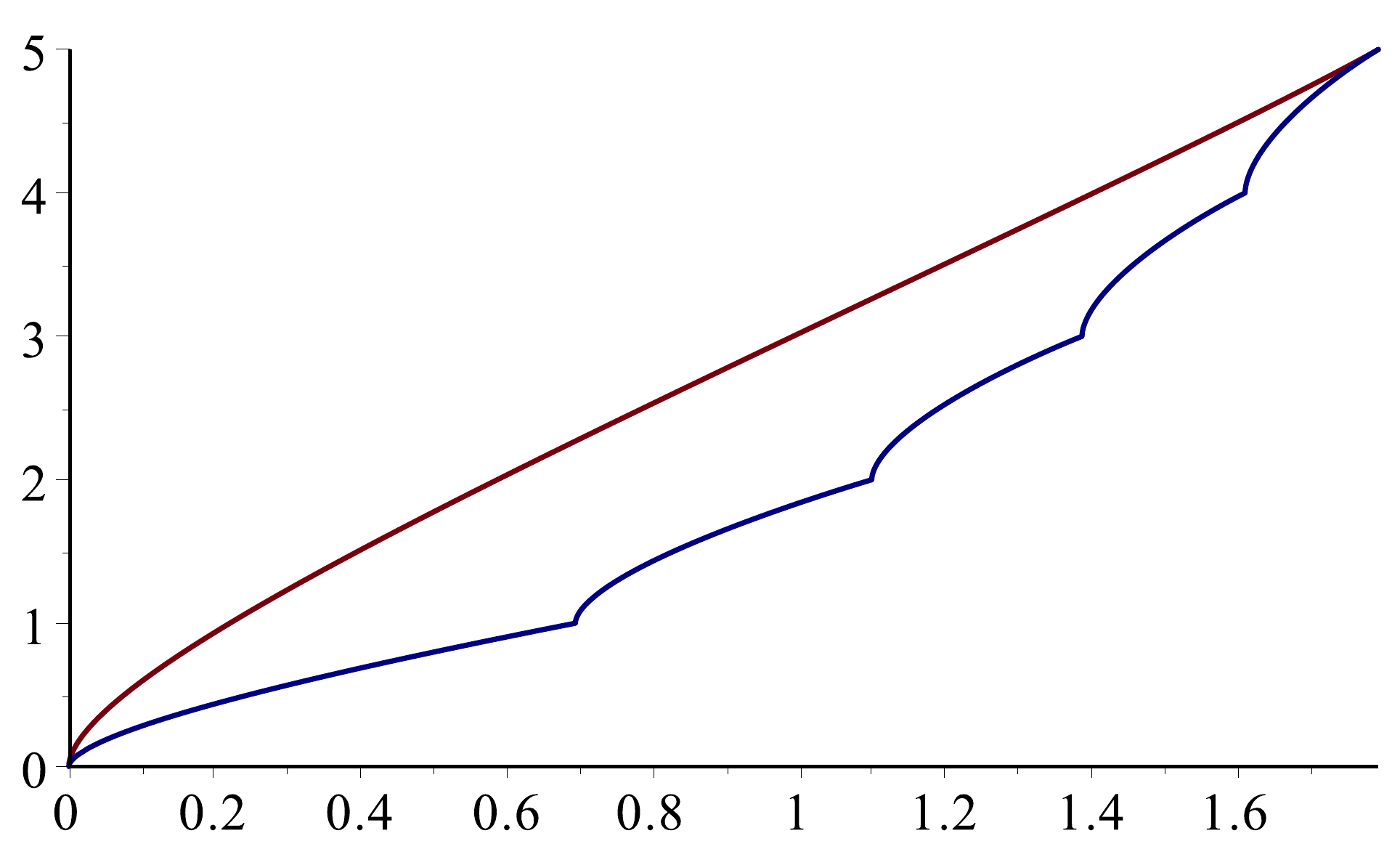}
\put(-5, 25){\rotatebox{90}{$H_{\gamma}( \bvec{p} )$}}
\put(75, -2.5){$H( \bvec{p} )$}
\put(95.5, 1.5){\scriptsize [nats]}
\put(40, 38){\color{burgundy} $\bvec{v}_{n}( \cdot )$}
\put(72, 27){\color{navyblue} $\bvec{w}_{n}( \cdot )$}
\end{overpic}
}
\caption{
Plots of the boundaries of $\{ (H( \bvec{p} ), H_{\gamma}( \bvec{p} )) \mid \bvec{p} \in \mathcal{P}_{n} \}$ with $n = 6$.
If $0 < \alpha < 1$, then the upper- and lower-boundaries correspond to distributions $\bvec{v}_{n}( \cdot )$ and $\bvec{w}_{n}( \cdot )$, respectively.
If $\alpha > 1$, then these correspondences are reversed.}
\end{figure}

\begin{remark}
Harremo\"{e}s and Tops{\o}e \cite{topsoe} showed that the exact region of $\Delta_{n} = \{ (H( \bvec{p} ), IC( \bvec{p} )) \mid \bvec{p} \in \mathcal{P}_{n} \}$ for $n \ge 3$, where $IC( \bvec{p} ) \triangleq \| \bvec{p} \|_{2}^{2}$ denotes the index of coincidence.
Then, we can see that Corollary \ref{cor:extremes} contains its result by $f( x ) = x^{2}$.
\end{remark}

\subsection{Applications for uniformly focusing channels}
\label{subsect:focusing}

In this subsection, we consider applications of Corollary \ref{cor:extremes} for a particular class of discrete memoryless channels (DMCs), i.e., uniformly focusing channels \cite{massey}.
Let the R\'{e}nyi divergence \cite{renyi} of order $\alpha \in (0, 1) \cup (1, \infty)$ is denoted by
\begin{align}
D_{\alpha}(\bvec{p} \; \| \; \bvec{q})
\triangleq
\frac{1}{\alpha - 1} \ln \sum_{i=1}^{n} p_{i}^{\alpha} q_{i}^{1-\alpha} ,
\end{align}
for $\bvec{p}, \bvec{q} \in \mathcal{P}_{n}$.
Since $\lim_{\alpha \to 1} D_{\alpha}(\bvec{p} \, \| \, \bvec{q}) = D(\bvec{p} \, \| \, \bvec{q})$ by L'H\^{o}pital's rule, we write $D_{1}(\bvec{p} \, \| \, \bvec{q}) \triangleq D(\bvec{p} \, \| \, \bvec{q})$, where
\begin{align}
D(\bvec{p} \, \| \, \bvec{q})
\triangleq
\sum_{i=1}^{n} p_{i} \ln \frac{p_{i}}{q_{i}}
\end{align}
denotes the relative entropy.
Since
\begin{align}
D_{\alpha}(\bvec{p} \; \| \; \bvec{u}_{n})
& =
\ln n - H_{\alpha}( \bvec{p} )
\label{eq:RenyiDiv_unif}
%\\
%D(\bvec{p} \ \| \ \bvec{u}_{n})
%& =
%\ln n - H( \bvec{p} ) ,
\end{align}
for $\alpha \in (0, \infty)$, we can obtain Corollary \ref{cor:RenyiDiv} from \eqref{eq:Renyi_bound1} and \eqref{eq:Renyi_bound2}.

\begin{corollary}
\label{cor:RenyiDiv}
If $0 < \alpha < 1$, then
\begin{align}
D_{\alpha}(\bar{\bvec{v}}_{n}( \bvec{p} ) \; \| \; \bvec{u}_{n})
\le
D_{\alpha}(\bvec{p} \; \| \; \bvec{u}_{n})
\le
D_{\alpha}(\bar{\bvec{w}}_{n}( \bvec{p} ) \; \| \; \bvec{u}_{n})
\end{align}
for any $n \ge 2$ and any $\bvec{p} \in \mathcal{P}_{n}$.
Moreover, if $\alpha > 1$, then
\begin{align}
D_{\alpha}(\bar{\bvec{w}}_{n}( \bvec{p} ) \; \| \; \bvec{u}_{n})
\le
D_{\alpha}(\bvec{p} \; \| \; \bvec{u}_{n})
\le
D_{\alpha}(\bar{\bvec{v}}_{n}( \bvec{p} ) \; \| \; \bvec{u}_{n})
\end{align}
for any $n \ge 2$ and any $\bvec{p} \in \mathcal{P}_{n}$.
\end{corollary}

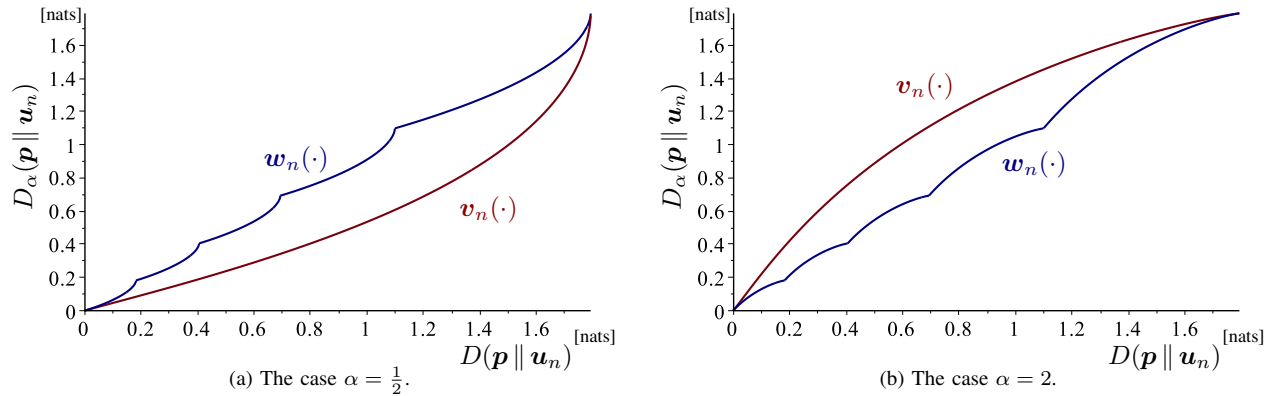
\begin{figure}[!t]
\centering
\subfloat[The case $\alpha = \frac{1}{2}$.]{
\begin{overpic}[width = 0.45\hsize, clip]{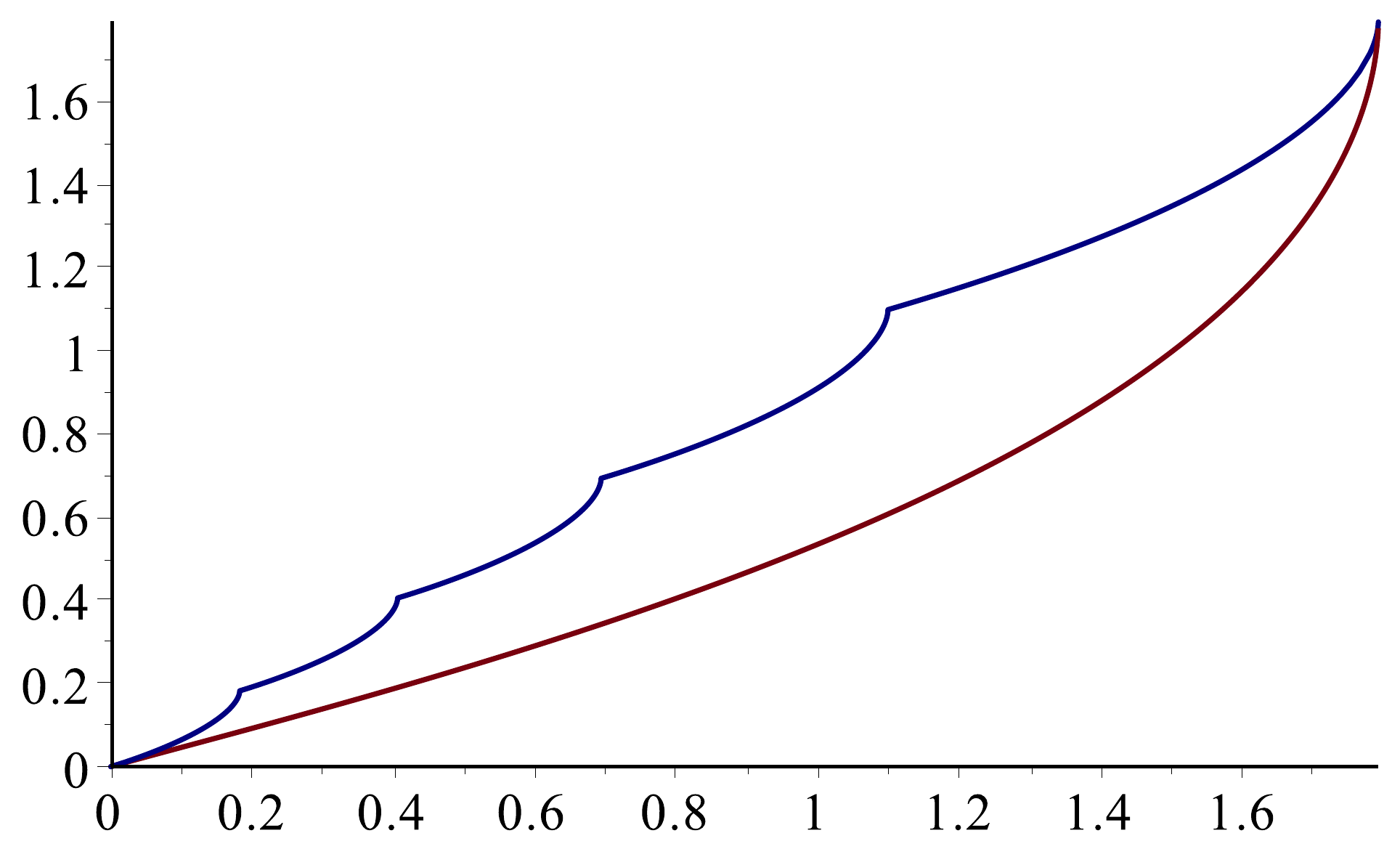}
\put(-5, 25){\rotatebox{90}{$D_{\alpha}( \bvec{p} \, \| \, \bvec{u}_{n} )$}}
\put(0, 59){\scriptsize [nats]}
\put(75, -2.5){$D( \bvec{p} \, \| \, \bvec{u}_{n} )$}
\put(95.5, 1.5){\scriptsize [nats]}
\put(75, 24){\color{burgundy} $\bvec{v}_{n}( \cdot )$}
\put(40, 33){\color{navyblue} $\bvec{w}_{n}( \cdot )$}
\end{overpic}
}\hspace{0.05\hsize}
\subfloat[The case $\alpha = 2$.]{
\begin{overpic}[width = 0.45\hsize, clip]{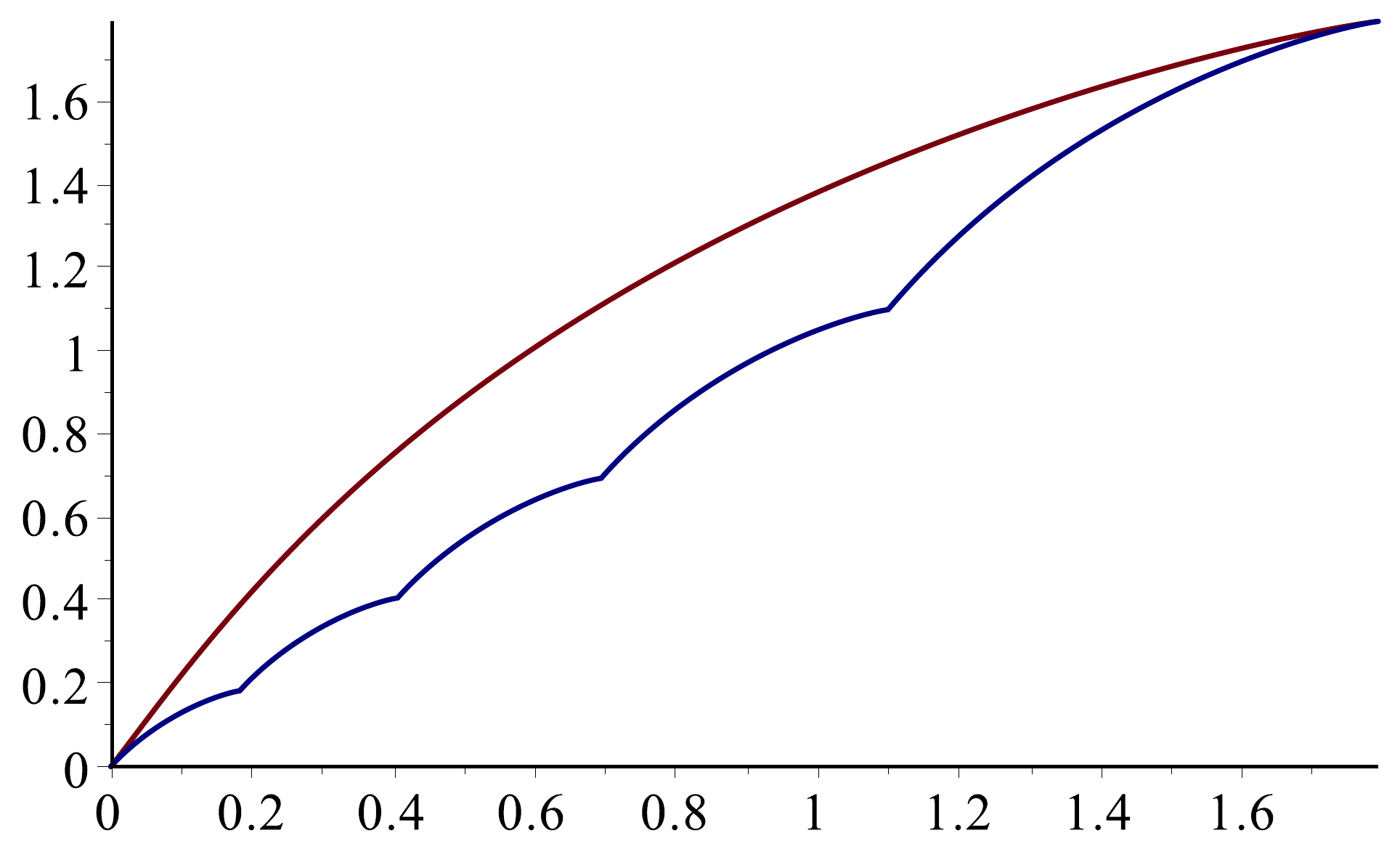}
\put(-5, 25){\rotatebox{90}{$D_{\alpha}( \bvec{p} \, \| \, \bvec{u}_{n} )$}}
\put(0, 59){\scriptsize [nats]}
\put(75, -2.5){$D( \bvec{p} \, \| \, \bvec{u}_{n} )$}
\put(95.5, 1.5){\scriptsize [nats]}
\put(37, 46){\color{burgundy} $\bvec{v}_{n}( \cdot )$}
\put(56, 32){\color{navyblue} $\bvec{w}_{n}( \cdot )$}
\end{overpic}
}
\caption{
Plots of the boundaries of $\{ (D( \bvec{p} \, \| \, \bvec{u}_{n} ), D_{\alpha}( \bvec{p} \, \| \, \bvec{u}_{n} )) \mid \bvec{p} \in \mathcal{P}_{n} \}$ with $n = 6$.
If $0 < \alpha < 1$, then the upper- and lower-boundaries correspond to distributions $\bvec{w}_{n}( \cdot )$ and $\bvec{v}_{n}( \cdot )$, respectively.
If $\alpha > 1$, then these correspondences are reversed.}
\label{fig:RenyiDiv}
\end{figure}

Since $D(\bvec{p} \, \| \, \bvec{u}_{n}) = \ln n - H( \bvec{p} )$, we note that Corollary \ref{cor:RenyiDiv} shows the tight bounds of R\'{e}nyi divergence from a uniform distribution with a fixed relative entropy from a uniform distribution.
Namely, Corollary \ref{cor:RenyiDiv} implies the boundary of
\begin{align}
\{ (D( \bvec{p} \, \| \, \bvec{u}_{n} ), D_{\alpha}( \bvec{p} \, \| \, \bvec{u}_{n} )) \mid \bvec{p} \in \mathcal{P}_{n} \}
\end{align}
for any $n \ge 2$ and any $\alpha \in (0, 1) \cup (1, \infty)$.
We illustrate boundaries of its region in Fig. \ref{fig:RenyiDiv}.

We now define DMCs as follows:
Let the discrete random variables $X \in \mathcal{X}$ and $Y \in \mathcal{Y}$ denote the input and output of a DMC, respectively, where $\mathcal{X}$ and $\mathcal{Y}$ denote the finite input and output alphabets, respectively.
%Let $P_{X}( \cdot )$ denotes the input distribution of a DMC $(X, Y)$. %, i.e., $X \sim P_{X}$.
Let
%\footnote{By convention, we write transition probabilities of a DMC $(X, Y)$ as $W(\cdot \mid \cdot)$ rather than $P_{Y|X}(\cdot \mid \cdot)$ in this paper.}
$P_{Y|X}(y \mid x)$ denote the transition probability of a DMC $(X, Y)$ for $(x, y) \in \mathcal{X} \times \mathcal{Y}$.
Then, we define the following three classes of DMCs.

\begin{definition}%[uniformly dispersive channels]
\label{def:dispersive}
A channel $(X, Y)$ is said to be \emph{uniformly dispersive \cite{massey}} or \emph{uniform from the input \cite{fano2}} if there exists a permutation $\pi_{x} : \mathcal{Y} \to \mathcal{Y}$ for each $x \in \mathcal{X}$ such that $P_{Y|X}(x \mid \pi_{x}( y )) = P_{Y|X}(x^{\prime} \mid \pi_{x^{\prime}}( y ))$ for all $(x, x^{\prime}, y) \in \mathcal{X}^{2} \times \mathcal{Y}$.
\end{definition}

\begin{definition}%[uniformly focusing channels]
\label{def:focusing}
A channel $(X, Y)$ is said to be \emph{uniformly focusing \cite{massey}} or \emph{uniform from the output \cite{fano2}} if there exists a permutation $\pi_{y} : \mathcal{X} \to \mathcal{X}$ for each $y \in \mathcal{Y}$ such that $P_{Y|X}(\pi_{y}( x ) \mid y) = P_{Y|X}(\pi_{y^{\prime}}( x ) \mid y^{\prime})$ for all $(x, y, y^{\prime}) \in \mathcal{X} \times \mathcal{Y}^{2}$.
\end{definition}

\begin{definition}%[strongly symmetric channels]
\label{def:strongly}
A channel is said to be \emph{strongly symmetric \cite{massey}} or \emph{doubly uniform \cite{fano2}} if it is both uniformly dispersive and uniformly focusing.
\end{definition}

For a uniformly dispersive channel $(X, Y)$, it is known that
\begin{align}
H(Y \mid X)
=
H(Y \mid X = x)
%\label{eq:dispersive_condH}
\end{align}
for any $x \in \mathcal{X}$ (see \cite[Eq. (5.18)]{fano2} or \cite[Lemma 4.1]{massey}),
where the conditional Shannon entropy \cite{shannon} of $(X, Y) \sim P_{X|Y} P_{Y}$ is defined by
\begin{align}
H(X \mid Y)
\triangleq
\mathbb{E}[ H( P_{X|Y}( \cdot \mid Y ) ) ]
\end{align}
and $\mathbb{E}[ \cdot ]$ denotes the expected value of the random variable.
Moreover, let the conditional R\'{e}nyi entropy \cite{arimoto} of order $\alpha \in (0, 1) \cup (1, \infty)$ be denoted by
\begin{align}
H_{\alpha}( X \mid Y )
\triangleq
\frac{ \alpha }{ 1 - \alpha } \ln \mathbb{E}[ \| P_{X|Y}( \cdot \mid Y ) \|_{\alpha} ]
\end{align}
for $(X, Y) \sim P_{X|Y} P_{Y}$.
By convention, we write $H_{1}(X \mid Y) \triangleq H(X \mid Y)$.
As with uniformly focusing channels, for uniformly focusing channels, we can provide the following lemma.

\begin{lemma}
\label{lem:focusing}
If a channel $(X, Y)$ is uniformly focusing and the input $X$ follows a uniform distribution, then
\begin{align}
H_{\alpha}(X \mid Y)
=
H_{\alpha}(X \mid Y = y)
%\label{eq:Halpha_focusing}
\end{align}
for any $y \in \mathcal{Y}$ and any $\alpha \in (0, \infty)$.
\end{lemma}

\begin{IEEEproof}[Proof of Lemma \ref{lem:focusing}]
Consider a uniformly focusing channel $(X, Y)$.
Assume that the input $X$ follows a uniform distribution, i.e., $P_{X}( x ) = \frac{1}{|\mathcal{X}|}$ for all $x \in \mathcal{X}$.
Note from \cite[p. 127]{fano2} or \cite[Vol. I, Lemma 4.2]{massey} that, if the input $X$ follows a uniform distribution, then the output $Y$ also follows a uniform distribution, i.e., $P_{Y}( y ) = \frac{1}{|\mathcal{Y}|}$ for all $y \in \mathcal{Y}$.
Then, since the a posteriori probability of $(X, Y)$ is written as
\begin{align}
P_{X|Y}(x \mid y)
=
\frac{ P_{X}( x ) P_{Y|X}(y \mid x) }{ P_{Y}( y ) }
\end{align}
for $(x, y) \in \mathcal{X} \times \mathcal{Y}$ by Bayes' rule and the fraction $\frac{ P_{X}( x ) }{ P_{Y}( y ) }$ is constant for $(x, y) \in \mathcal{X} \times \mathcal{Y}$, it follows from Definition \ref{def:focusing} that there exists a permutation $\pi_{y} : \mathcal{X} \to \mathcal{X}$ for each $y \in \mathcal{Y}$ such that
\begin{align}
P_{X|Y}(\pi_{y}( x ) \mid y)
=
P_{X|Y}(\pi_{y^{\prime}}( x ) \mid y^{\prime})
\label{eq:a_posteriori_pi}
\end{align}
for all $(x, y, y^{\prime}) \in \mathcal{X} \times \mathcal{Y}^{2}$.
Hence, we get
\begin{align}
H(X \mid Y)
& =
\sum_{y \in \mathcal{Y}} P_{Y}( y ) H(X \mid Y = y)
\\
& =
\sum_{y \in \mathcal{Y}} P_{Y}( y ) \left( - \sum_{x \in \mathcal{X}} P_{X|Y}(x \mid y) \ln P_{X|Y}(x \mid y) \right)
\\
& =
\sum_{y \in \mathcal{Y}} P_{Y}( y ) \left( - \sum_{x \in \mathcal{X}} P_{X|Y}(\pi_{y}( x ) \mid y) \ln P_{X|Y}(\pi_{y}( x ) \mid y) \right)
\\
& \overset{\eqref{eq:a_posteriori_pi}}{=}
\left( \sum_{y \in \mathcal{Y}} P_{Y}( y ) \right) \left( - \sum_{x \in \mathcal{X}} P_{X|Y}(\pi_{y^{\prime}}( x ) \mid y^{\prime}) \ln P_{X|Y}(\pi_{y^{\prime}}( x ) \mid y^{\prime}) \right)
\\
& =
- \sum_{x \in \mathcal{X}} P_{X|Y}(\pi_{y^{\prime}}( x ) \mid y^{\prime}) \ln P_{X|Y}(\pi_{y^{\prime}}( x ) \mid y^{\prime})
\\
& =
- \sum_{x \in \mathcal{X}} P_{X|Y}(x \mid y^{\prime}) \ln P_{X|Y}(x \mid y^{\prime})
\\
& =
H(X \mid Y = y^{\prime} )
\label{eq:cond_H_focusing}
\end{align}
for any $y^{\prime} \in \mathcal{Y}$.
Similarly, we also get
\begin{align}
\mathbb{E}[ \| P_{X|Y}(\cdot \mid Y) \|_{\alpha} ]
& =
\sum_{y \in \mathcal{Y}} P_{Y}( y ) \| P_{X|Y}(\cdot \mid y) \|_{\alpha}
\\
& =
\sum_{y \in \mathcal{Y}} P_{Y}( y ) \left( \sum_{x \in \mathcal{X}} P_{X|Y}(x \mid y)^{\alpha} \right)^{\frac{1}{\alpha}}
\\
& =
\sum_{y \in \mathcal{Y}} P_{Y}( y ) \left( \sum_{x \in \mathcal{X}} P_{X|Y}(\pi_{y}( x ) \mid y)^{\alpha} \right)^{\frac{1}{\alpha}}
\\
& \overset{\eqref{eq:a_posteriori_pi}}{=}
\left( \sum_{y \in \mathcal{Y}} P_{Y}( y ) \right) \left( \sum_{x \in \mathcal{X}} P_{X|Y}(\pi_{y^{\prime}}( x ) \mid y^{\prime})^{\alpha} \right)^{\frac{1}{\alpha}}
\\
& =
\left( \sum_{x \in \mathcal{X}} P_{X|Y}(\pi_{y^{\prime}}( x ) \mid y^{\prime})^{\alpha} \right)^{\frac{1}{\alpha}}
\\
& =
\left( \sum_{x \in \mathcal{X}} P_{X|Y}(x \mid y^{\prime})^{\alpha} \right)^{\frac{1}{\alpha}}
\\
& =
\| P_{X|Y}(\cdot \mid y^{\prime}) \|_{\alpha}
\label{eq:cond_N_focusing}
\end{align}
for any $x \in \mathcal{X}$ and any $\alpha \in (0, \infty)$.
Since $H_{\alpha}(X \mid Y) \triangleq \frac{ \alpha }{ 1 - \alpha } \ln \mathbb{E}[ \| P_{X|Y}(\cdot \mid Y) \|_{\alpha} ]$ for $\alpha \in (0, 1) \cup (1, \infty)$ and $H_{1}(X \mid Y) = H(X \mid Y)$, Eqs. \eqref{eq:cond_H_focusing} and \eqref{eq:cond_N_focusing} imply Lemma \ref{lem:focusing}.
\end{IEEEproof}

Therefore, it follows from Lemma \ref{lem:focusing} that the results of Corollary \ref{cor:extremes} can be applied to uniformly focusing channels $(X, Y)$ if the input $X$ follows a uniform distribution, as with \eqref{eq:Renyi_bound1} and \eqref{eq:Renyi_bound2}.
For a channel $(X, Y)$, let the mutual information of order $\alpha \in (0, \infty)$ \cite{arimoto} between $X$ and $Y$ be denoted by
\begin{align}
I_{\alpha}(X; Y)
\triangleq
H_{\alpha}(X) - H_{\alpha}(X \mid Y)
\end{align}
for $\alpha \in (0, \infty)$.
Note that $I_{1}(X; Y) \triangleq I(X; Y)$ denotes the (ordinary) mutual information between $X$ and $Y$.
In this paragraph, we assume that a channel $(X, Y)$ is uniformly focusing and the input $X$ follows a uniform distribution.
Since $H_{\alpha}( \bvec{u}_{n} ) = \ln n$ for $\alpha \in (0, \infty)$, it follows from Lemma \ref{lem:focusing} that
\begin{align}
I_{\alpha}(X; Y)
& =
\ln |\mathcal{X}| - H_{\alpha}(X \mid Y = y)
\\
& \overset{\eqref{eq:RenyiDiv_unif}}{=}
D_{\alpha}(P_{X|Y}(\cdot \mid y) \ \| \ \bvec{u}_{|\mathcal{X}|})
\label{eq:Ialpha_focusing}
\end{align}
for any $y \in \mathcal{Y}$ and any $\alpha \in (0, \infty)$, where $| \cdot |$ denotes the cardinality of the finite set.
Therefore, it follows that the tight bounds of $I_{\alpha}(X; Y)$ with a fixed $I(X; Y)$ are equivalent to the bounds of Corollary \ref{cor:RenyiDiv} under the hypotheses.

Furthermore, we consider Gallager's $E_{0}$ function \cite{gallager} of a channel $(X, Y)$, defined by
\begin{align}
E_{0}(\rho, X, Y)
& =
E_{0}(\rho, P_{X}, P_{Y|X})
\\
& \triangleq
- \ln \sum_{y \in \mathcal{Y}} \! \left( \sum_{x \in \mathcal{X}} P_{X}( x ) P_{Y|X}(y \mid x)^{\frac{1}{1+\rho}} \! \right)^{\!\!1+\rho}
\end{align}
for $\rho \in (-1, \infty)$.
Then, we can obtain the following theorem.

\begin{theorem}
\label{th:E0_focusing}
For a uniformly focusing channel $(X, Y)$, let
\begin{align}
E_{0}^{(\sbvec{v}_{n})}(\rho, X, Y)
& \triangleq
\rho \, D_{\frac{1}{1+\rho}}( \hat{\bvec{v}}_{n}(X \mid Y) \ \| \ \bvec{u}_{n} ) ,
\label{def:Div_vn} \\
E_{0}^{(\sbvec{w}_{n})}(\rho, X, Y)
& \triangleq
\rho \, D_{\frac{1}{1+\rho}}( \hat{\bvec{w}}_{n}(X \mid Y) \ \| \ \bvec{u}_{n} ) ,
\label{def:Div_wn}
\end{align}
where $\hat{\bvec{v}}_{n}(X \mid Y) \triangleq \bvec{v}_{n}( H_{\sbvec{v}_{n}}^{-1}( H(X \mid Y) ) )$, $\hat{\bvec{w}}_{n}(X \mid Y) \triangleq \bvec{w}_{n}( H_{\sbvec{w}_{n}}^{-1}( H(X \mid Y) ) )$, and $n = | \mathcal{X} |$.
If the input $X$ follows a uniform distribution, then we observe that
\begin{align}
E_{0}^{(\sbvec{v}_{n})}(\rho, X, Y)
\le
E_{0}(\rho, X, Y)
\le
E_{0}^{(\sbvec{w}_{n})}(\rho, X, Y)
\label{eq:E0_focusing}
\end{align}
for any $\rho \in (-1, \infty)$.
\end{theorem}

\begin{IEEEproof}[Proof of Theorem \ref{th:E0_focusing}]
We can see from \cite[Eq. (16)]{arimoto} that
\begin{align}
\frac{ E_{0}(\rho, P_{X^{\alpha}}, P_{Y|X}) }{ \rho }
=
I_{\frac{1}{1+\rho}}(X; Y) ,
\label{eq:E0_Ialpha}
\end{align}
where
\begin{align}
P_{X^{\alpha}}( x )
\triangleq
\frac{ P_{X}( x )^{\alpha} }{ \sum_{x^{\prime} \in \mathcal{X}} P_{X}( x^{\prime} )^{\alpha} }
\end{align}
denotes the tilted distribution.
We can see from \eqref{eq:E0_Ialpha} that the $E_{0}$ function is closely related to the mutual information of order $\alpha$.
Note that, if the distribution $P_{X}$ is a uniform distribution, then its tilted distribution $P_{X^{\alpha}}$ is also a uniform distribution for any $\alpha \in (0, \infty)$.
Thus, if a channel $(X, Y)$ is uniformly focusing and the input $X$ follows a uniform distribution, then it follows from \eqref{eq:Ialpha_focusing} and \eqref{eq:E0_Ialpha} that
\begin{align}
E_{0}(\rho, X, Y)
=
\rho \, D_{\frac{1}{1+\rho}}( P_{X|Y}(\cdot \mid y) \ \| \ \bvec{u}_{|\mathcal{X}|} )
\end{align}
for any $\rho \in (-1, \infty)$ and any $y \in \mathcal{Y}$.
Hence, noting the relations
\begin{align}
-1 < \rho < 0
& \iff
1 < \alpha < \infty ,
\\
0 < \rho < \infty
& \iff
0 < \alpha < 1 ,
\end{align}
the $E_{0}$ function can also be evaluate as with Corollary \ref{cor:RenyiDiv}.
\end{IEEEproof}

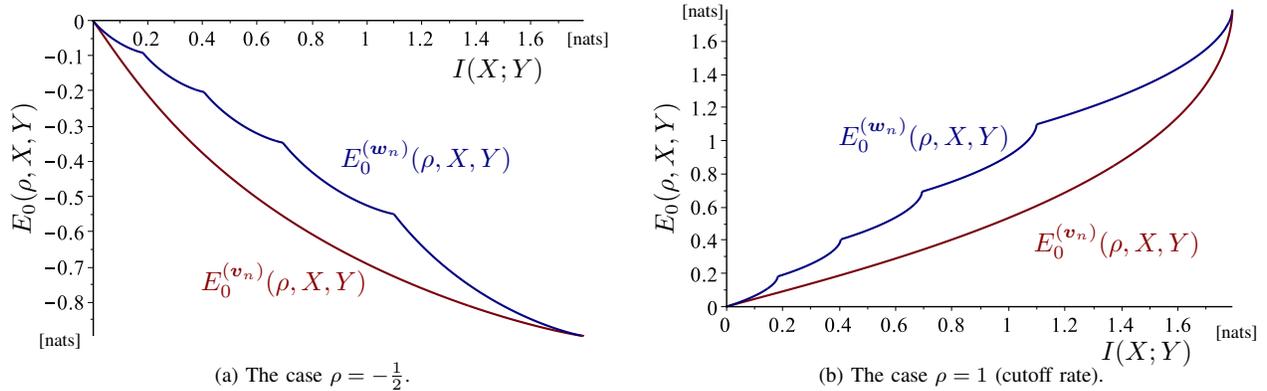
\begin{figure}[!t]
\centering
\subfloat[The case $\rho = - \frac{1}{2}$.]{
\begin{overpic}[width = 0.45\hsize, clip]{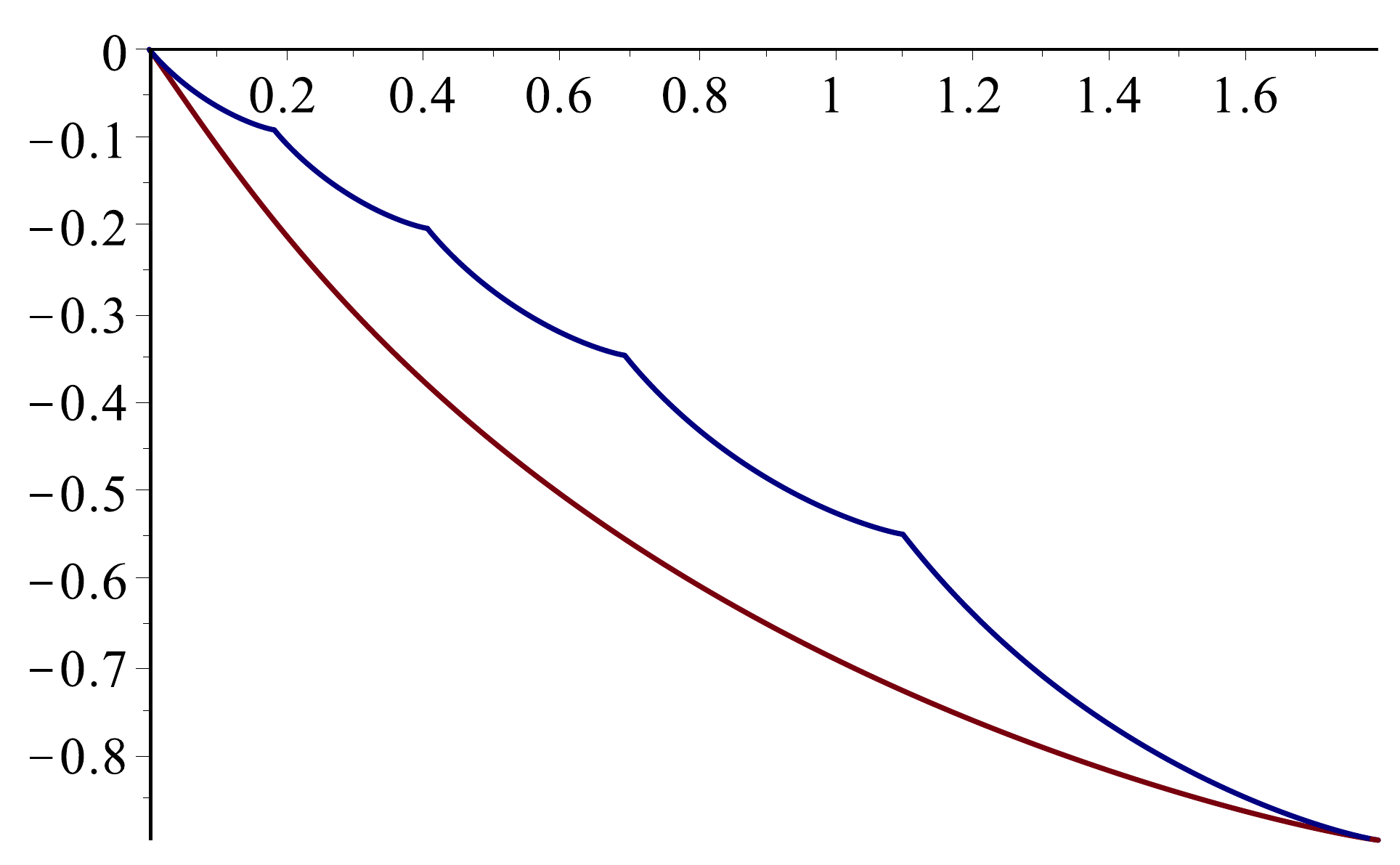}
\put(-4, 20){\rotatebox{90}{$E_{0}(\rho, X, Y)$}}
\put(1, 0){\scriptsize [nats]}
\put(75, 48){$I(X; Y)$}
\put(95.5, 54){\scriptsize [nats]}
\put(30, 10){\color{burgundy} $E_{0}^{(\sbvec{v}_{n})}(\rho, X, Y)$}
\put(55, 32){\color{navyblue} $E_{0}^{(\sbvec{w}_{n})}(\rho, X, Y)$}
\end{overpic}
}\hspace{0.05\hsize}
\subfloat[The case $\rho = 1$ (cutoff rate).]{
\begin{overpic}[width = 0.45\hsize, clip]{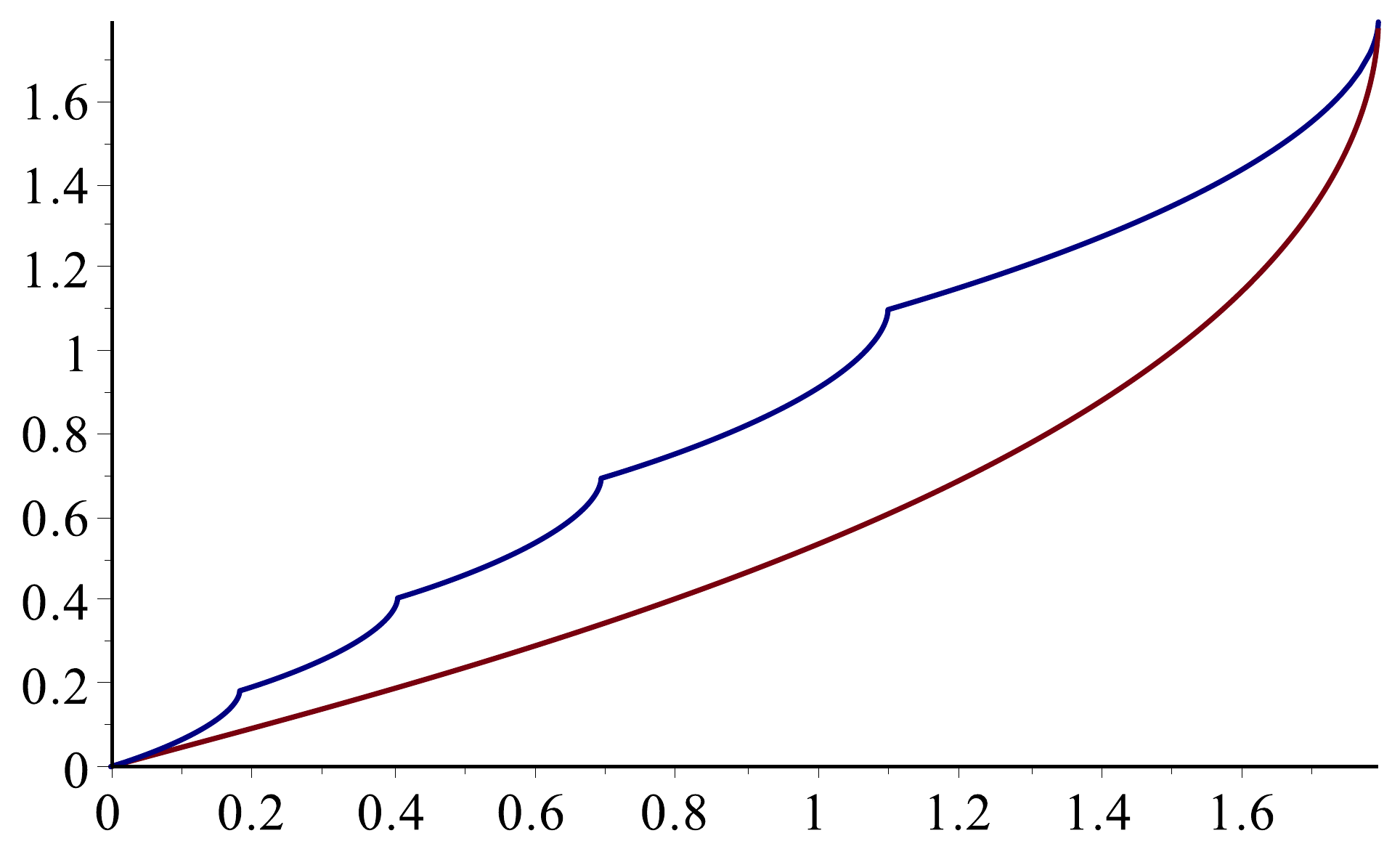}
\put(-5, 20){\rotatebox{90}{$E_{0}(\rho, X, Y)$}}
\put(0, 59){\scriptsize [nats]}
\put(75, -2.5){$I(X; Y)$}
\put(95.5, 1.5){\scriptsize [nats]}
\put(63, 17){\color{burgundy} $E_{0}^{(\sbvec{v}_{n})}(\rho, X, Y)$}
\put(28, 36){\color{navyblue} $E_{0}^{(\sbvec{w}_{n})}(\rho, X, Y)$}
\end{overpic}
}
\caption{
Plots of the bounds between $I(X; Y)$ and $E_{0}(\rho, X, Y)$ for all uniformly focusing channels $(X, Y)$ with $|\mathcal{X}| = 6$ and a uniform input distribution $P_{X}$.
The upper and lower bounds of $E_{0}(\rho, X, Y)$ with a fixed $I(X; Y)$ correspond to $E_{0}^{(\sbvec{w}_{n})}(\rho, X, Y)$ and $E_{0}^{(\sbvec{v}_{n})}(\rho, X, Y)$, respectively.}
\label{fig:E0_focusing}
\end{figure}

Note that the distributions $\hat{\bvec{v}}_{n}(X \mid Y)$ and $\hat{\bvec{w}}_{n}(X \mid Y)$ denote $\bvec{v}_{n}( p )$ and $\bvec{w}_{n}( q )$, respectively, such that $H_{\sbvec{v}_{n}}( p ) = H_{\sbvec{w}_{n}}( q ) = H(X \mid Y)$ for a given channel $(X, Y)$.
Since $I(X; Y) = \ln |\mathcal{X}| - H(X \mid Y)$ under a uniform input distribution, Theorem \ref{th:E0_focusing} shows bounds of the $E_{0}$ function with a fixed mutual information.
Note that, since \eqref{def:Div_vn} and \eqref{def:Div_wn} are defined by the $\bvec{v}_{n}(\cdot)$ and $\bvec{w}_{n}(\cdot)$, respectively, there exist two strongly symmetric channels which attain each equality of the bounds \eqref{eq:E0_focusing}.
%More precisely, we can create the extremal channels as follows:
%If a strongly symmetric channel $(X, Y)$ is defined to satisfy $P_{Y|X}(\cdot \mid x)_{\downarrow} = \bvec{v}_{|\mathcal{X}|}( p )$ for all $x \in \mathcal{X}$ and some $p \in [0, |\mathcal{X}|^{-1}]$, then it follows that
%$
%E_{0}(\rho, X, Y)
%=
%E_{0}^{(\sbvec{v}_{n})}(\rho, X, Y)
%$.
%Similarly, if a strongly symmetric channel $(X, Y)$ is defined to satisfy $P_{Y|X}(\cdot \mid x)_{\downarrow} = \bvec{w}_{|\mathcal{X}|}( p )$ for all $x \in \mathcal{X}$ and some $p \in [|\mathcal{X}|^{-1}, (|\mathcal{X}| - 1)^{-1}]$, then it follows that
%$
%E_{0}(\rho, X, Y)
%=
%E_{0}^{(\sbvec{w}_{n})}(\rho, X, Y)
%$.
Namely, Theorem \ref{th:E0_focusing} provides tight bounds \eqref{eq:E0_focusing}. %of the $E_{0}$ function for uniformly focusing channels with a fixed mutual information under a uniform input distribution.
We illustrate graphical representations of Theorem \ref{th:E0_focusing} in Fig. \ref{fig:E0_focusing}, as with Figs. \ref{fig:region_P6_half} and \ref{fig:Renyi}.
Theorem \ref{th:E0_focusing} is a generalization of \cite[Theorem 2]{isit2015} from ternary-input strongly symmetric channels to $n$-ary input uniformly focusing channels under a uniform input distribution.

Finally, we consider the hypothesis of a uniform input distribution.
If a channel $(X, Y)$ is symmetric%
\footnote{Symmetric channels are defined in \cite[p. 94]{gallager}.},
then the mutual information of order $\alpha$ is maximized by a uniform input distribution%
\footnote{This fact can be verified by using, e.g., \cite[Theorem 7.2]{jelinek}.} %, i.e.,
%$
%\argmax_{P_{X}} I_{\alpha}(X; Y)
%=
%\bvec{u}_{|\mathcal{X}|}
%\label{eq:symmetric_maximize}
%$
for $\alpha \in (0, \infty)$.
Therefore, since a strongly symmetric channel is symmetric, the hypothesis is optimal if the channel $(X, Y)$ is strongly symmetric.

\section{Conclusion}
\label{sect:conclusion}

In this study, we established the tight bounds of the $\ell_{\alpha}$-norm with a fixed Shannon entropy in Theorem \ref{th:extremes}, and vise versa in Theorem \ref{th:extremes2}.
Previously, the tight bounds of the Shannon entropy with a fixed error probability were derived \cite{fano, kovalevsky, tebbe, feder, verdu, ben-bassat}.
Since the error probability is closely related to the $\ell_{\infty}$-norm, this study is a generalization of previous studies \cite{fano, kovalevsky, tebbe, feder, verdu, ben-bassat}.
Note that the set of all $n$-ary probability vectors, which are sorted in decreasing order, with a fixed $\ell_{\alpha}$-norm is convex set.
The previous works \cite{fano, kovalevsky, tebbe, feder, verdu, ben-bassat} used the concavity of the Shannon entropy in probability vectors to examine the Shannon entropy with a fixed $\ell_{\alpha}$-norm.
However, since
%\footnote{These convexity and concavity can be derived by Minkowski's inequality.
%This proof is cited in, e.g., \cite[p. 143]{boekee}.}
$\| \bvec{p} \|_{\alpha}$ is strictly concave in $\bvec{p} \in \mathcal{P}_{n}$ when $\alpha \in (0, 1)$ and is strictly convex in $\bvec{p} \in \mathcal{P}_{n}$ when $\alpha \in (1, \infty)$, the concavity of the Shannon entropy in probability vectors turns out to be hard-to-use when the $\ell_{\alpha}$-norm is fixed.
In this study, we derived Theorems \ref{th:extremes} and \ref{th:extremes2} by using elementary calculus without using the concavity of the Shannon entropy.
%Since Theorems \ref{th:extremes} and \ref{th:extremes2} imply the boundary of $\mathcal{R}_{n}( \alpha ) = \{ (H( \bvec{p} ), \| \bvec{p} \|_{\alpha}) \mid \bvec{p} \in \mathcal{P}_{n} \}$, we illustrated the boundaries in Fig. \ref{fig:region_P6_half}.

\if0
As application, we extend the bounds of Theorem \ref{th:extremes} from the $\ell_{\alpha}$-norm to several information measures, which are determined by the $\ell_{\alpha}$-norm, in Corollary \ref{cor:extremes}.
As instances, we showed some applications of Corollary \ref{cor:extremes} in Table \ref{table:extremes};
in particular, we illustrated the boundary of $\mathcal{R}_{n}^{\text{R\'{e}nyi}}( \alpha )$ in Fig. \ref{fig:Renyi}.
In addition, we can apply Corollary \ref{cor:extremes} to several diversity indices, such as the index of coincidence.
Moreover, we presented further applications of Corollary \ref{cor:extremes} to uniformly focusing channels, defined in Definition \ref{def:focusing}, in Section \ref{subsect:focusing}.
%Then, we showed tight bounds of (i) R\'{e}nyi divergence from a uniform distribution in Corollary \ref{cor:RenyiDiv}, (ii) the mutual information of order $\alpha$ under a uniform input distribution, and (iii) Gallager's $E_{0}$ function under a uniform input distribution in Theorem \ref{th:E0_focusing}.
\fi

\section*{Acknowledgment}

This study was partially supported by the Ministry of Education, Science, Sports and Culture, Grant-in-Aid for Scientific Research (C) 26420352.




%\bibliographystyle{IEEEtran}

\begin{thebibliography}{99}

\bibitem{shannon}
C. E. Shannon,
``A mathematical theory of communication,''
\emph{Bell Syst. Tech. J.},
vol. 27, pp. 379--423 and 623--656, July and Oct. 1948.

\bibitem{fano}
R. M. Fano,
``Class notes for Transmission of Information,''
Course 6.574, MIT, Cambridge, MA, 1952.

\bibitem{kovalevsky}
V. A. Kovalevsky,
``The problem of character recognition from the point of view of mathematical statistics,''
\emph{Character Readers and Pattern Recognition.}
New York: Spartan, pp. 3--30, 1968. (Russian edition in 1965).

\bibitem{tebbe}
D. L. Tebbe and S. J. Dwyer III,
``Uncertainty and probability of error,''
\emph{IEEE Trans. Inf. Theory},
vol. 14, no. 3, pp. 516--518, May 1968.

\bibitem{feder}
M. Feder and N. Merhav,
``Relations between entropy and error probability,''
\emph{IEEE Trans. Inf. Theory},
vol. 40, no. 1, pp. 259--266, Jan. 1994.

\bibitem{verdu}
S.-W. Ho and S. Verd\'{u},
``On the interplay between conditional entropy and error probability,''
\emph{IEEE Trans. Inf. Theory}
vol. 56, no. 12, pp. 5930--5942, Dec. 2010.

\bibitem{marshall}
A. W. Marshall and I. Olkin,
\emph{Inequalities: Theory of Majorization and Its Applications.}
New York: Academic, 1979.

\bibitem{topsoe}
P. Harremo\"{e}s and F. Tops{\o}e,
``Inequalities between entropy and index of coincidence derived from information diagrams,''
\emph{IEEE Trans. Inf. Theory},
vol. 47, no. 7, pp. 2944--2960, Nov. 2001.

\bibitem{renyi}
A. R\'{e}nyi,
``On measures of information and entropy,''
\emph{Proc. 4th Berkeley Symp. Math. Statist. Prob.},
Berkeley, Calif., vol. 1, Univ. of Calif. Press, pp. 547--561, 1961.

\bibitem{tsallis2}
C. Tsallis,
``Possible generalization of Boltzmann-Gibbs statistics,''
\emph{J. Statist. Phys.},
vol. 52, no. 1--2, pp. 479--487, 1988.

\bibitem{havrda}
J. Havrda and F. Charv\'{a}t,
``Quantification method of classification processes. Concept of structural $a$-entropy,''
\emph{Kybernetika},
vol. 3, no. 1, pp. 30--35, 1967.

\bibitem{daroczy}
Z. Dar\'{o}czy,
``Generalized information functions,''
\emph{Inf. Control},
vol. 16, no. 1, pp. 36--51, Mar. 1970.

\bibitem{behara}
M. Behara and J. S. Chawla,
``Generalized $\gamma$-entropy,''
\emph{Entropy and Ergodic Theory: Selecta Statistica Canadiana.}
vol. 2, pp. 15--38, 1974.

\bibitem{boekee}
D. E. Boekee and J. C. A. Van der Lubbe,
``The $R$-norm information measure,''
\emph{Inf. Control},
vol. 45, no. 2, pp. 136--155, May 1980.

\bibitem{massey}
J. L. Massey,
\emph{Applied digital information theory I and II.}
Lecture notes, Signal and Information Processing Laboratory, ETH Zurich, 1995--1996.
[Online]. Available at \url{http://www.isiweb.ee.ethz.ch/archive/massey_scr/}.

\bibitem{fano2}
R. M. Fano,
\emph{Transmission of Information: A Statistical Theory of Communications.}
New York: M.I.T. Press, 1961.

\bibitem{arimoto}
S. Arimoto,
``Information measures and capacity of order $\alpha$ for discrete memoryless channels,''
in \emph{Topics in Information Theory, 2nd Colloq. Math. Soc. J. Bolyai},
Keszthely, Hungary, vol. 16, pp. 41--52, 1977.

\bibitem{gallager}
R. G. Gallager,
\emph{Information Theory and Reliable Communication.}
New York: Wiley, 1968.

\bibitem{tsallis}
C. Tsallis,
``What are the numbers that experiments provide?''
\emph{Qu\'{i}mica Nova},
vol. 17, no. 6, pp. 468--471, 1994.

\bibitem{fabregas}
A. Guill\'{e}n i F\`{a}bregas, I. Land, and A. Martinez,
``Extremes of error exponents,''
\emph{IEEE Trans. Inf. Theory},
vol. 59, no. 4, pp. 2201--2207, Apr. 2013.

%\bibitem{teixeira}
%A. Teixeira, A. Matos, and L. Antunes,
%``Conditional R\'{e}nyi Entropies,''
%\emph{IEEE Trans. Inf. Theory},
%vol. 58, no. 7, pp. 4273--4277, July 2012.

\bibitem{ben-bassat}
M. Ben-Bassat,
``$f$-entropies, probability of error, and feature selection,''
\emph{Inf. Control},
vol. 39, no. 3, pp. 227--242, Dec. 1978.

%\bibitem{cicalese}
%F. Cicalese and U. Vaccaro,
%``Bounding the average length of optimal source codes via majorization theory,''
%\emph{IEEE Trans. Inf. Theory},
%vol. 50, no. 4, pp. 633--637, Apr. 2004.

%\bibitem{sason}
%I. Sason and S. Verd\'{u},
%``Bounds among $f$-divergences,''
%Dec. 2015. [Online]. Available at
%\url{http://arxiv.org/abs/1508.00335v3}.

%\bibitem{alsan}
%M. Alsan,
%``Extremality for Gallager's reliability function $E_{0}$,''
%\emph{IEEE Trans. Inf. Theory},
%vol. 61, no. 8, pp. 4277--4292, Aug. 2015.

\bibitem{isit2015}
Y. Sakai and K. Iwata,
``Feasible regions of symmetric capacity and Gallager's $E_{0}$ function for ternary-input discrete memoryless channels,''
\emph{Proc. IEEE Int. Symp. Inf. Theory} (ISIT'2015),
Hong Kong, pp. 81--85, June 2015.

%\bibitem{wolfowitz}
%J. Wolfowitz,
%\emph{Coding Theorems of Information Theory.} 3rd ed.,
%New York: Springer-Verlag, 1978.

%\bibitem{hardy}
%G. H. Hardy, J. E. Littlewood, and G. P\'{o}lya,
%\emph{Inequalities},
%2nd. ed., Cambridge: Cambridge Univ. Press, 1988.

%\bibitem{rockafellar}
%R. T. Rockafellar,
%\emph{Convex Analysis.}
%Princeton, NJ: Princeton Univ. Press, 1970.

\bibitem{jelinek}
F. Jelinek
\emph{Probabilistic Information Theory: Discrete and Memoryless Models.}
New York: McGraw-Hill, 1968.

\bibitem{part2}
Y. Sakai and K. Iwata,
``Relations between conditional Shannon entropy and expectation of $\ell_{\alpha}$-norm,''
submitted to \emph{IEEE Int. Symp. Inf. Theory}, (ISIT'2016)
Barcelona, Spain, July 2016.


\end{thebibliography}

\end{document}